\documentclass[aps,prd,twocolumn,epsf,nofootinbib,superscriptaddress]{revtex4}

\usepackage{graphicx}
\usepackage{amsmath,amssymb,latexsym}
\usepackage{bm}

\def\lsim{\mathrel{\raise.3ex\hbox{$<$\kern-.75em\lower1ex\hbox{$\sim$}}}}
\def\gsim{\mathrel{\raise.3ex\hbox{$>$\kern-.75em\lower1ex\hbox{$\sim$}}}}

\def\lbldef#1#2{\expandafter\gdef\csname #1\endcsname {#2}}

\def\href#1#2{#2}

\newcommand{\bwide}{\begin{widetext}}
\newcommand{\ewide}{\end{widetext}}
\newcommand{\beq}[1]{\begin{equation} \label{(#1)}}
\newcommand{\eeq}{\end{equation}}
\newcommand{\ba}[1]{\begin{eqnarray} \label{(#1)}}
\newcommand{\ea}{\end{eqnarray}}

\begin{document}
\hspace*{130mm}{\large \tt FERMILAB-PUB-09-470-A}

\title{On The Heavy Chemical Composition of the Ultra-High Energy Cosmic Rays}

\author{Dan Hooper}
\affiliation{Fermi National Accelerator Laboratory, 
             Theoretical Astrophysics, Batavia, IL 60510, USA}
\affiliation{University of Chicago, Department of Astronomy and Astrophysics, Chicago, IL  60637, USA} 


\author{Andrew  M. Taylor}
\affiliation{Max-Planck-Institut f\"ur Kernphysik, 
             Postfach 103980, D-69029 Heidelberg, GERMANY}
\affiliation{ISDC, Chemin d'Ecogia 16, Versoix, CH-1290, SWITZERLAND}


\begin{abstract}

The Pierre Auger Observatory's (PAO) shower profile measurements can be used to constrain the chemical composition of the ultra-high energy cosmic ray (UHECR) spectrum. In particular, the PAO's measurements of the average depth of shower maximum and the fluctuations of the depth of shower maximum indicate that the cosmic ray spectrum is dominated by a fairly narrow distribution (in charge) of heavy or intermediate mass nuclei at the highest measured energies ($E \gsim 10^{19}$ eV), and contains mostly lighter nuclei or protons at lower energies ($E \sim 10^{18}$ eV). In this article, we study the propagation of UHECR nuclei with the goal of using these measurements, along with those of the shape of the spectrum, to constrain the chemical composition of the particles accelerated by the sources of the UHECRs. We find that with modest intergalactic magnetic fields, 0.3~nG in strength with 1~Mpc coherent lengths, good fits to the combined PAO data can be found for the case in which the sources accelerate primarily intermediate mass nuclei (such as nitrogen or silicon). Without intergalactic magnetic fields, we do not find any composition scenarios that can accommodate the PAO data. For a spectrum dominated by heavy or intermediate mass nuclei, the Galactic (and intergalactic) magnetic fields are expected to erase any significant angular correlation between the sources and arrival directions of UHECRs.

\end{abstract}


\maketitle

\section{Introduction}

The chemical composition of the ultra-high energy cosmic ray (UHECR) spectrum has long been a topic of great interest~\cite{Stecker:1969fw,Puget:1976nz,Stecker:1998ib,us,debate,magnetic}. Until recently, however, very little was known about the nature of these particles. On one side of the debate, the so called Hillas criterion~\cite{hillas} gives a preference for the electromagnetic fields of cosmic ray sources to accelerate heavy nuclei to higher energies than protons or light nuclei. On the other side, it has been argued that the angular correlations reported by the Pierre Auger Observatory (PAO)~\cite{anisotropy}, as well as features in the shape of the UHECR spectrum~\cite{berezinsky}, suggest that these particles consist largely of protons. None of these arguments, however, has yet settled the question of what types of particles make up the UHECR spectrum.

Data from the Pierre Auger Observatory (PAO), however, is offering increasingly powerful insights into this question. Firstly, the spectral shape predicted for the UHECR all-particle spectrum depends not only on the injected spectrum and spatial distribution of the sources, but also on the chemical composition that is injected from the sources of the highest energy cosmic rays. As the PAO measures the UHECR spectrum with increasing precision~\cite{spectrum}, this information can be used to constrain the chemical composition of these particles~\cite{spectrumcomposition}. Furthermore, the PAO is capable of performing several measurements that can be used to directly or indirectly determine the chemical composition of UHECRs as they enter the Earth's atmosphere. Among these empirical tools are the measurements of the average depth of shower maximum, $\langle X_{\rm max} \rangle$, and the RMS variation of this quantity, $\mathrm{RMS}(X_{\rm max})$. On average, proton-induced showers reach their maximum development, $\langle X_{\rm max} \rangle$, deeper in the Earth's atmosphere than do showers of the same energy generated by heavier nuclei. Accompanying this result, the shower to shower fluctuation of $X_{\rm max}$ about the mean, $\mathrm{RMS}(X_{\rm max})$, is larger for proton-induced showers than for iron-induced showers of the same energy. As a result, measurements of both $\langle X_{\rm max} \rangle$ and $\mathrm{RMS}(X_{\rm max})$ can be used to infer the average chemical composition of the UHECRs as a function energy.

Very recently, the PAO collaboration has announced their first measurements of $\mathrm{RMS}(X_{\rm max})$~\cite{xmax,ungerSocor}. These measurements, along with those of $\langle X_{\rm max} \rangle$, imply that the UHECR spectrum contains a large fraction of heavy or intermediate mass nuclei, especially at the highest energies measured. Furthermore, the small values of $\mathrm{RMS}(X_{\rm max})$ measured by the PAO also imply that the composition of the UHECR spectrum is relatively narrowly distributed at the highest measured energies, containing little or no protons or light nuclei. In this way, the new $\mathrm{RMS}(X_{\rm max})$ measurements not only confirm and reinforce the conclusions drawn from earlier average depth of shower maximum measurements, but also provide complementary information that enables one to constrain the distribution of the various chemical species present within the UHECR spectrum. 

The remainder of this article is structured as follows. In Sec.~\ref{prop} we review the physics of UHECR nuclei propagation. In Sec.~\ref{results}, we calculate the spectrum and chemical composition of the UHECR spectrum for various choices of the injected chemical species and compare these results to measurements of the PAO, neglecting the effects of intergalactic magnetic fields. In Sec.~\ref{resultsb}, we include magnetic fields in our calculation, and show that fields with stengths and coherent scales $\sim$($B$/0.3~nG)$\times$($L_{\rm coh}$/1~Mpc)$^{1/2}$, are strongly preferred to accomodate the PAO's data. The best fits found are for cases in which the sources of the UHECRs inject mostly intermediate mass nuclei, such as nitrogen or silicon. Finally, in Sec.~\ref{conclusions}, we discuss the implications of our results and summarize our conclusions.


\section{The Propagation of Ultra-High Energy Cosmic Ray Nuclei}
\label{prop}

For cosmic rays in the form of ultra-high energy (UHE) protons, the dominant processes effecting their propagation over cosmological distances are their interactions with the cosmic microwave background (CMB) or cosmic infrared background (CIB), producing either pions or electron-positron pairs. Pair production ($p +\gamma \rightarrow p + e^+ + e^-$) dominates for energies below $10^{19.6}$ eV, and occurs with sufficiently small inelasticity that it can be treated as a continuous energy loss process \cite{Blumenthal}. Pion production, by contrast, occurs with large inelasticity, with individual occurrences of the processes $p + \gamma_{\rm CMB} \rightarrow p +\pi^0$ or $p + \gamma_{\rm CMB} \rightarrow n + \pi^+$ causing the primary proton to lose a considerable fraction of its energy. Thus pion production must be treated as a stocastic process, and following its onset leads to the catastrophic attenuation of the UHECR spectrum above $\sim 10^{19.6}$ eV known as the GZK cutoff~\cite{gzk}. Furthermore, if enough center-of-mass energy is available, multi-pion production can also become important.


The propagation of UHE cosmic ray nuclei is somewhat more complicated. In addition to energy losses from pair-production, cosmic ray nuclei undergo photodisintegration in scattering off the CMB
and/or CIB at a rate given by:
\begin{equation}
R_{A, Z, i_p, i_n} = \frac{A^2 m^2_p}{2 E^2} \int^{\infty}_{0} 
\frac{d \epsilon\, n (\epsilon)}{\epsilon^2} \int^{2 E \epsilon/A m_p}_{0} 
d \epsilon^\prime \epsilon^\prime \sigma_{A, Z, i_p, i_n} (\epsilon^\prime),
\end{equation}
where $A$ and $Z$ are the atomic number and charge of the nucleus,
$i_p$ and $i_n$ are the numbers of protons and neutrons broken off
from a nucleus in the interaction, $n (\epsilon)$ is the density of
background photons of energy $\epsilon$, and $\sigma_{A, Z, i_p,
i_n}(\epsilon^\prime)$ is the appropriate cross-section~\cite{Stecker:1969fw,Puget:1976nz,Stecker:1998ib,Khan:2004nd}. In our calculations, we use the infrared background model of Ref.~\cite{ms}, although other models~\cite{other} yield similar results. As a result of the process of photodisintegration, the chemical composition that reaches Earth can be significantly different than the composition that is initially injected from the sources of UHECRs.

In previous work, we have described our Monte Carlo code which models the propagation of UHE cosmic ray nuclei over cosmological distances~(see Ref.~\cite{us}). In this article, we use this code to determine the spectrum and composition of UHECRs that is predicted to reach Earth, and compare this to the measurments of the PAO for a variety of injected chemical compositions.

\section{Results In The Case of Negligible Magnetic Fields}
\label{results}

To begin, we consider the case in which only iron nuclei are accelerated by the sources of UHECRs, and assume that the deflection of UHECRs by intergalactic magnetic fields is of negligible importance. We adopt a homogeneous distribution of sources, and a simple cosmic ray injection spectrum of the form:
\begin{equation}
\frac{d N}{d E} \propto E^{-\alpha} e^{-E/E_{\rm max}}.
\end{equation}
It is important to recognize that this describes the spectrum that is produced by the sources of the UHECRs, which can be very different from the spectrum (and composition) that exists after the effects of propagation are taken into account. As a function of the spectral slope injected, $\alpha$, and the energy cutoff, $E_{\rm max}$, we can calculate the resulting spectrum and composition predicted to be observed at Earth, and compare this to the measurements of the PAO.


\begin{figure*}[!]
\begin{center}
{\includegraphics[angle=0,width=0.32\linewidth,type=pdf,ext=.pdf,read=.pdf]{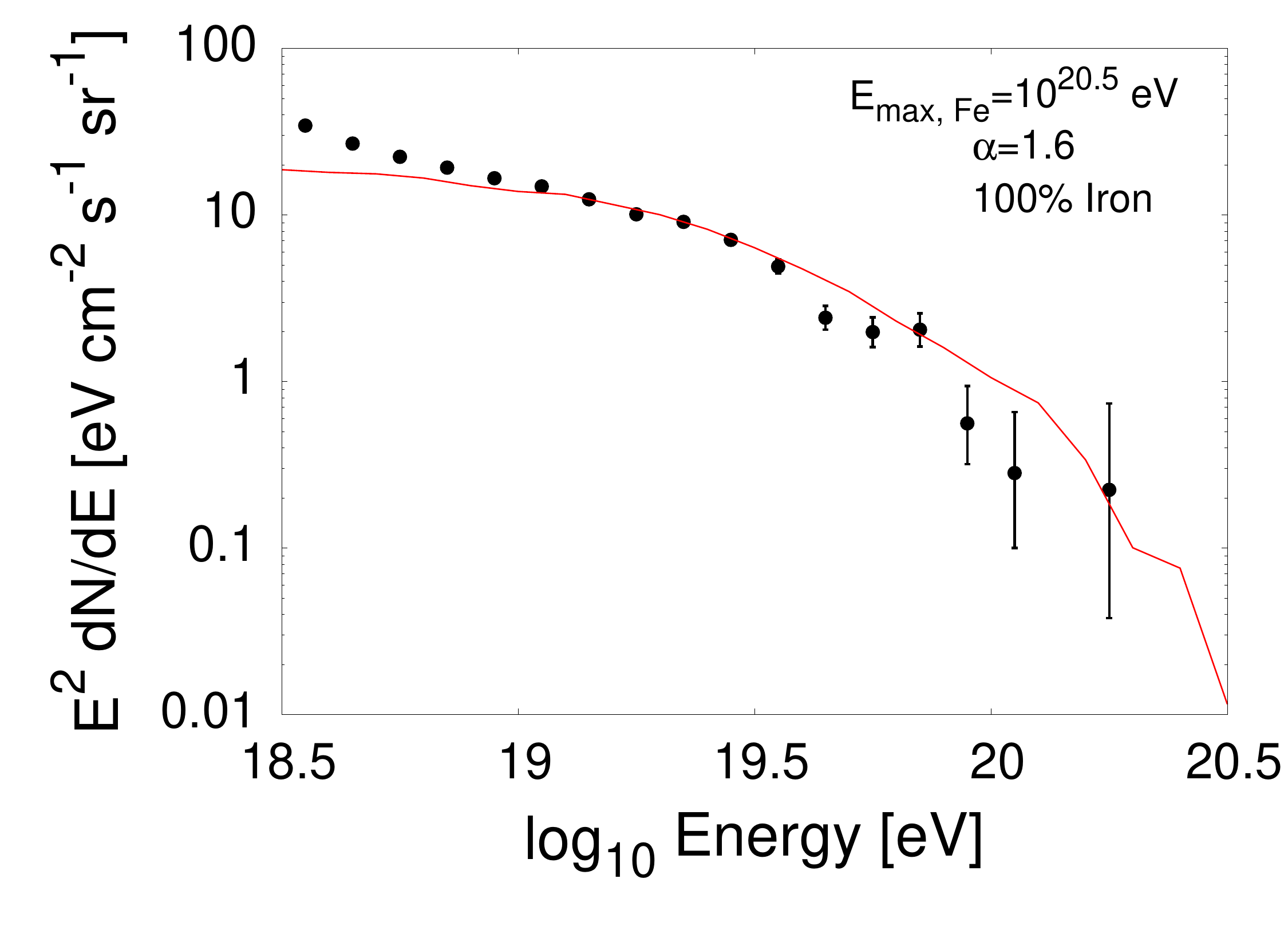}}
{\includegraphics[angle=0,width=0.32\linewidth,type=pdf,ext=.pdf,read=.pdf]{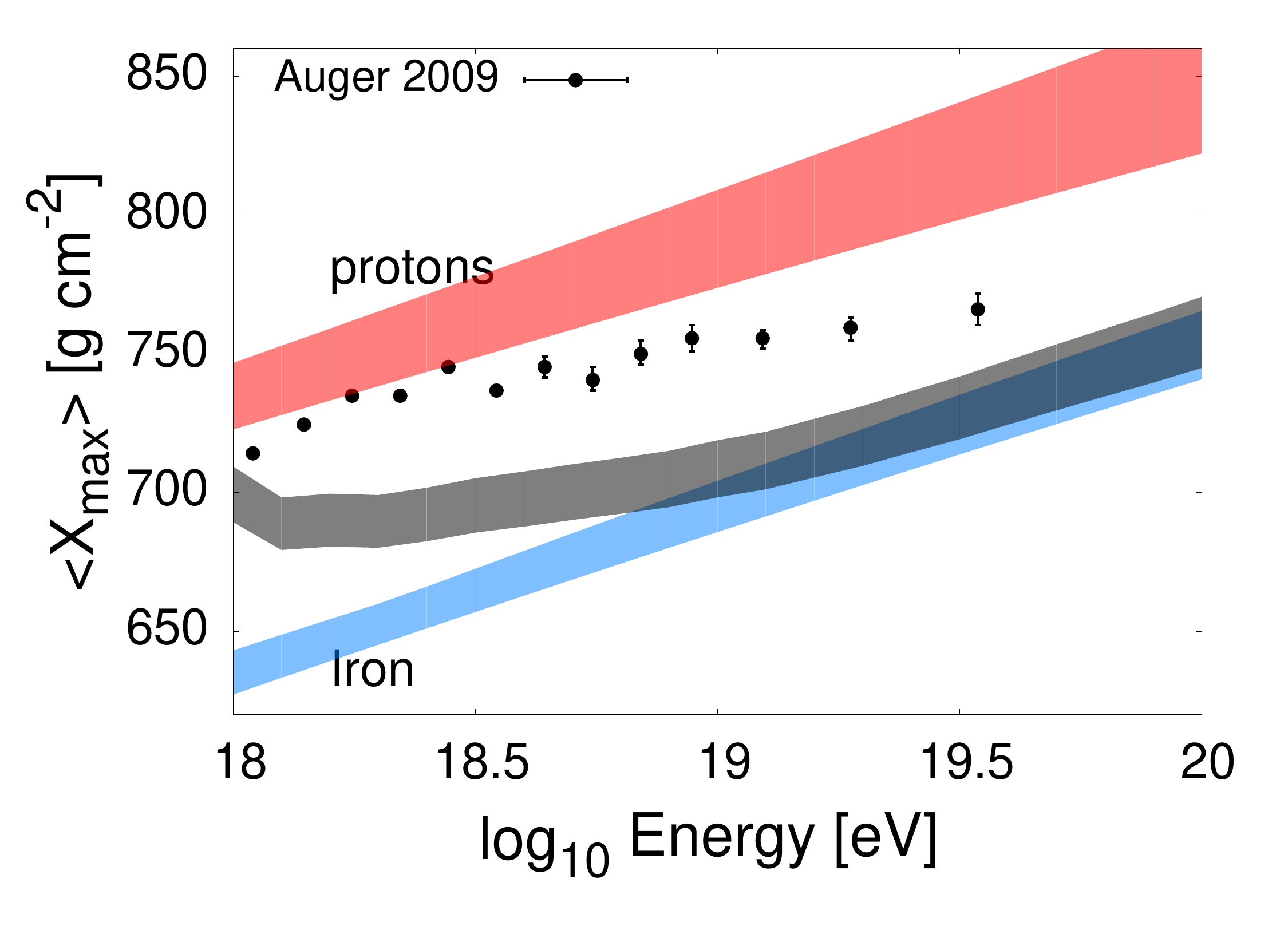}}
{\includegraphics[angle=0,width=0.32\linewidth,type=pdf,ext=.pdf,read=.pdf]{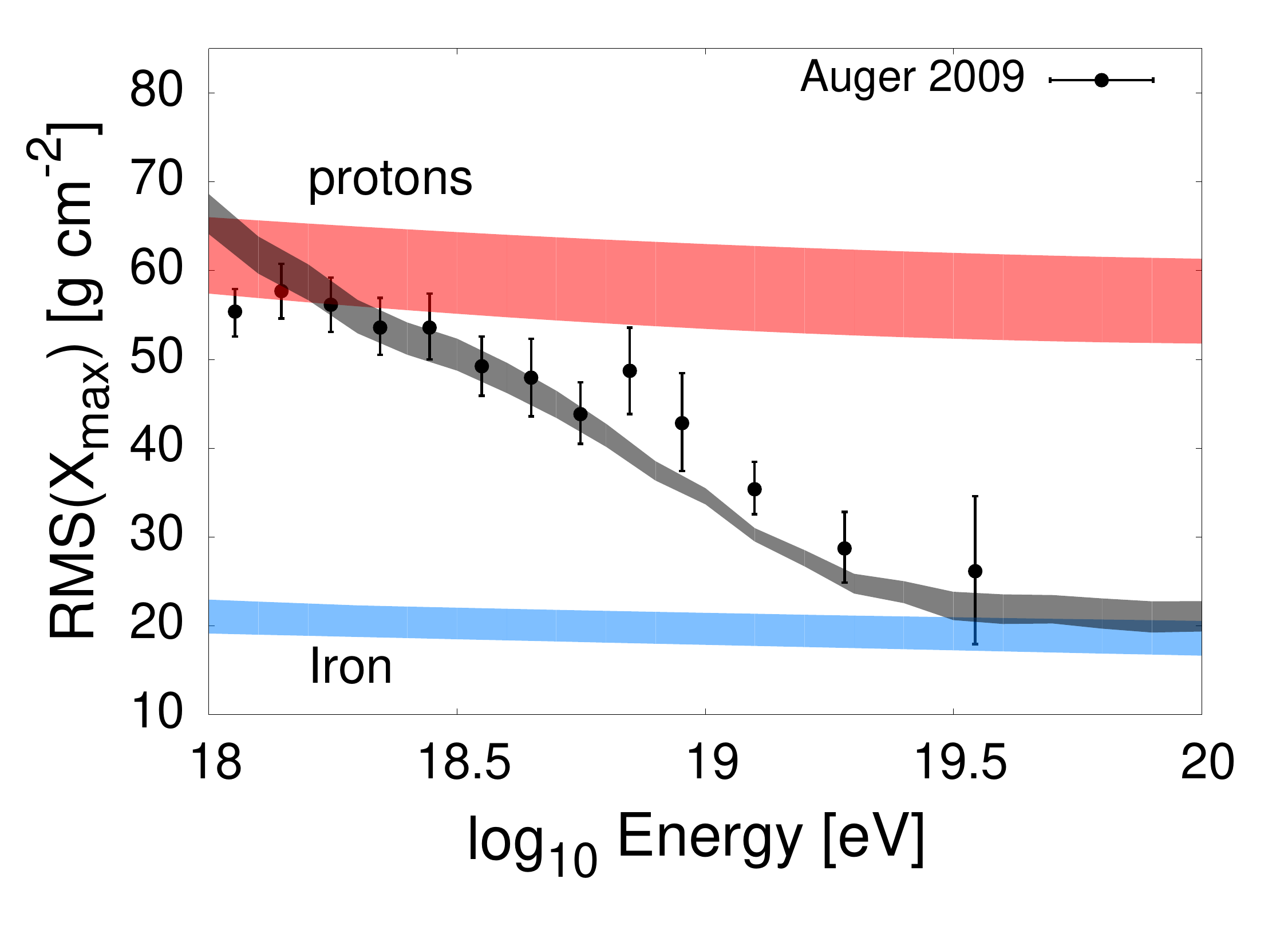}}\\
{\includegraphics[angle=0,width=0.32\linewidth,type=pdf,ext=.pdf,read=.pdf]{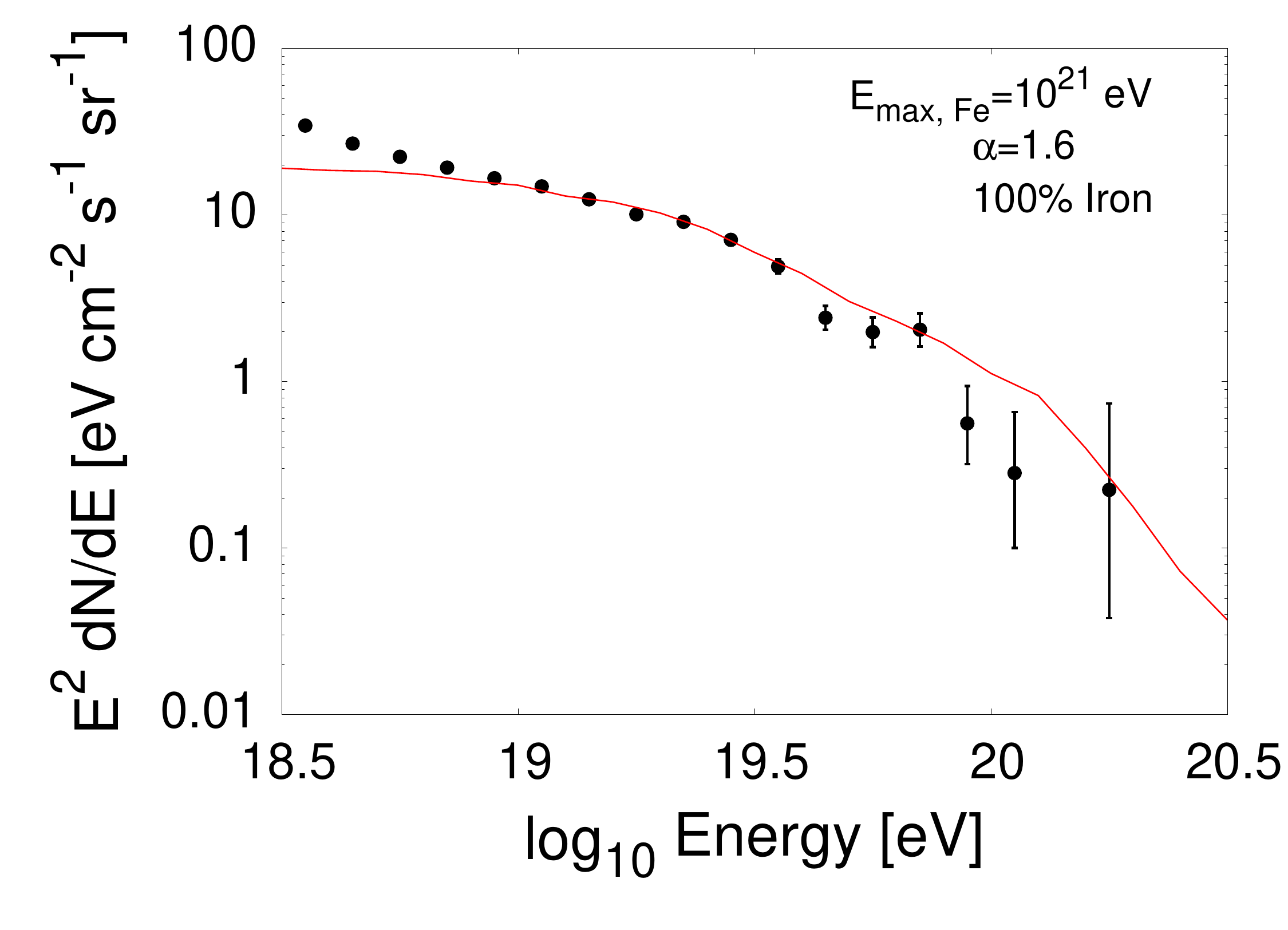}}
{\includegraphics[angle=0,width=0.32\linewidth,type=pdf,ext=.pdf,read=.pdf]{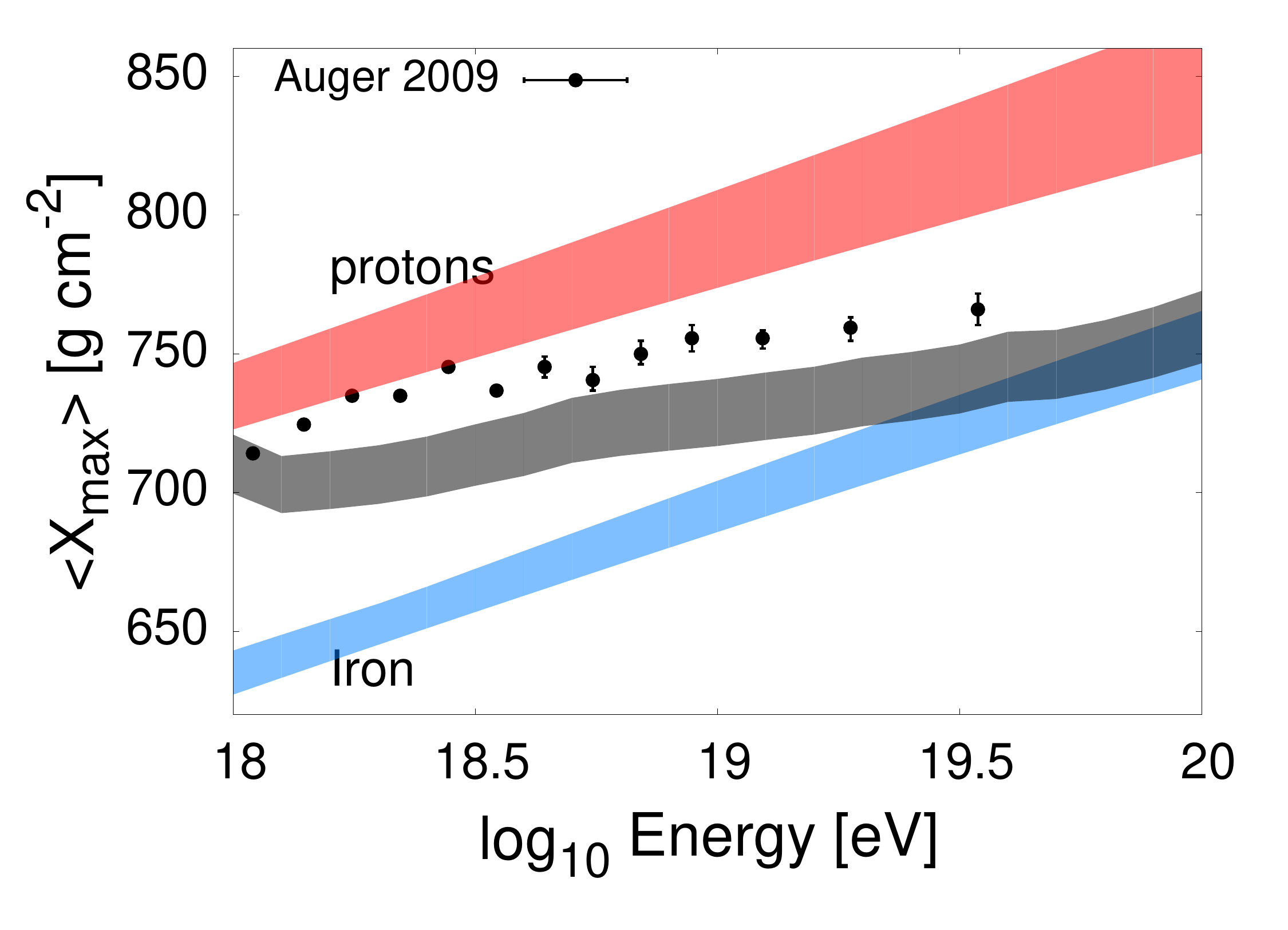}}
{\includegraphics[angle=0,width=0.32\linewidth,type=pdf,ext=.pdf,read=.pdf]{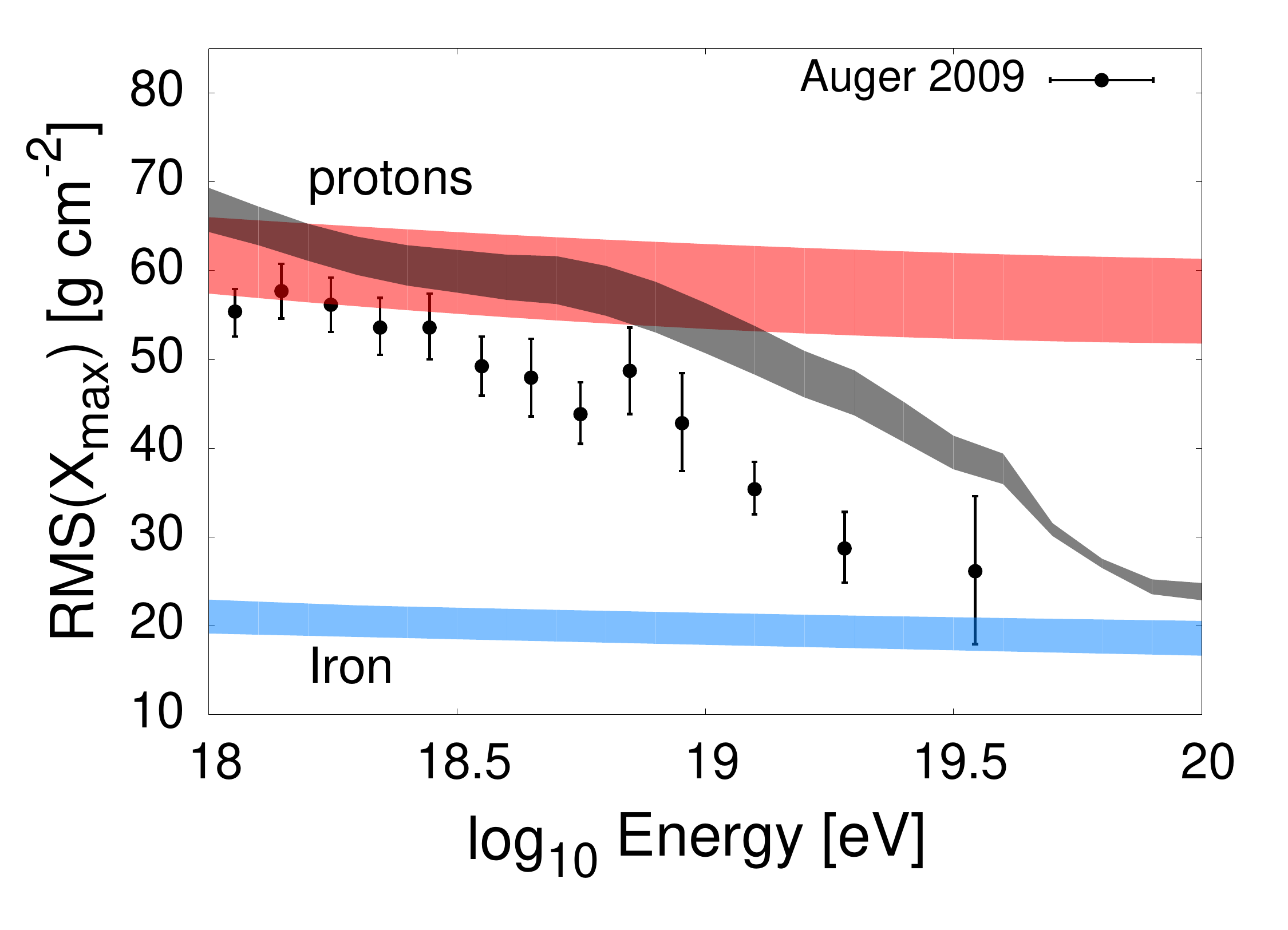}}\\
\caption{The spectrum (left), the average depth of shower maximum (center), and the RMS variation of the depth of shower maximum (right), as measured by the Pierre Auger Observatory (PAO) \cite{spectrum,xmax}, compared to the predictions in the case that the sources of the ultra-high energy cosmic rays accelerate only iron nuclei. The iron nuclei were assumed to be injected from their sources with a spectrum of $dN/dE \propto E^{-\alpha} e^{-E/E_{\rm max}}$, where $E_{\rm max}$ is $10^{20.5}$ or $10^{21}$ eV in the upper and lower frames, respectively. In each case, the spectral index $\alpha$ was selected to provide the best fit to the spectrum measured by the PAO ($\alpha=1.6$ in each case shown). In the center and right frames, we show the regions bounded by the predictions of four different hadronic models (QGSJET~\cite{qgsjet}, SIBYLL~\cite{sibyll}, and EPOS~\cite{epos}), and for comparison also show each simulation's prediction for a spectrum at Earth consisting of purely protons or purely iron nuclei. For each shaded region in the shower maximum plot the upper and lower boundaries are predominantly dictated by the EPOS and QGSJET models respectively. No simple rule, however, explains the model responsible for the upper and lower boundaries of the shaded regions in the RMS variation plot.}
\label{iron}
\end{center}
\end{figure*}

\begin{figure*}[t]
\begin{center}
{\includegraphics[angle=0,width=0.32\linewidth,type=pdf,ext=.pdf,read=.pdf]{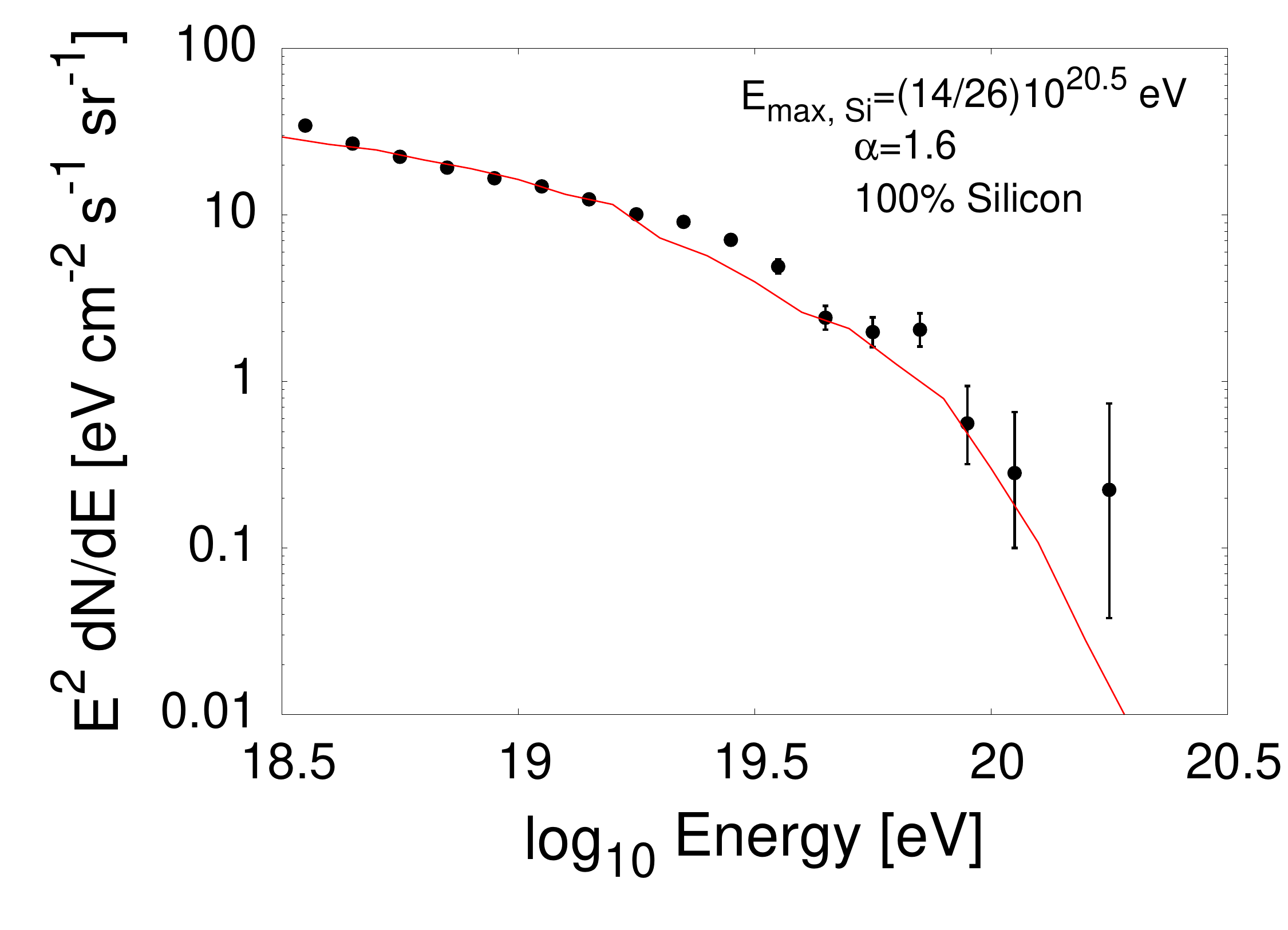}}
{\includegraphics[angle=0,width=0.32\linewidth,type=pdf,ext=.pdf,read=.pdf]{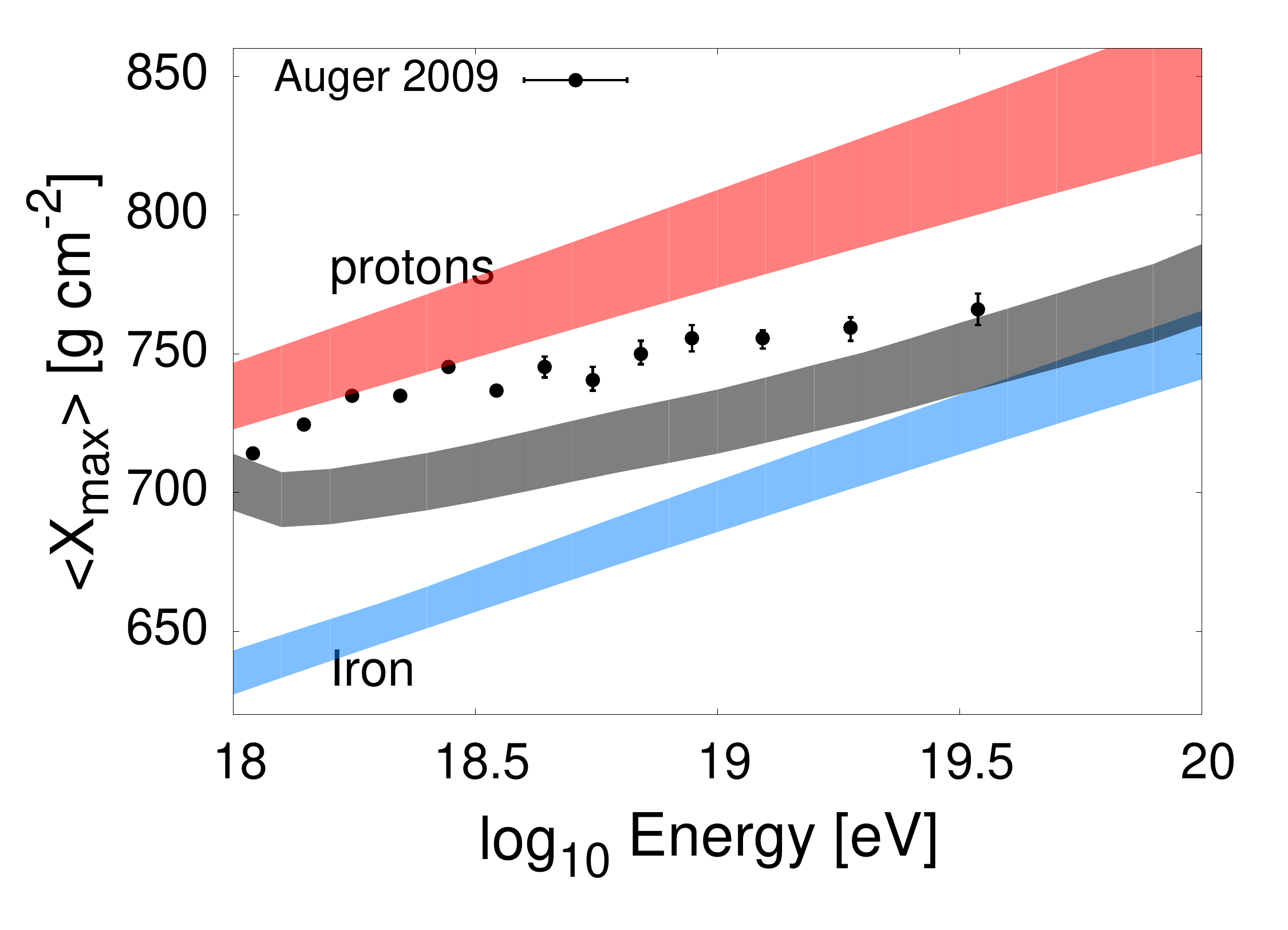}}
{\includegraphics[angle=0,width=0.32\linewidth,type=pdf,ext=.pdf,read=.pdf]{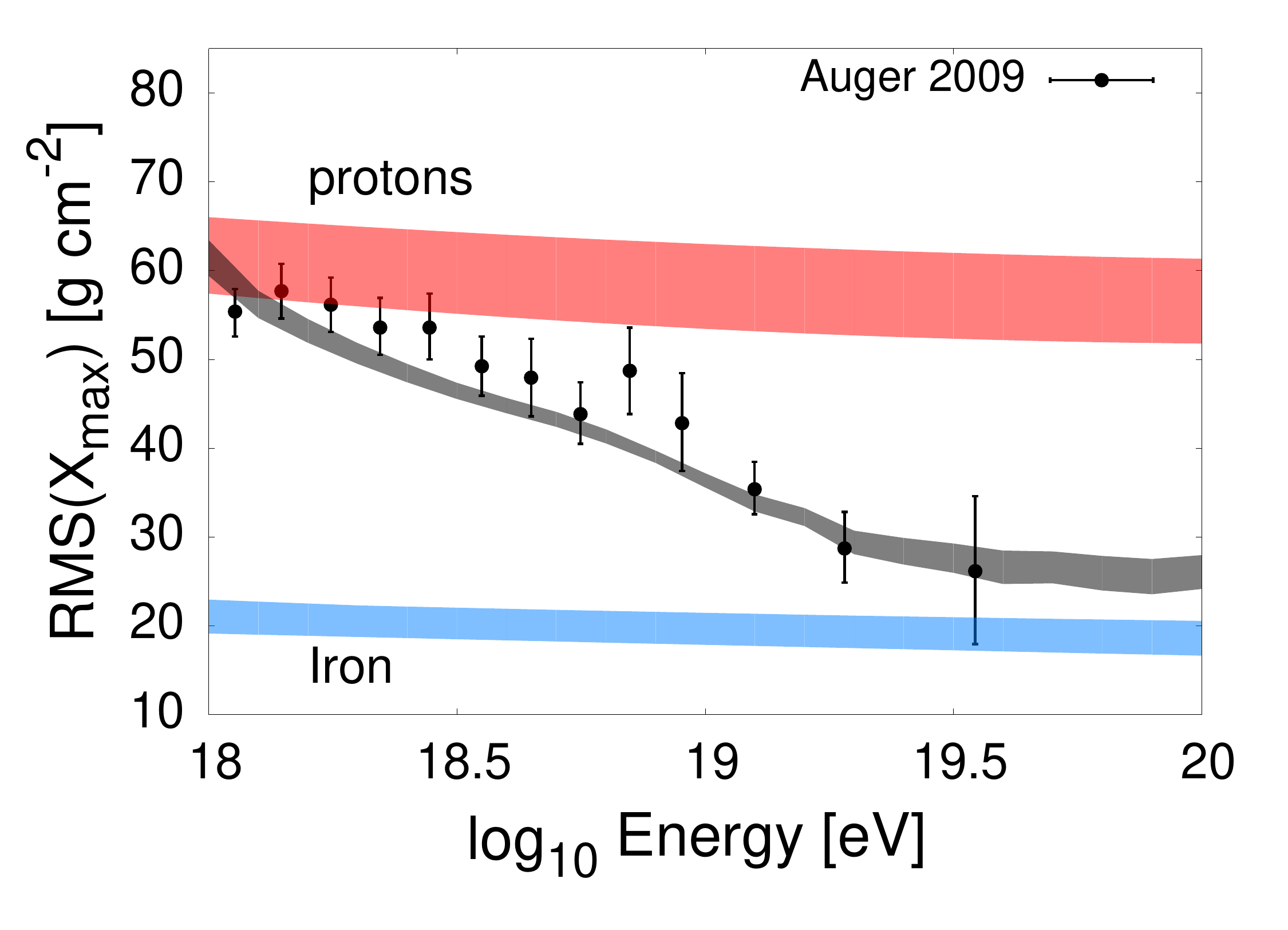}}\\
{\includegraphics[angle=0,width=0.32\linewidth,type=pdf,ext=.pdf,read=.pdf]{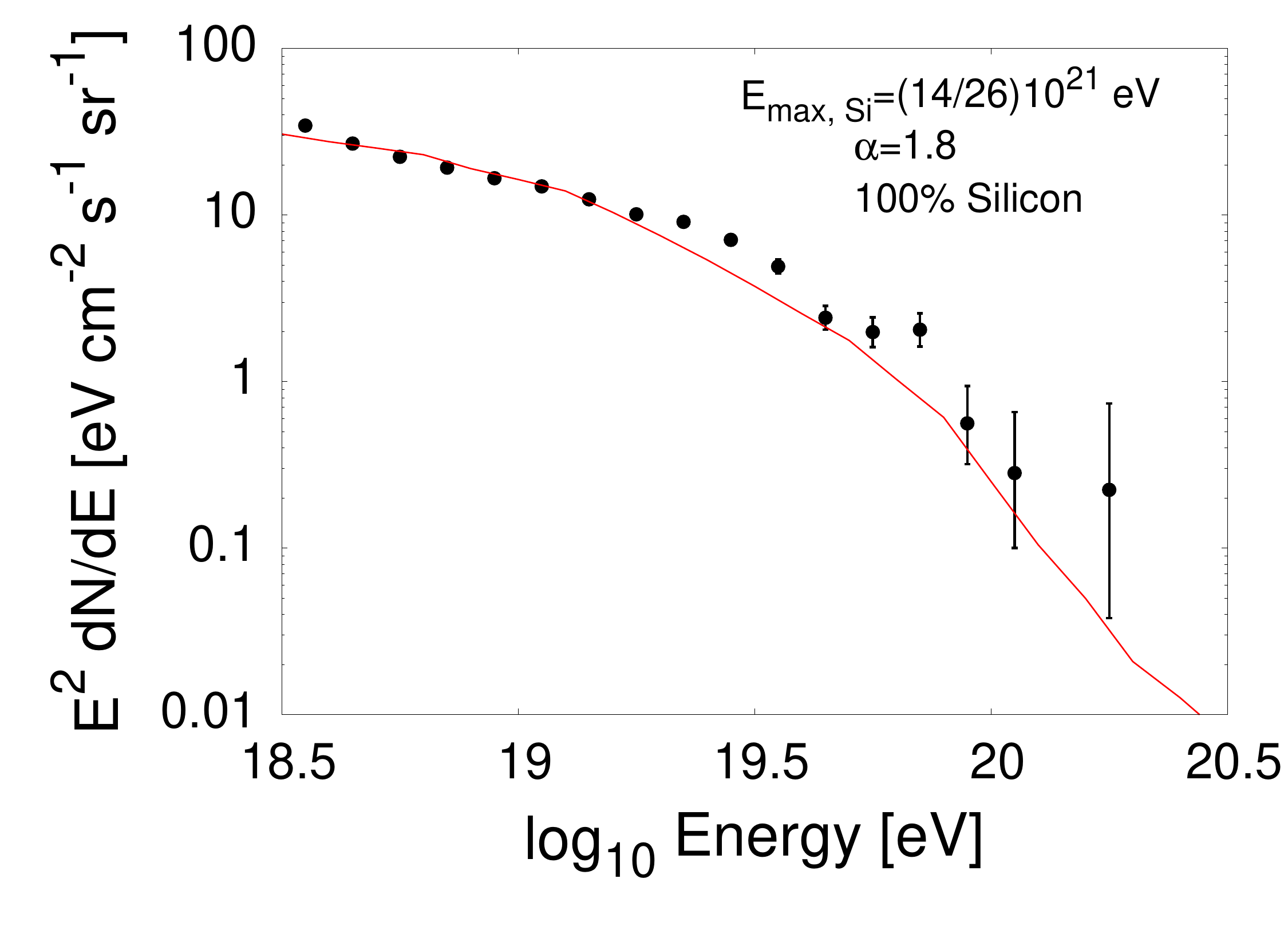}}
{\includegraphics[angle=0,width=0.32\linewidth,type=pdf,ext=.pdf,read=.pdf]{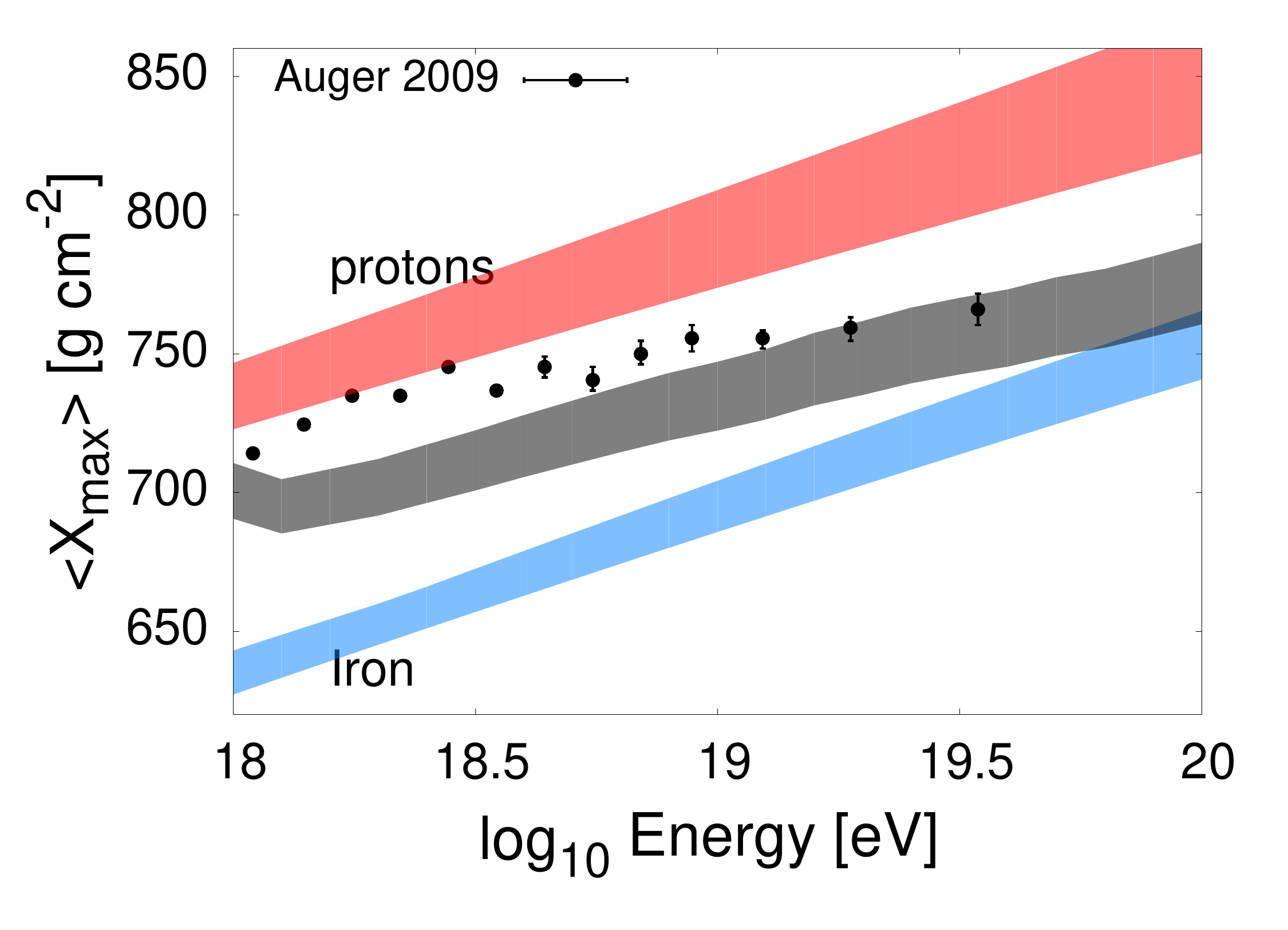}}
{\includegraphics[angle=0,width=0.32\linewidth,type=pdf,ext=.pdf,read=.pdf]{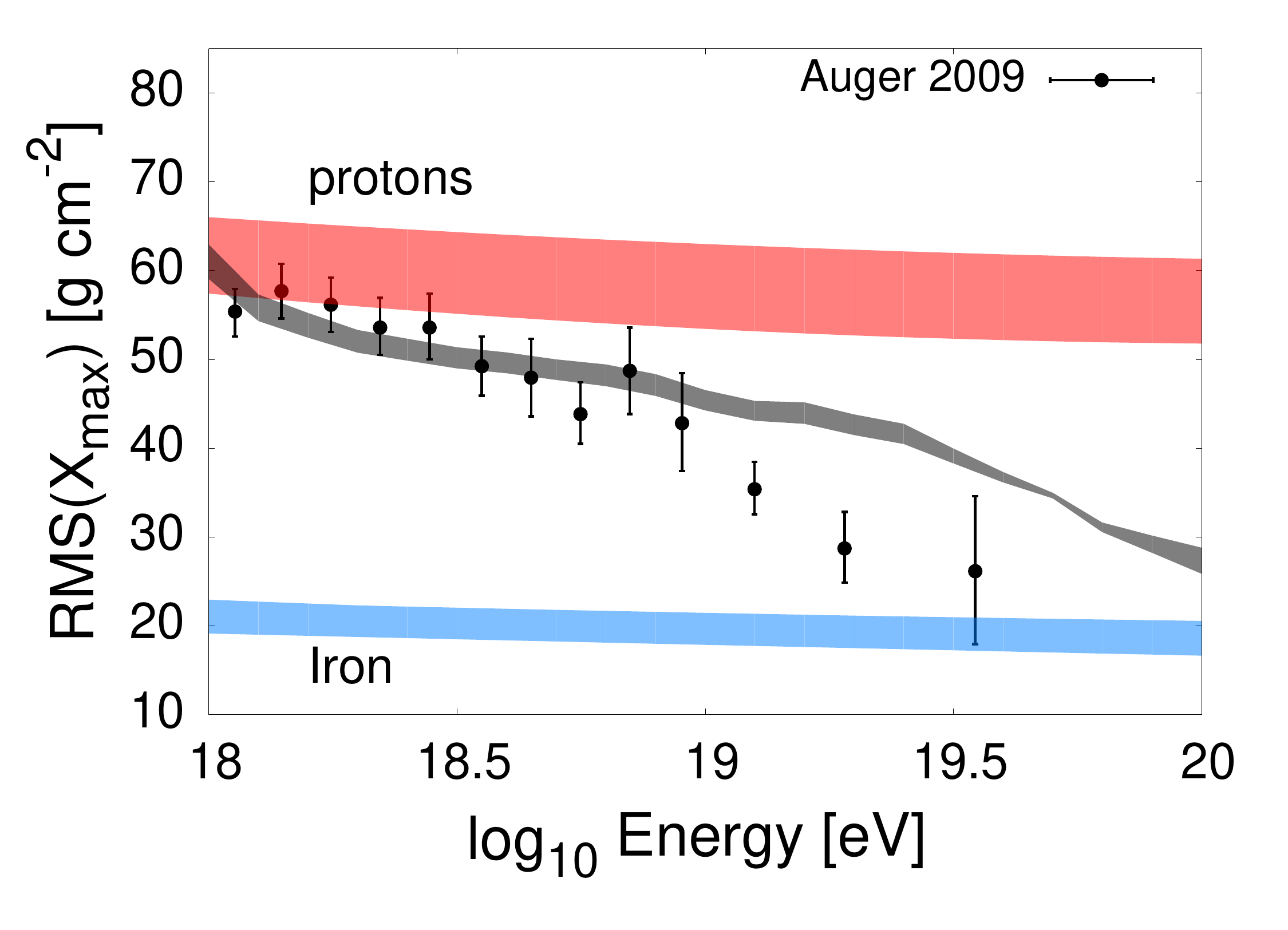}}\\
\caption{As in Fig.~\ref{iron}, but for the case in which only silicon nuclei are accelerated by the sources of the ultra-high energy cosmic rays.}  
\label{silicon}
%
{\includegraphics[angle=0,width=0.32\linewidth,type=pdf,ext=.pdf,read=.pdf]{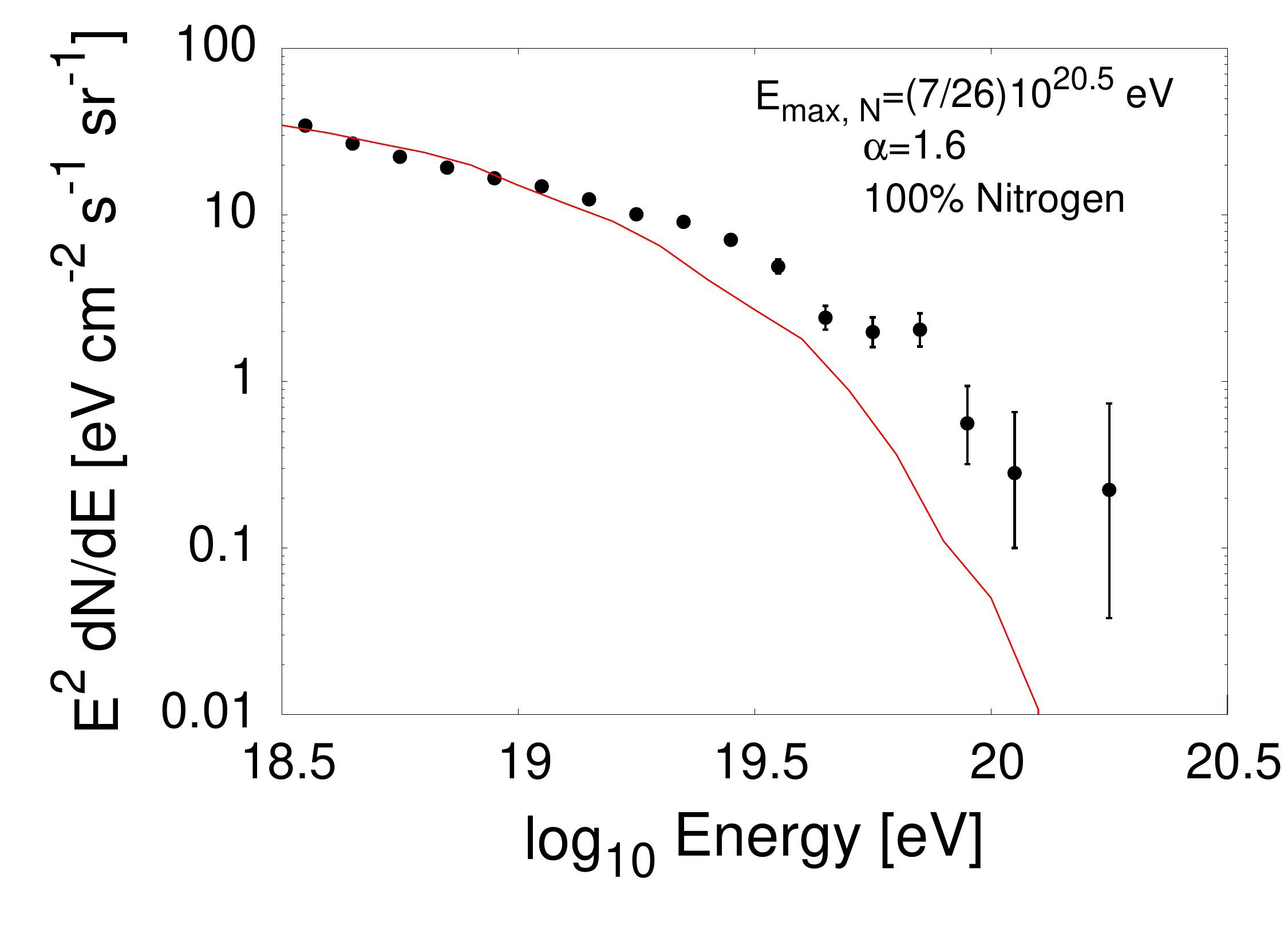}}
{\includegraphics[angle=0,width=0.32\linewidth,type=pdf,ext=.pdf,read=.pdf]{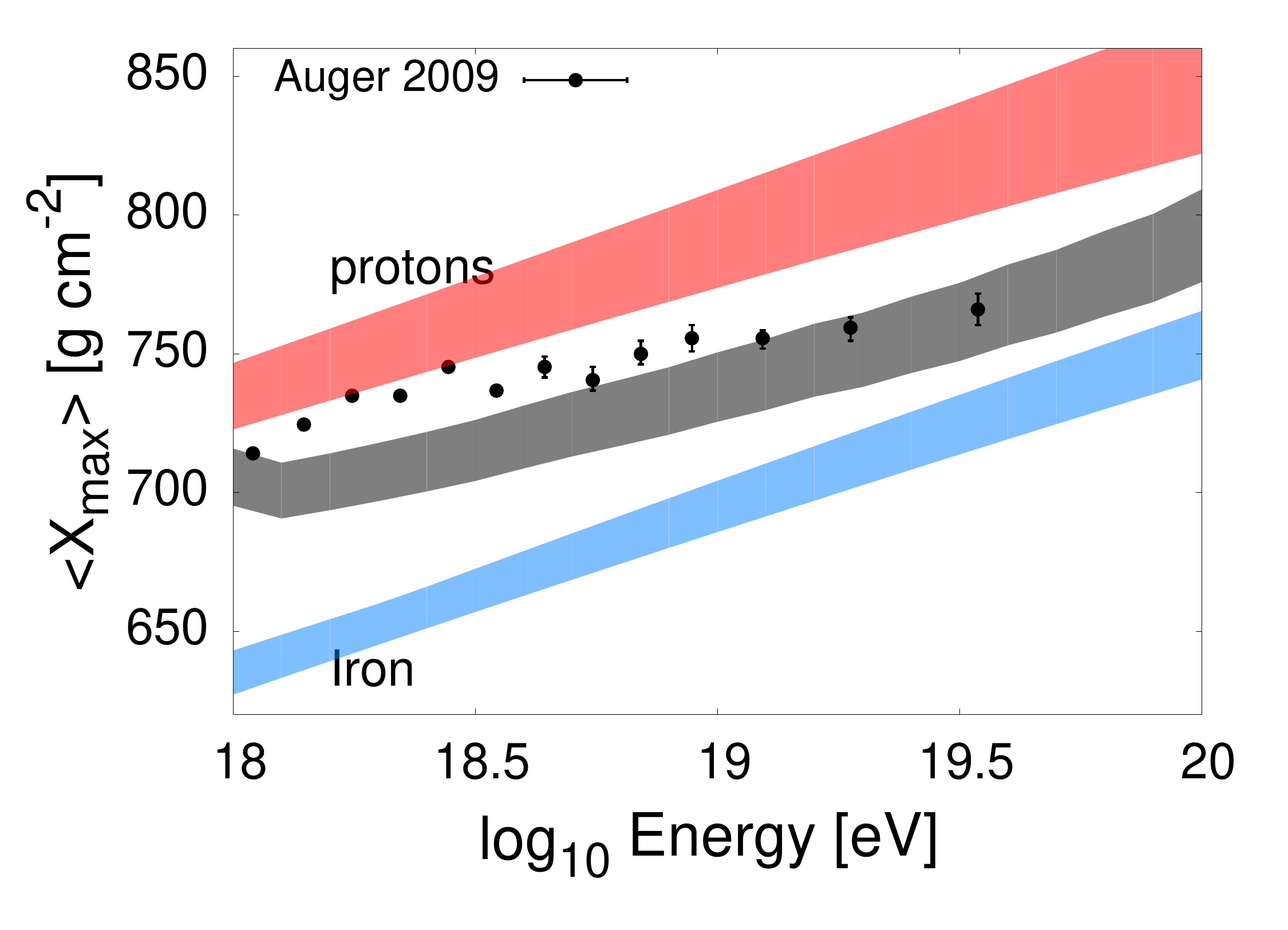}}
{\includegraphics[angle=0,width=0.32\linewidth,type=pdf,ext=.pdf,read=.pdf]{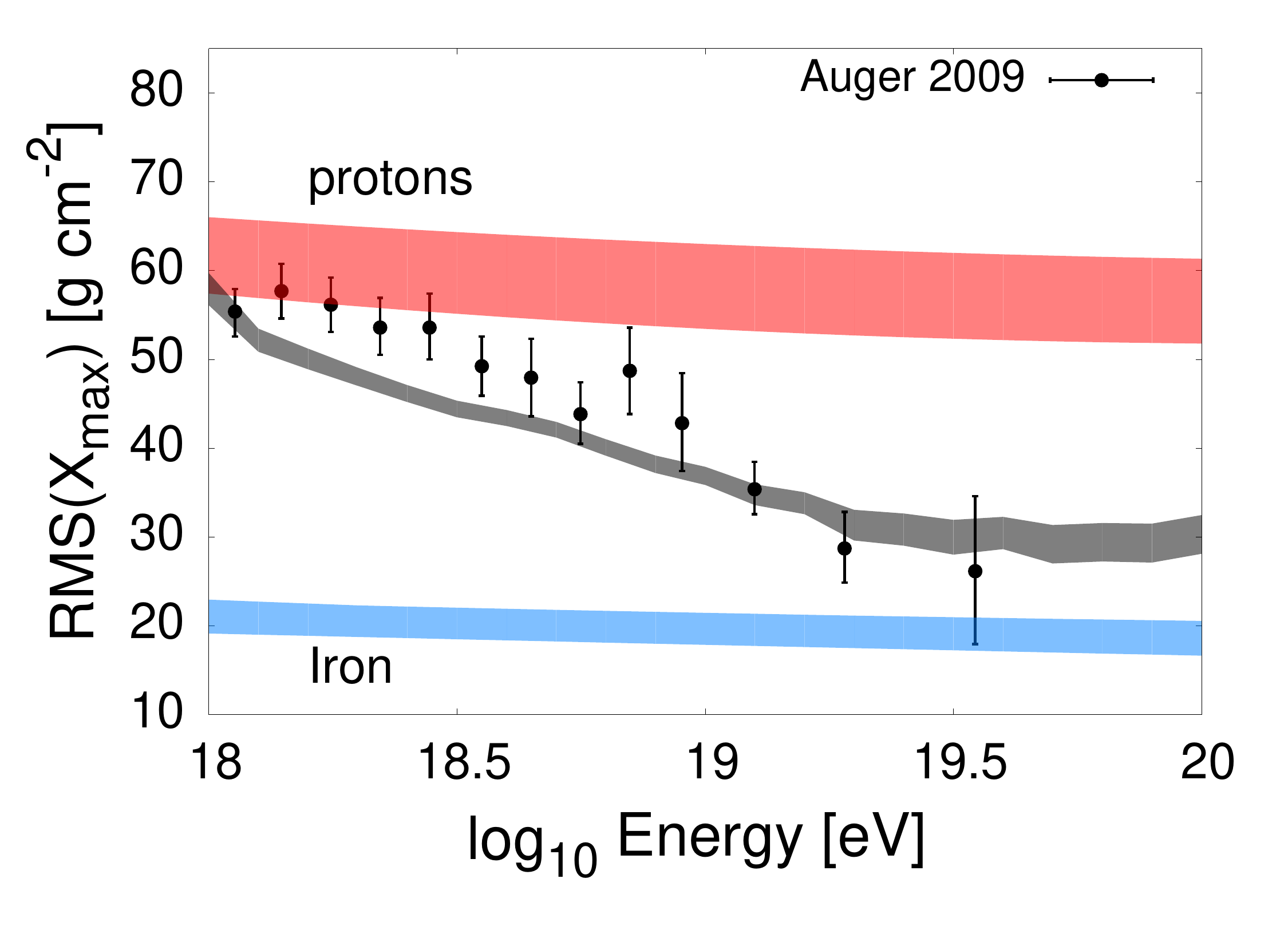}}\\
{\includegraphics[angle=0,width=0.32\linewidth,type=pdf,ext=.pdf,read=.pdf]{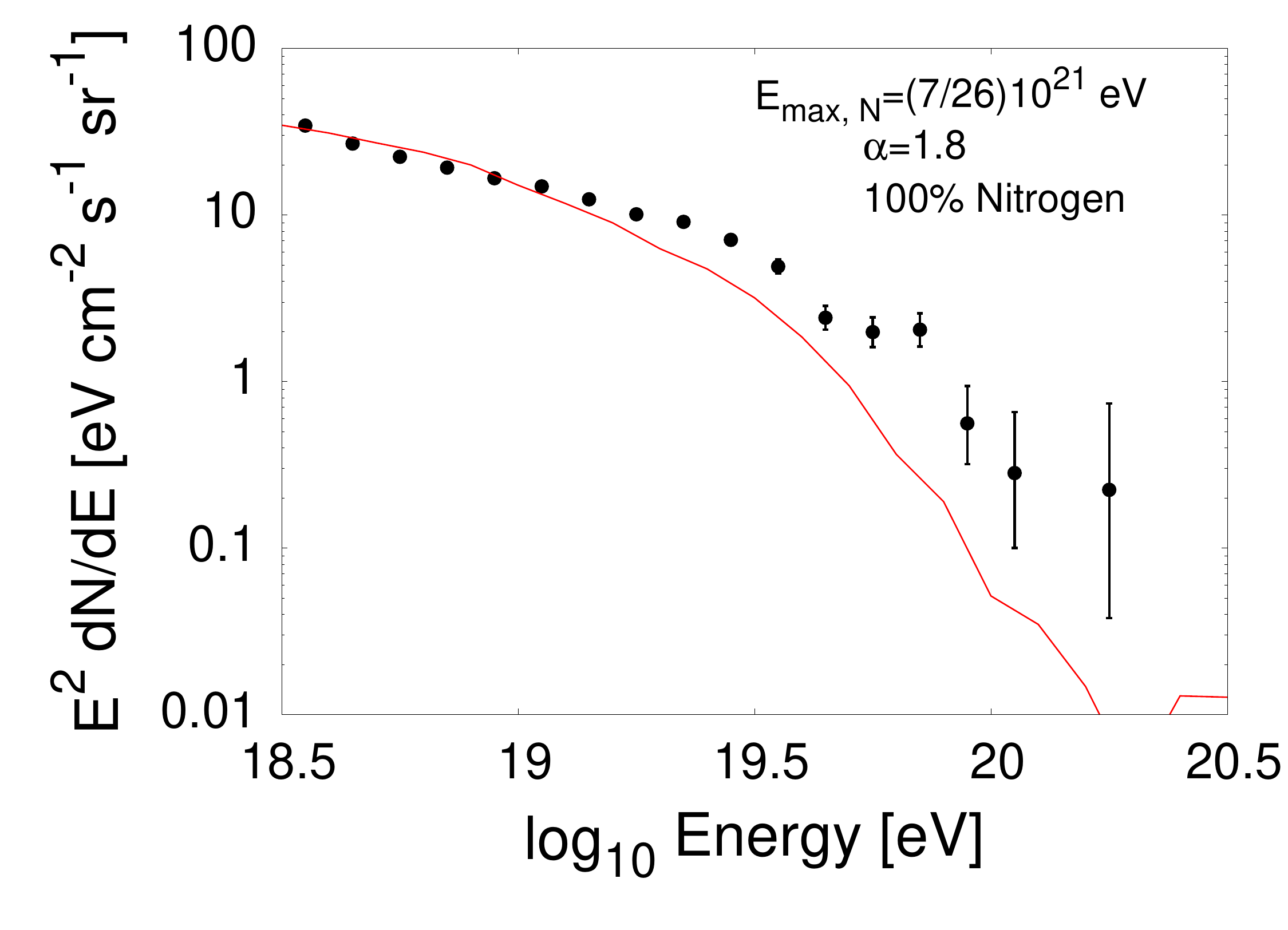}}
{\includegraphics[angle=0,width=0.32\linewidth,type=pdf,ext=.pdf,read=.pdf]{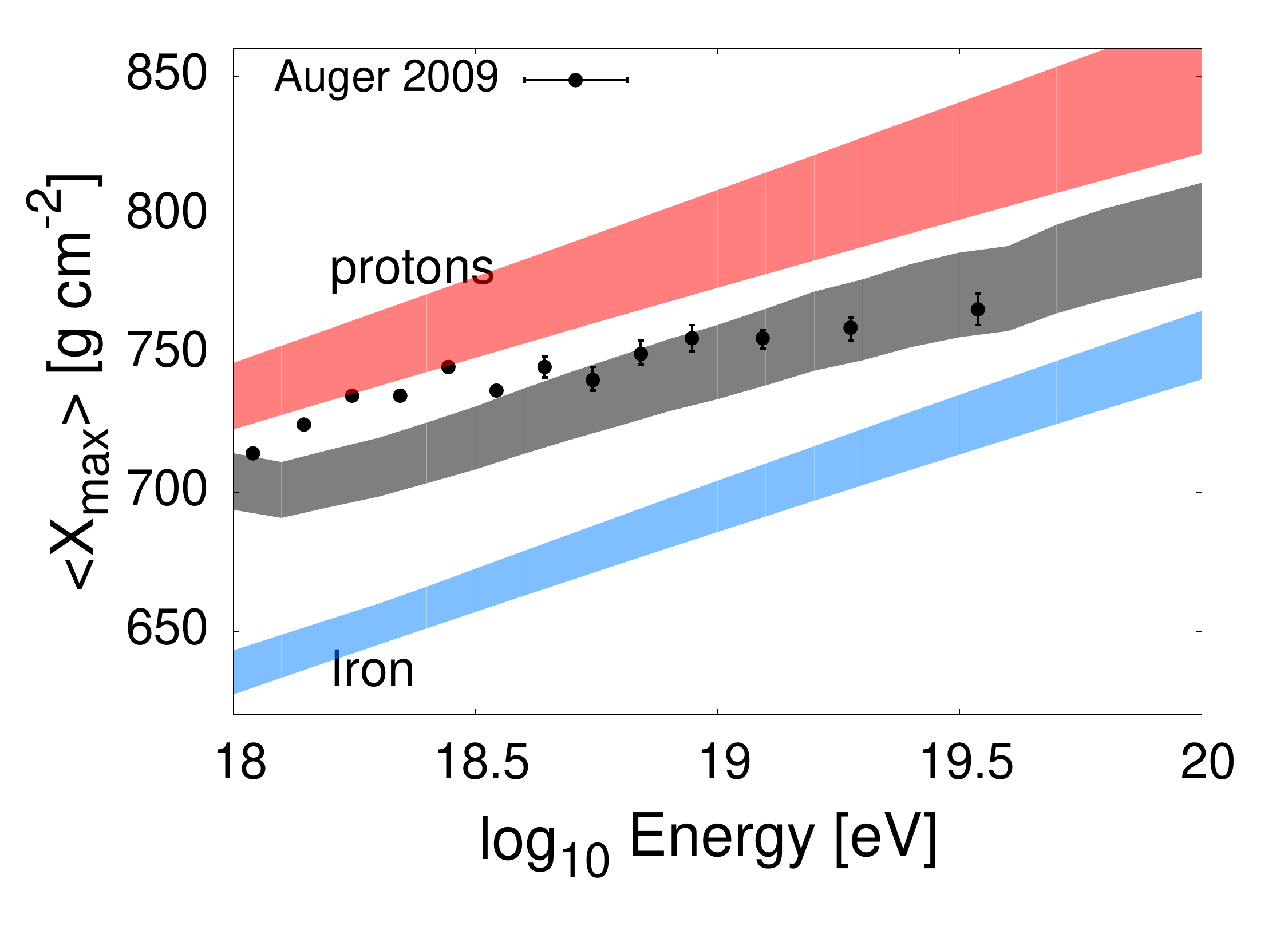}}
{\includegraphics[angle=0,width=0.32\linewidth,type=pdf,ext=.pdf,read=.pdf]{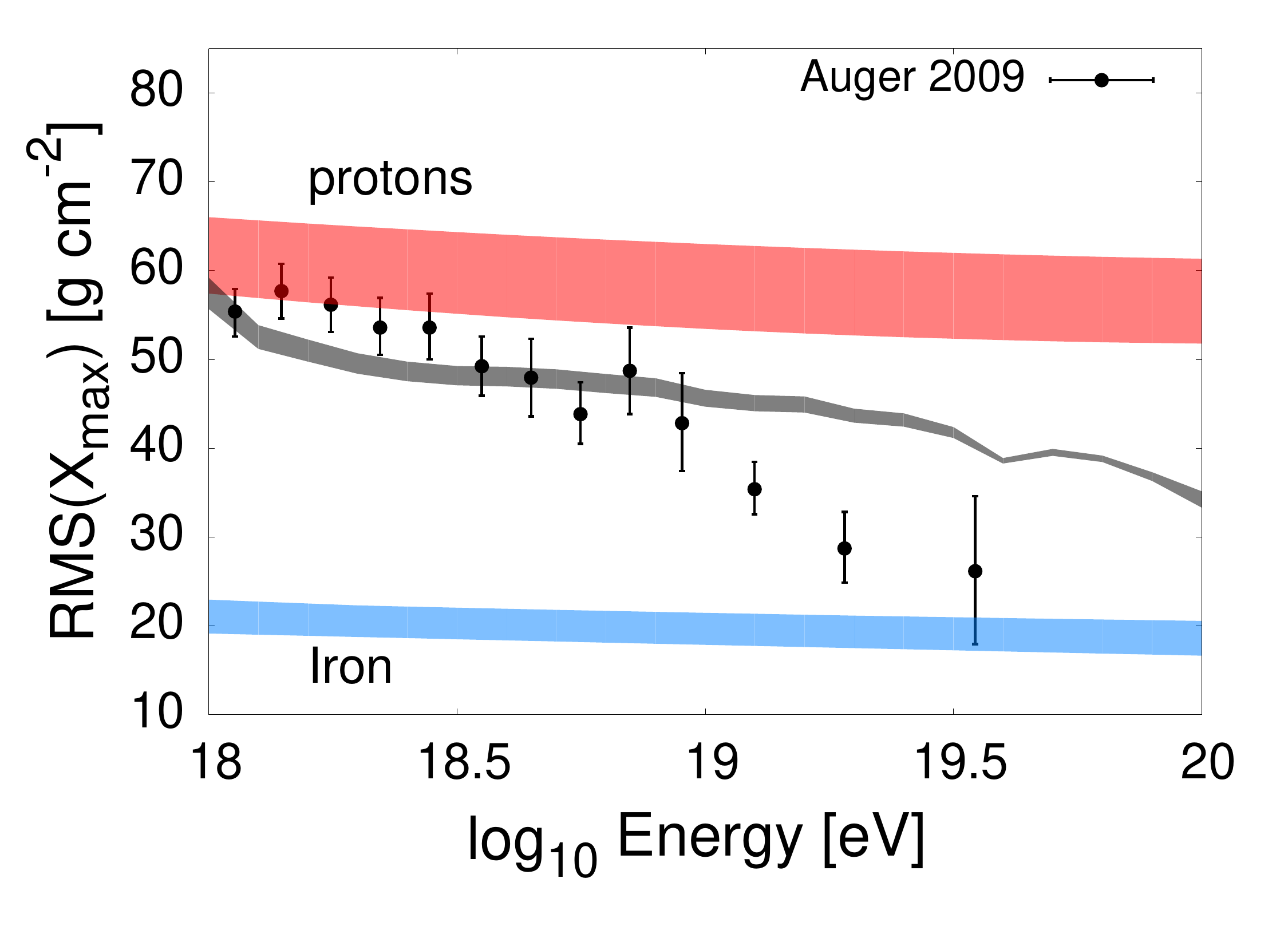}}\\
\caption{As in Figs.~\ref{iron} and~\ref{silicon}, but for the case in which only nitrogen nuclei are accelerated by the sources of the ultra-high energy cosmic rays.} 
\label{nitrogen}
\end{center}
\end{figure*}

In the following we obtain best fits to the arriving spectrum, $\langle X_{\rm max} \rangle$, and $\mathrm{RMS}(X_{\rm max})$ data by scanning over $E_{\rm max}$ from 10$^{20}$-10$^{22}$~eV, and $\alpha$ from $1.6$-$2.4$, and minimising the $\chi^{2}$ for these runs, keeping the best fit results. Only data for CR energies above 10$^{18.5}$~eV were used in these fits, with the assumption that events below this energy are Galactic in origin. All spectral results are shown for energies above  this assumed transition energy. However, to provide a comprehensive understanding of the change in composition following photodisintegration interactions, all composition related plots ($\langle X_{\rm max} \rangle$, and $\mathrm{RMS}(X_{\rm max})$) are shown down to energies of 10$^{18}$~eV. We reiterate to the reader that only data points with energies above 10$^{18.5}$~eV have been used to obtain the best fit results. In order to keep the composition related plots as simple as possible, we plot only the upper and lower $\langle X_{\rm max} \rangle$ and $\mathrm{RMS}(X_{\rm max})$ values allowed from the different hadronic models. For each shaded region in the $\langle X_{\rm max} \rangle$ plot the upper and lower boundaries are predominantly dictated by the EPOS and QGSJET models respectively. No simple rule, however, explains the model responsible for the upper and lower boundaries of the shaded regions in the $\mathrm{RMS}(X_{\rm max})$ plot.

In Fig.~\ref{iron}, we show our results for the cases of $E_{\rm max}=10^{20.5}$ and $10^{21}$ eV ($\alpha=$1.6 in each case). We note that although such hard spectra are not well motivated by non-relativistic diffusive shock acceleration theory, such spectra have been suggested by relativistic shock acceleration theory \cite{Stecker:2007zj}. Furthermore, the reader should bear in mind that such hard spectra obtain good agreement through the considerable lightening of the arriving composition that they allow, caused through the additional contribution of liberated protons following photodisintegration interactions. Thus hard injection spectra provide the possibility of a lightened composition at lower energies. This lightening of the composition, however, comes at the price of a reduced spectral fit with the PAO data. Such good fits, for the case of heavy nuclei, therefore exist through the balance between these two effects. In the middle and right frames, we have used the results of hadronic simulations to convert the chemical composition of the spectrum found by our Monte Carlo into the related quantities observed by the PAO -- the average depth of shower maximum, and the RMS variation of this quantity~\cite{PC}. The shaded region in the plots is that bounded by the results obtained using four different commonly used parameterizations~\cite{ungerSocor} of air shower simulations~\cite{conex}: QGSJET II, QGSJET~\cite{qgsjet}, SIBYLL 2.1~\cite{sibyll}, and EPOS~\cite{epos}. For comparison, we also show the results that are obtained for the cases in which either only protons or iron nuclei are present in the cosmic ray spectrum at Earth. At some level, the variations between the results of the different hadronic simulations provide us with an estimate of the systematic errors associated with our ignorance of the non-perturbative QCD necessary for a complete description of the shower development. Furthermore, we highlight to the reader the fact that these models only provide an incomplete scan of the full parameter space \cite{Ulrich:2009yq,Ulrich:2009hm,Wibig:2009zza,Wibig:2009zz}, leaving room for the possibility that the model actually realised by nature sits outside this range.

From Fig.~\ref{iron}, we see that in the case of all-iron injecting sources with $E_{\rm max} \lsim 10^{20.5}$ eV, the measured average depth of shower maximum consistently exceed the predicted values. This is easy to understand from the fact that $10^{20.5}$ eV iron nuclei fragment into nucleons with only $\sim 6 \times 10^{18}$ eV of energy each, thus leaving only heavy nuclei to make up the spectrum at higher energies. As the maximum energy is increased, however, more protons are present among the UHECRs, causing the average composition to be considerably lighter. If we increase $E_{\rm max}$, we can better accomodate the measurements of $\langle X_{\rm max} \rangle$, but also begin to significantly exceed those of $\mathrm{RMS}(X_{\rm max})$. For an all-iron source composition, we do not find any combination of $E_{\rm max}$ and $\alpha$ that can simultaneously accomodate the PAO's measurements of $\langle X_{\rm max} \rangle$ and $\mathrm{RMS}(X_{\rm max})$.

To see whether less heavy species of cosmic ray nuclei might be able to better provide a reasonable fit to the PAO data, we show results in Figs.~\ref{silicon} and~\ref{nitrogen} for the cases of a pure silicon or pure nitrogen injection spectrum. For the all-silicon case, the best fit corresponds to the choices of $E_{\rm max}\approx 10^{21}$ eV and $\alpha \approx 1.8$.\footnote{Throughout this study, the quantity $E_{\rm max}$ refers to the cutoff energy for iron nuclei. This is related to the cutoff for other species of charge $Z$ by $E_{{\rm max}, Z} = (Z/26)\times E_{\rm max}$.} Even in this best case, however, the fit to the $\langle X_{\rm max} \rangle$ data is quite poor, although better than that found in the all-iron case.


The all-nitrogen case even more clearly fails to accommodate the observations, in particular with regards to the observed shape of the UHECR spectrum which, unlike the composition measurements, are not impacted by uncertainties associated with the hadronic models. This is in agreement with the conclusions drawn in Ref.~\cite{spectrumcomposition}. Furthermore, the all-nitrogen case does not predict a sufficiently rapid change of ${\mathrm{RMS}}(X_{\rm max})$ with energy. We thus conclude that while the PAO observations do not appear to be very well fit by any of the scenarios we have considered so far, the best ({\it ie.} least bad) fits are in those cases in which the observed UHECR spectrum originates largely from sources accelerating intermediate mass or heavy nuclei ($A \sim 20$).

If cosmic ray sources accelerate a mixture of different species of nuclei, this will impact $\langle X_{\rm max} \rangle$ and ${\mathrm{RMS}}(X_{\rm max})$ in somewhat different ways. In particular, whereas the average depth of shower maximum at a given energy simply scales with $\langle X_{\rm max} \rangle \propto 1/\left<\ln A \right>$, the ${\mathrm{RMS}}(X_{\rm max})$ for an ensemble of different
nuclear species is given by,
\begin{eqnarray}
\sigma_{\rm tot}^{2}=\sum_{A=1}^{56}\left(f_{A}\sigma_{A}^{2}+f_{A}(X_{\rm max, A}-\left<X_{\rm max}\right>)^{2}\right),\label{RMS}
\end{eqnarray}
where $f_{A}$ is the fraction of the total population with atomic mass $A$, and $\sigma_A$ is the RMS of $X_{\rm max}$ predicted for cosmic rays of that atomic mass. From this, we see that the relatively low value of ${\mathrm{RMS}}(X_{\rm max})$ observed by the PAO not only indicates that a large proportion of the arriving nuclei are heavy at the highest energies, but also that their composition is fairly narrowly distributed in charge-- {\it ie.} the fraction of protons or light nuclei in the UHECR spectrum must be small in order to prevent large contributions from the $f_{i}(X_{{\rm max},i}-\left<X_{\rm max}\right>)^{2}$ terms in the sum. In many cases, this leads to a tension between the PAO's measurements of $\langle X_{\rm max} \rangle$ and ${\mathrm{RMS}}(X_{\rm max})$, which can be made worse by considering cosmic ray sources which inject particles with a mixed chemical composition. We find that mixed (more than 2 component) composition scenarios, suggested to be indicated by the recent PAO data \cite{Aloisio:2009sj}, do not lead to sufficiently improved fits to justify their consideration.

\begin{figure*}[!]
\begin{center}
{\includegraphics[angle=0,width=0.32\linewidth,type=pdf,ext=.pdf,read=.pdf]{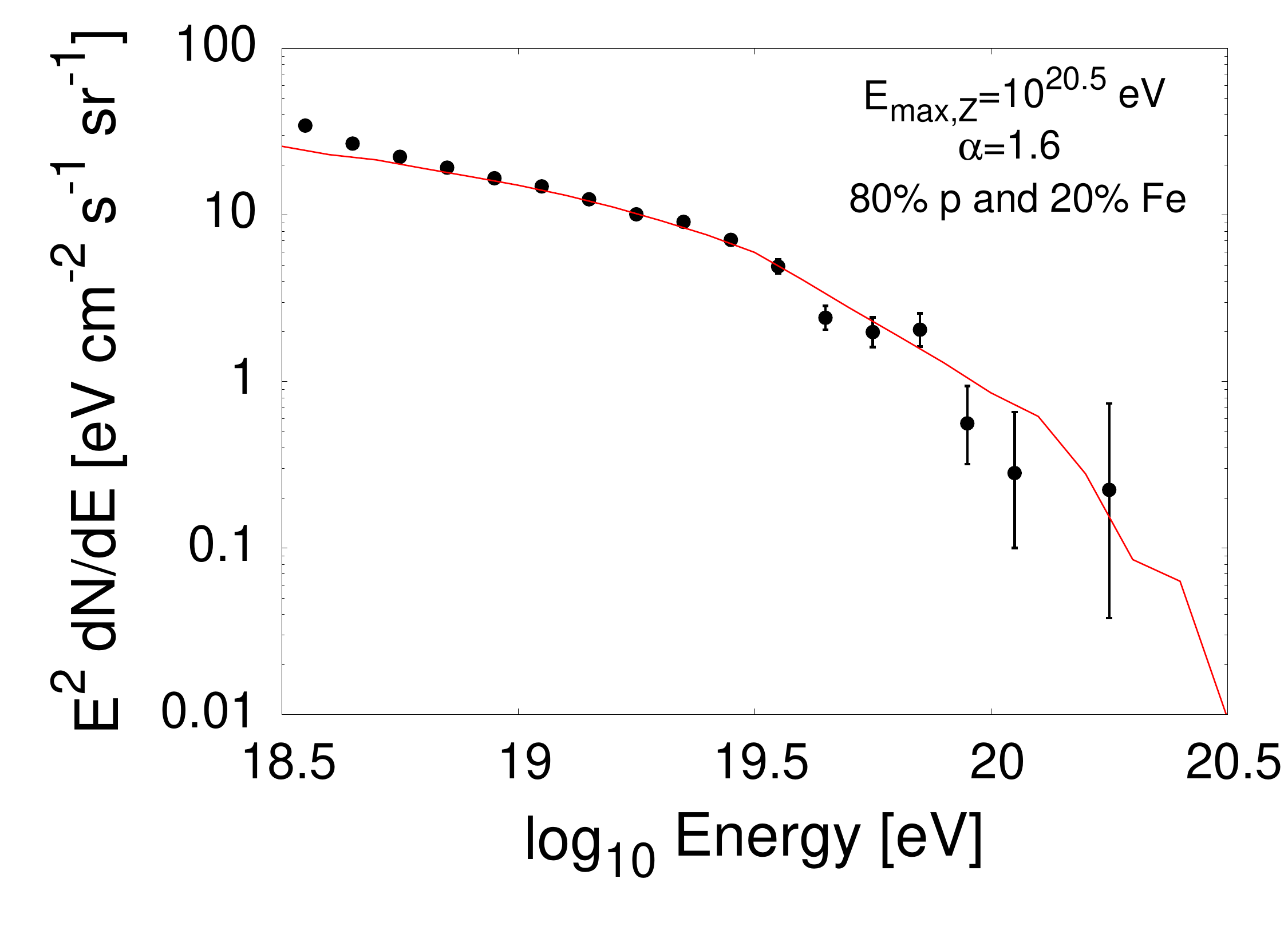}}
{\includegraphics[angle=0,width=0.32\linewidth,type=pdf,ext=.pdf,read=.pdf]{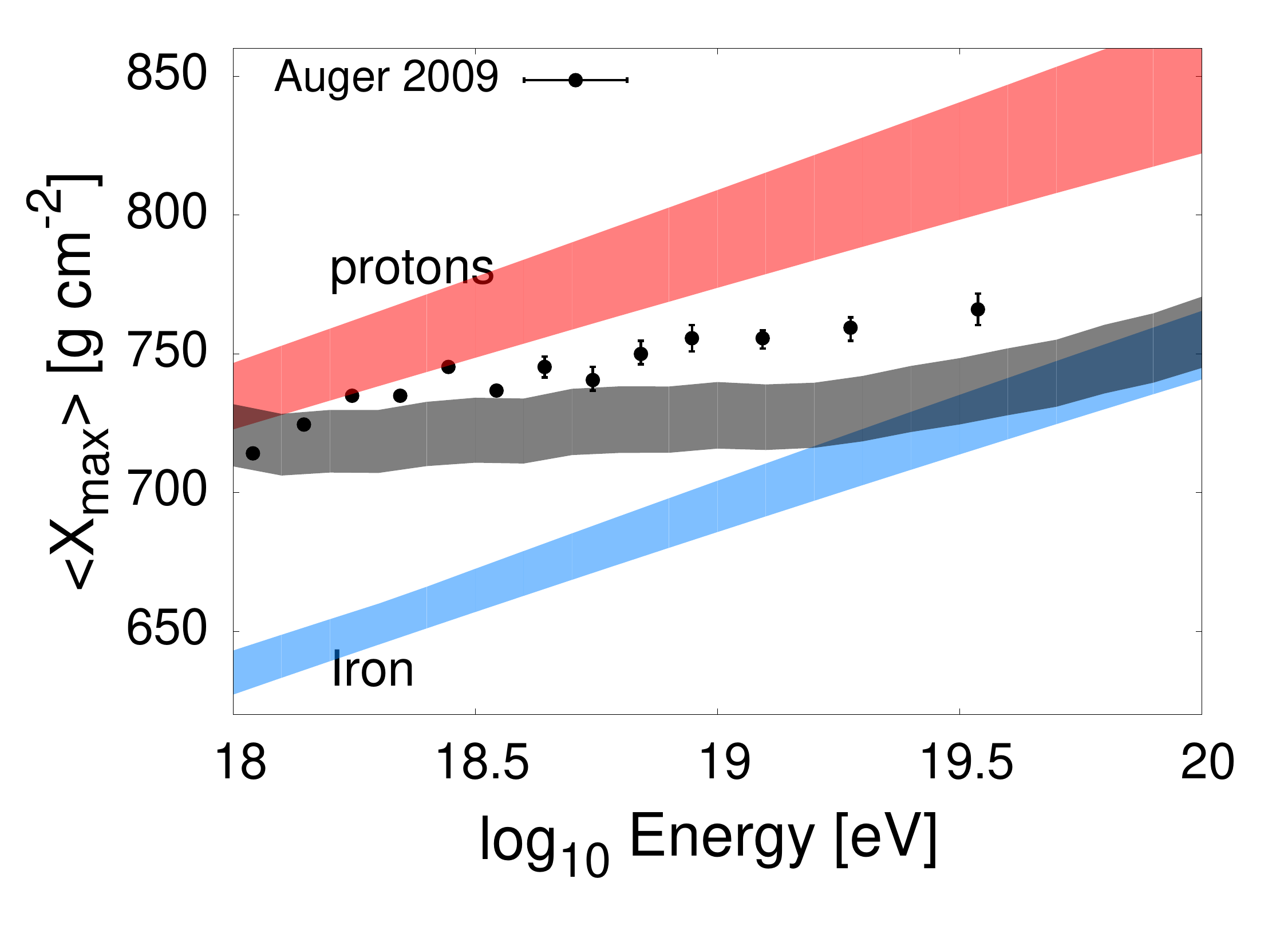}}
{\includegraphics[angle=0,width=0.32\linewidth,type=pdf,ext=.pdf,read=.pdf]{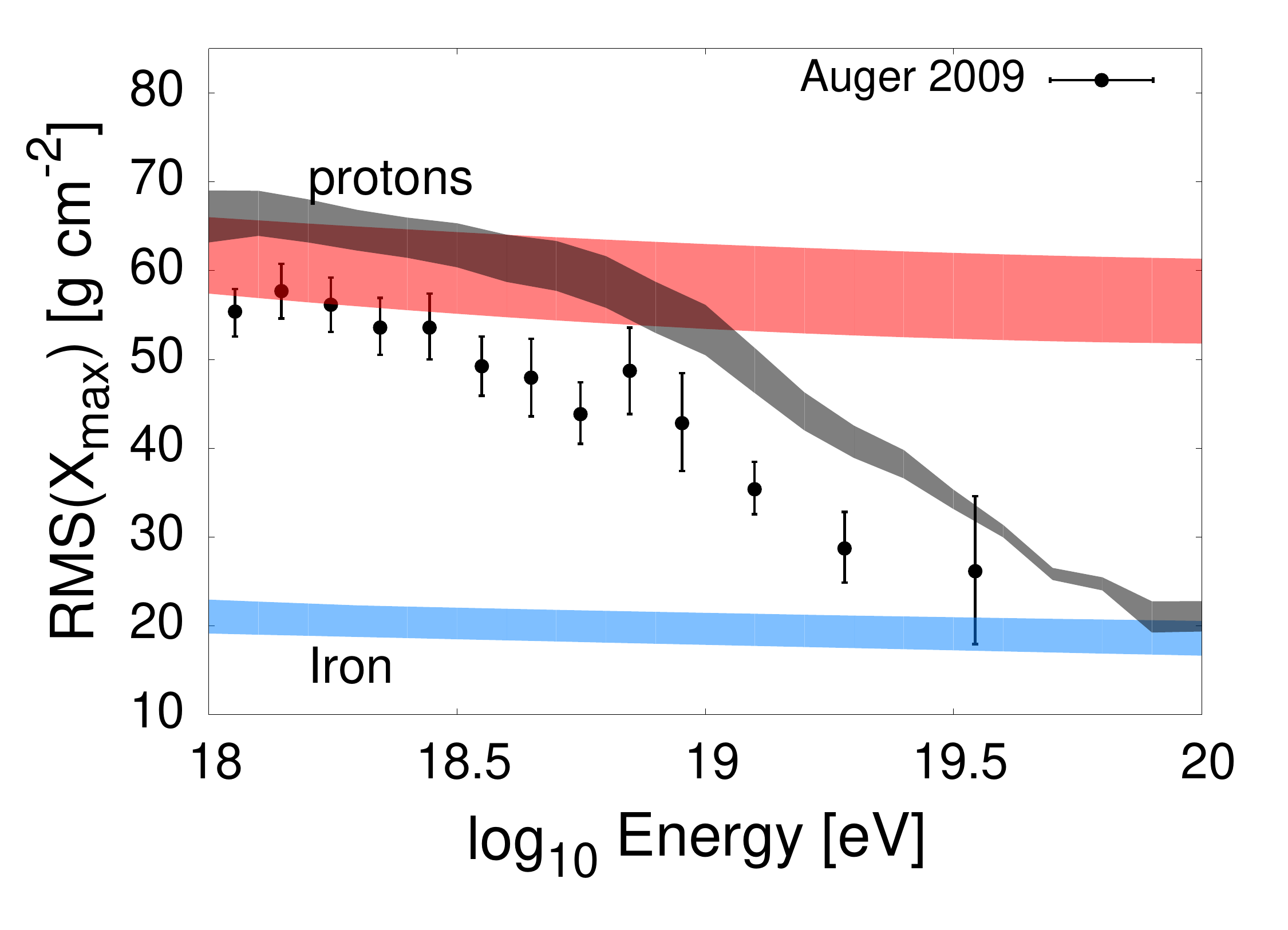}}\\

{\includegraphics[angle=0,width=0.32\linewidth,type=pdf,ext=.pdf,read=.pdf]{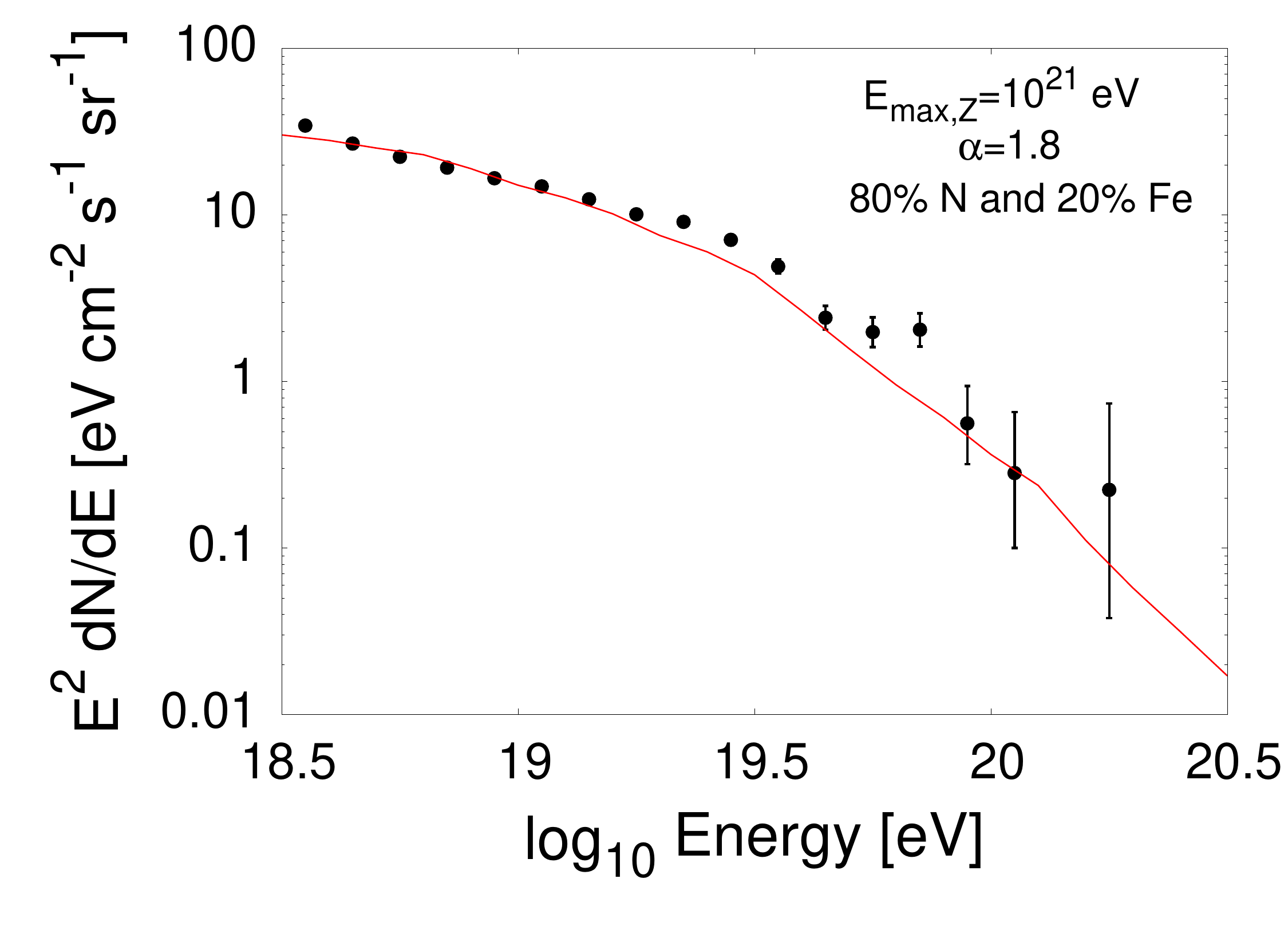}}
{\includegraphics[angle=0,width=0.32\linewidth,type=pdf,ext=.pdf,read=.pdf]{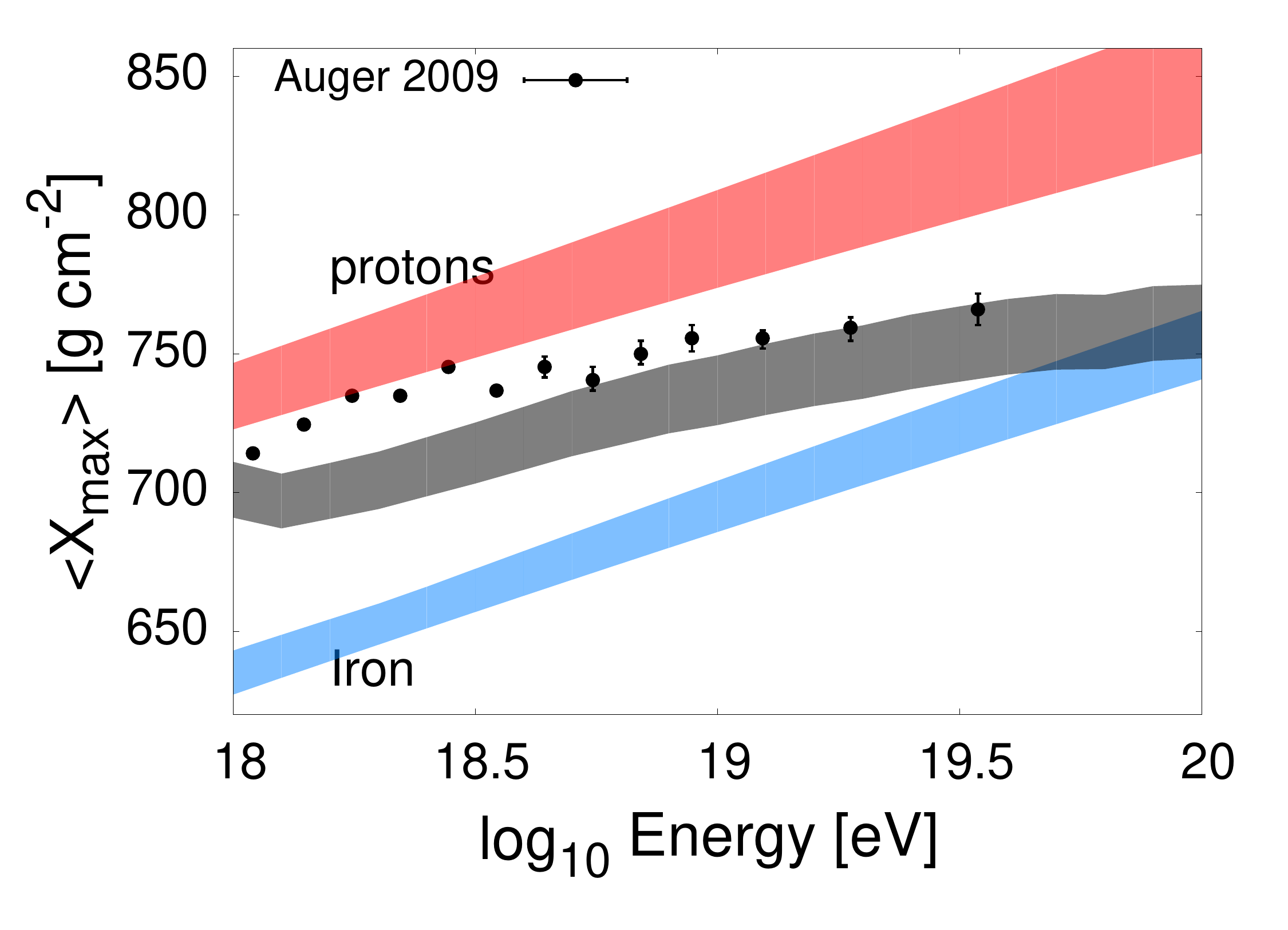}}
{\includegraphics[angle=0,width=0.32\linewidth,type=pdf,ext=.pdf,read=.pdf]{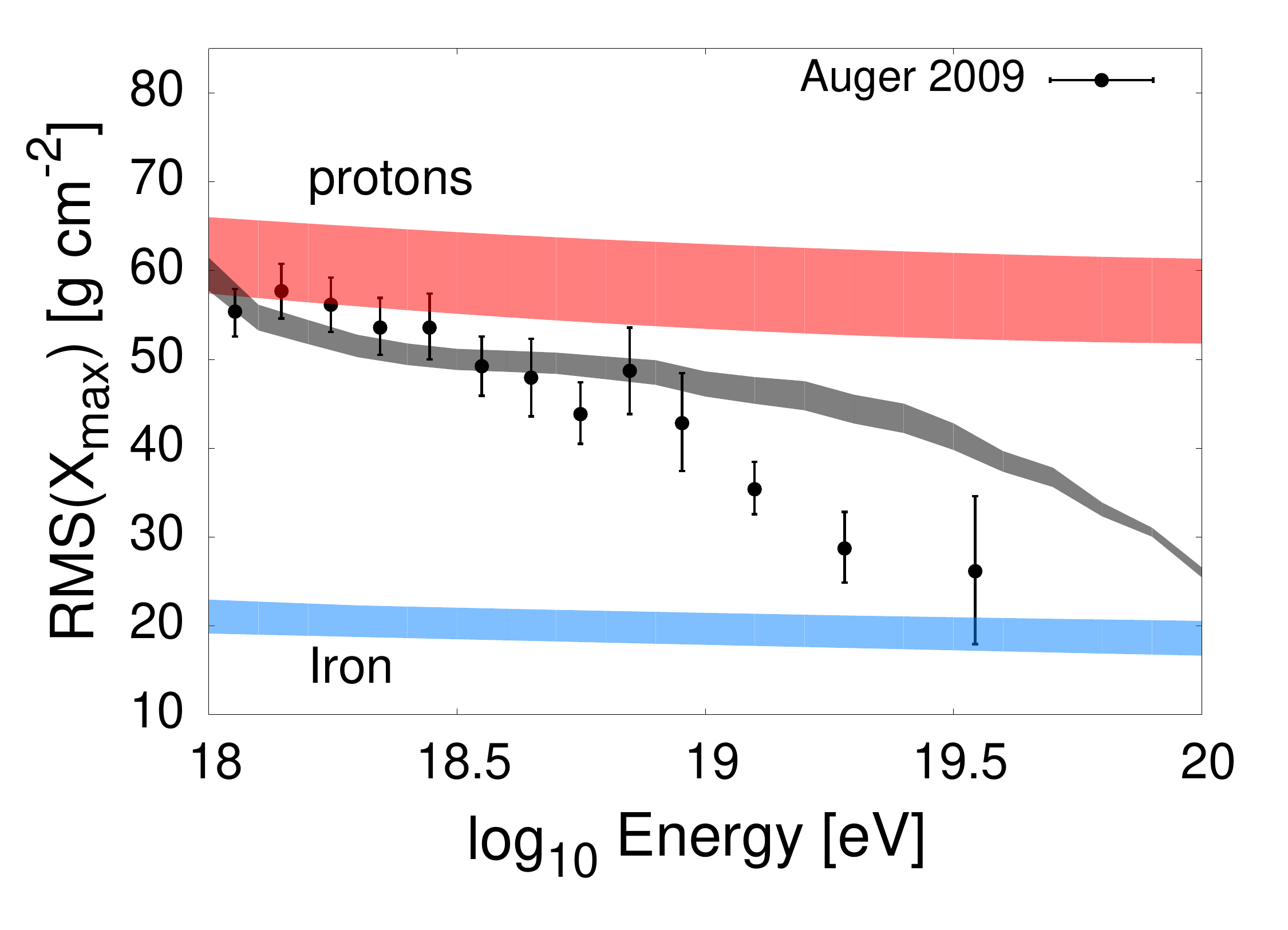}}\\

{\includegraphics[angle=0,width=0.32\linewidth,type=pdf,ext=.pdf,read=.pdf]{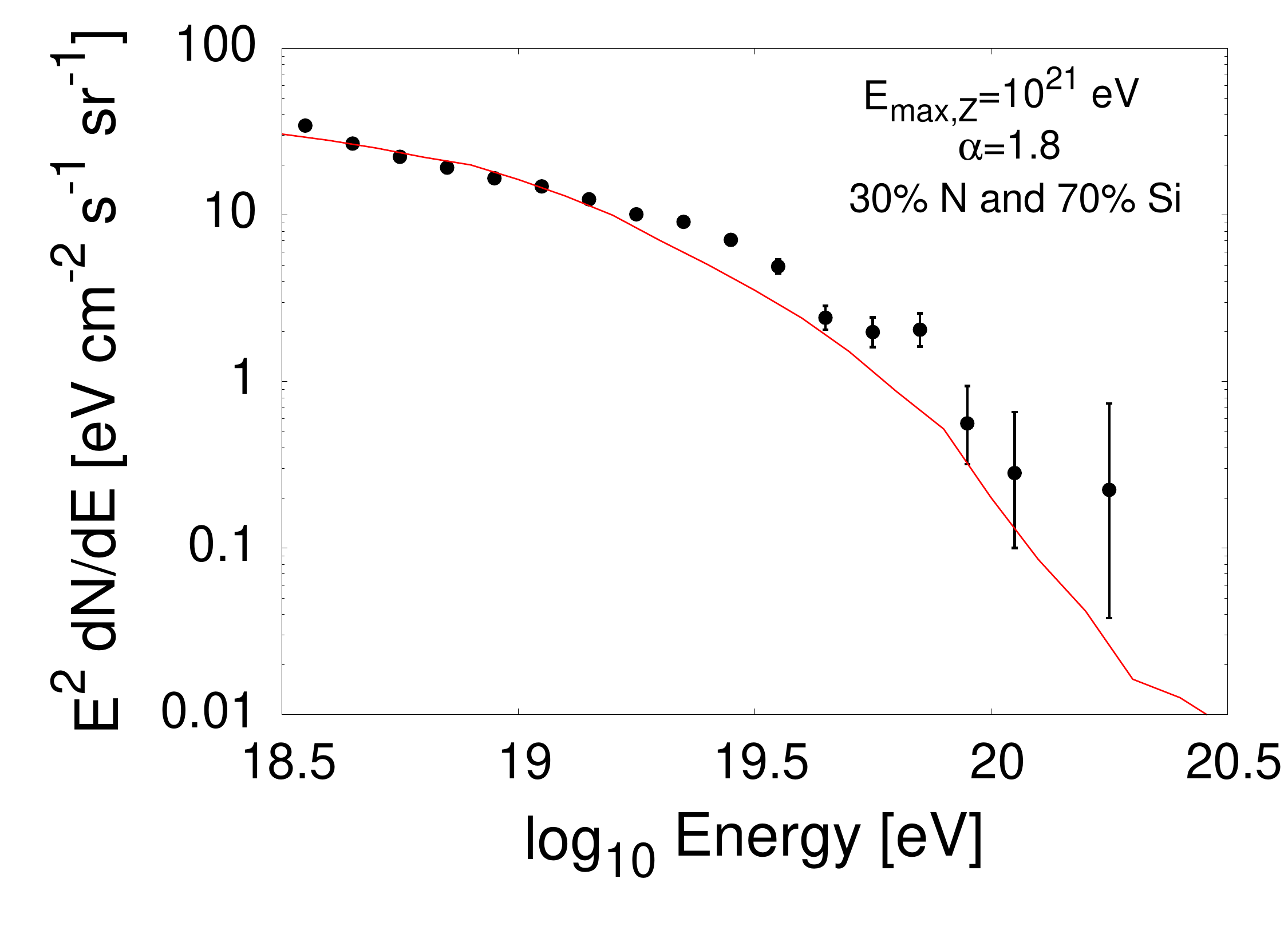}}
{\includegraphics[angle=0,width=0.32\linewidth,type=pdf,ext=.pdf,read=.pdf]{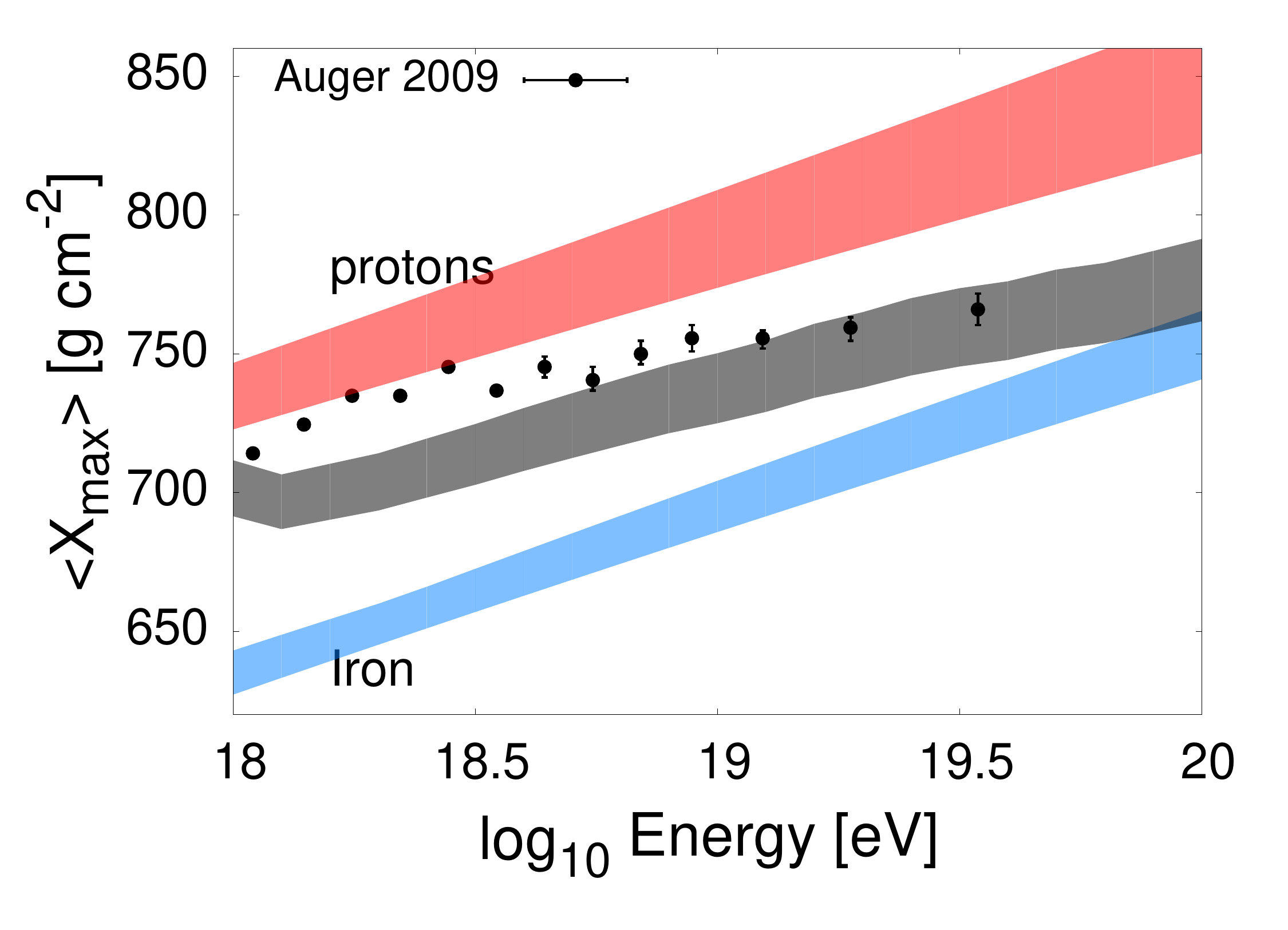}}
{\includegraphics[angle=0,width=0.32\linewidth,type=pdf,ext=.pdf,read=.pdf]{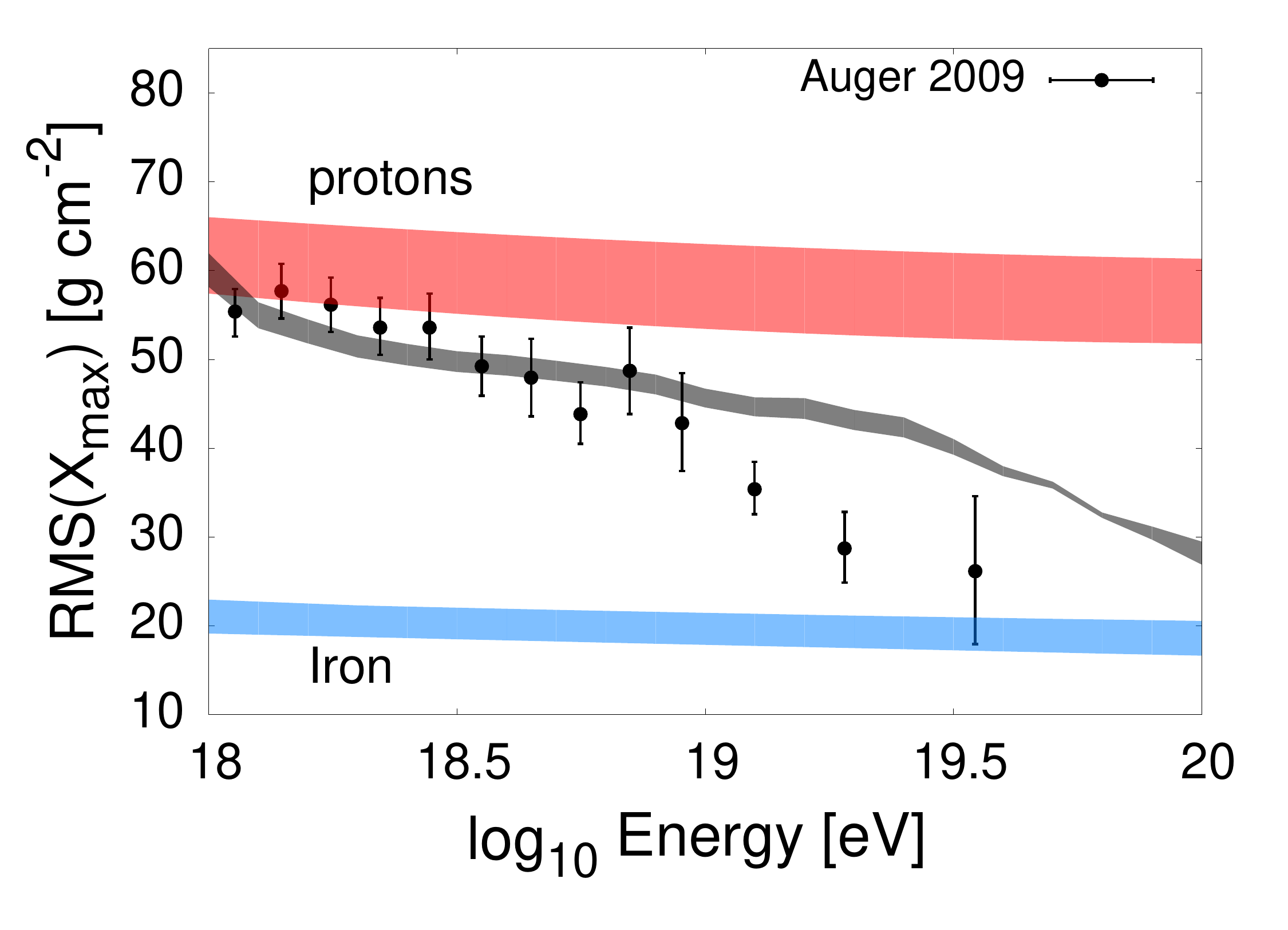}}\\

{\includegraphics[angle=0,width=0.32\linewidth,type=pdf,ext=.pdf,read=.pdf]{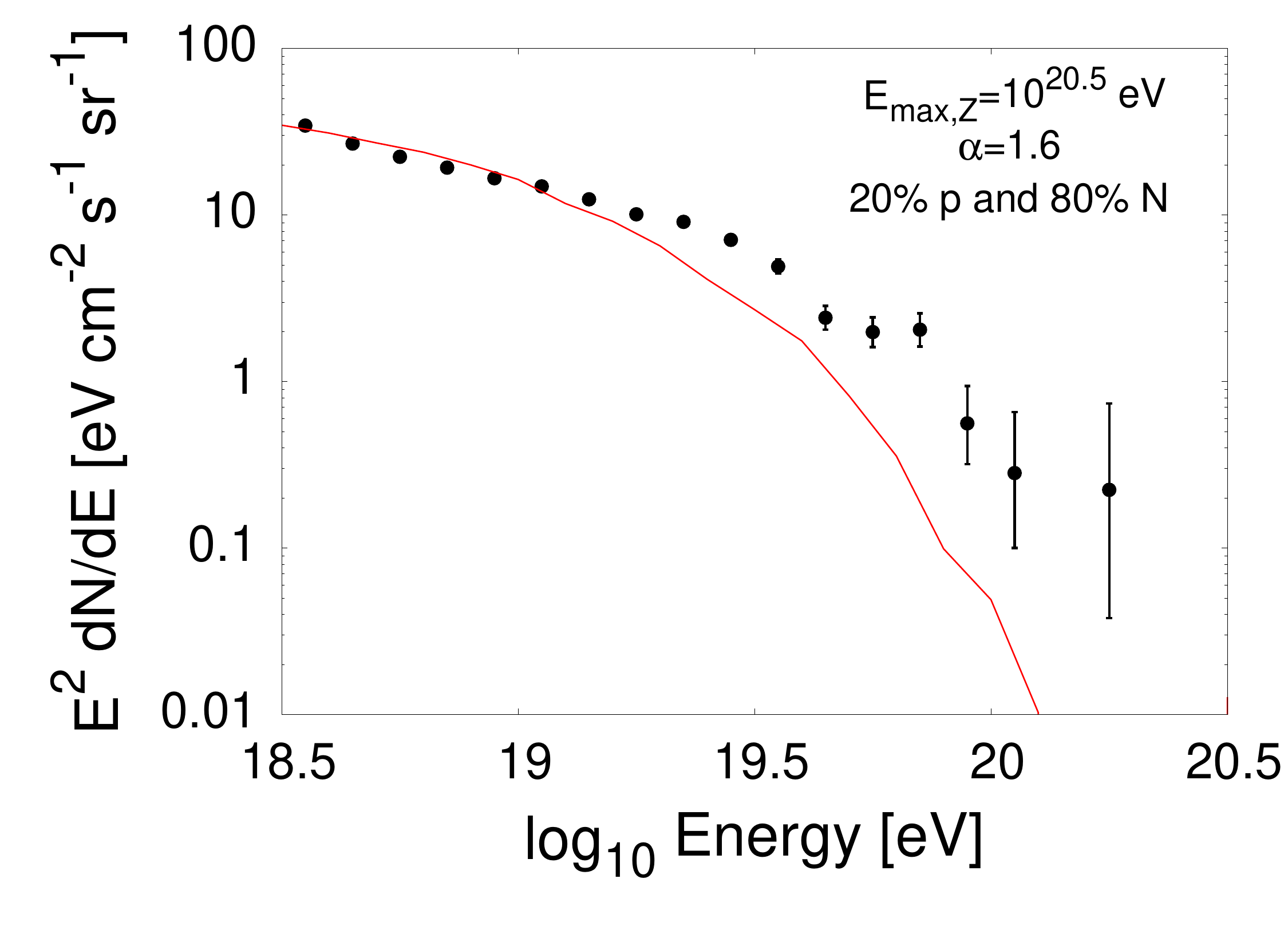}}
{\includegraphics[angle=0,width=0.32\linewidth,type=pdf,ext=.pdf,read=.pdf]{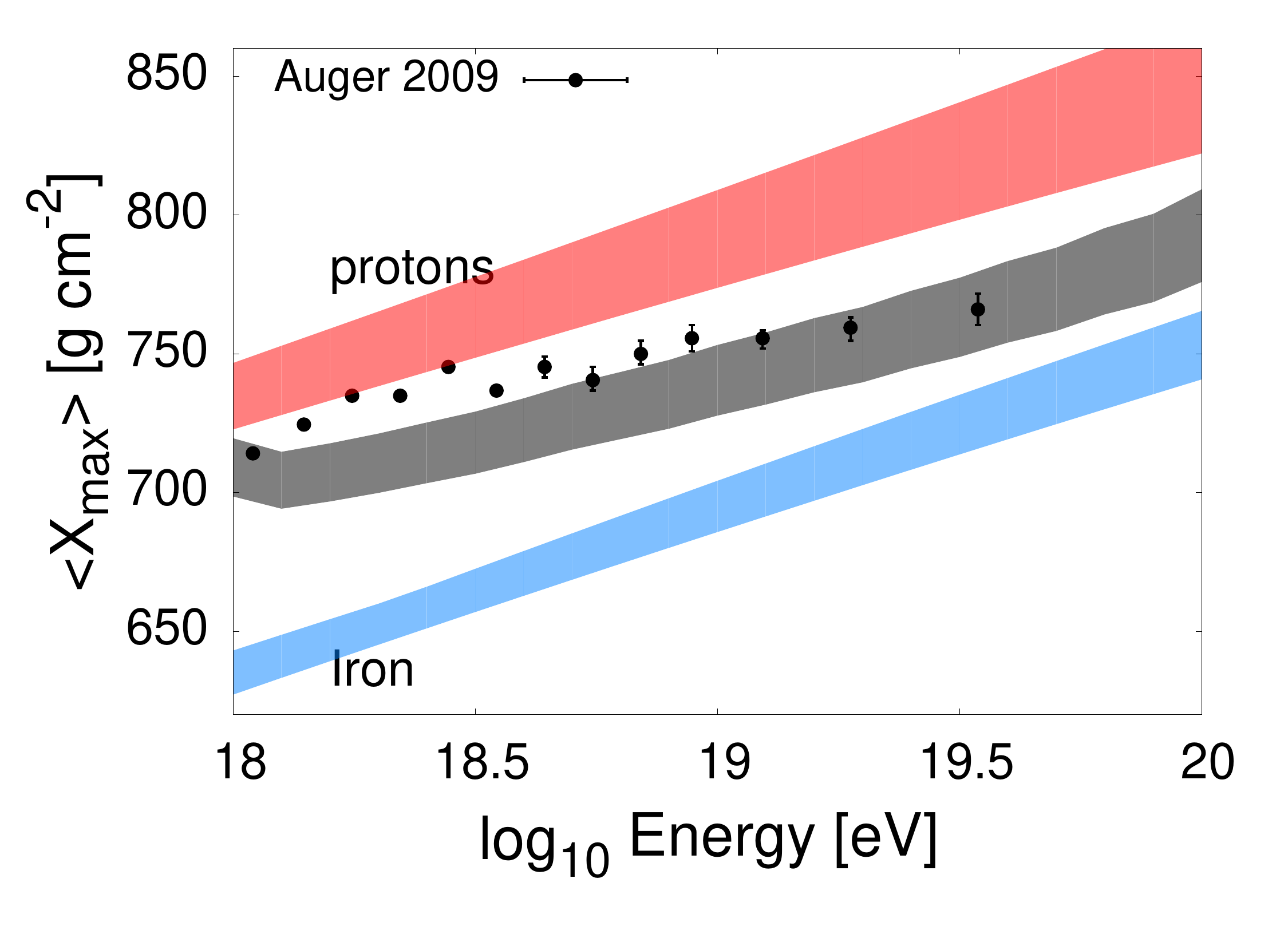}}
{\includegraphics[angle=0,width=0.32\linewidth,type=pdf,ext=.pdf,read=.pdf]{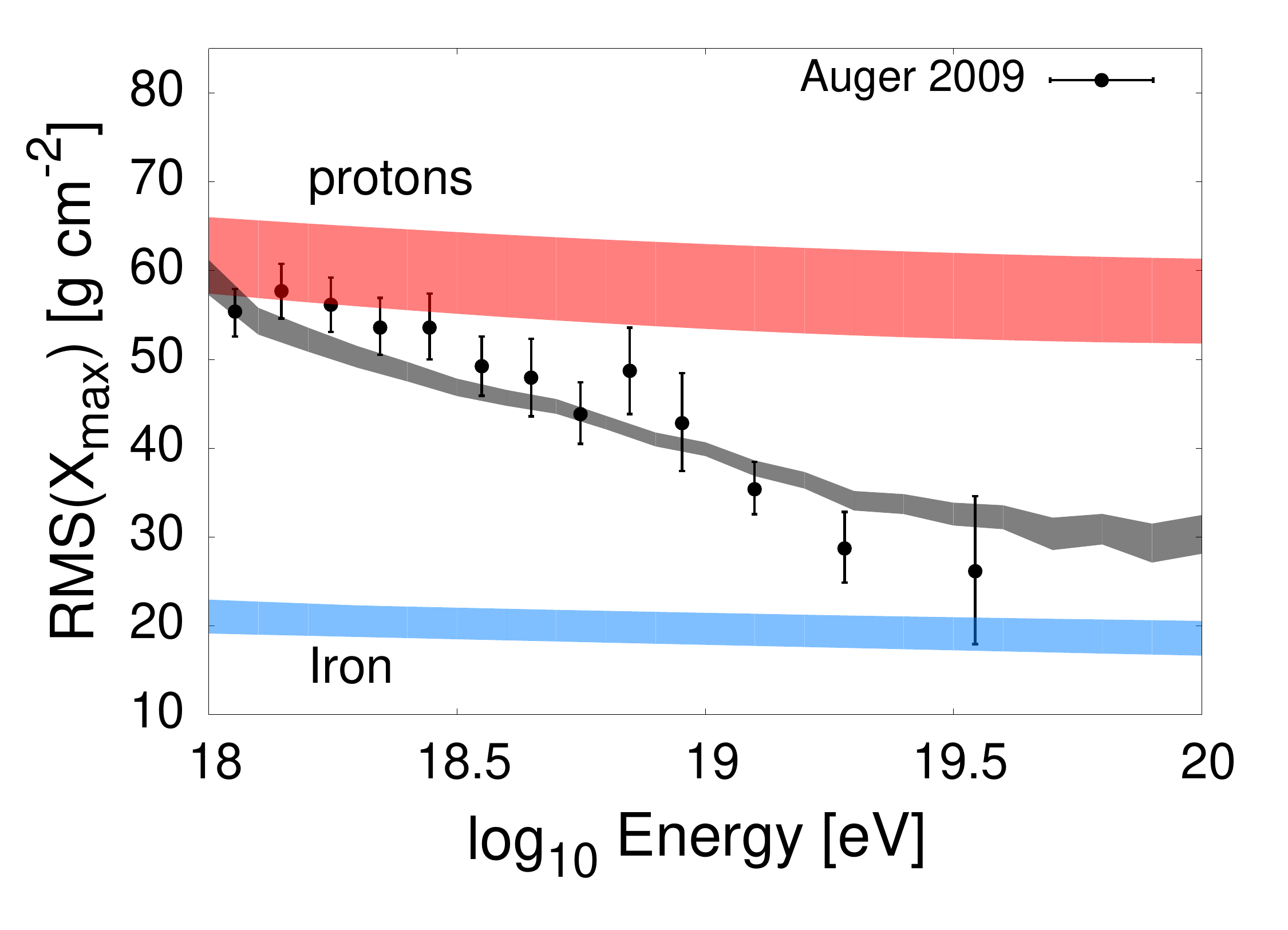}}\\

{\includegraphics[angle=0,width=0.32\linewidth,type=pdf,ext=.pdf,read=.pdf]{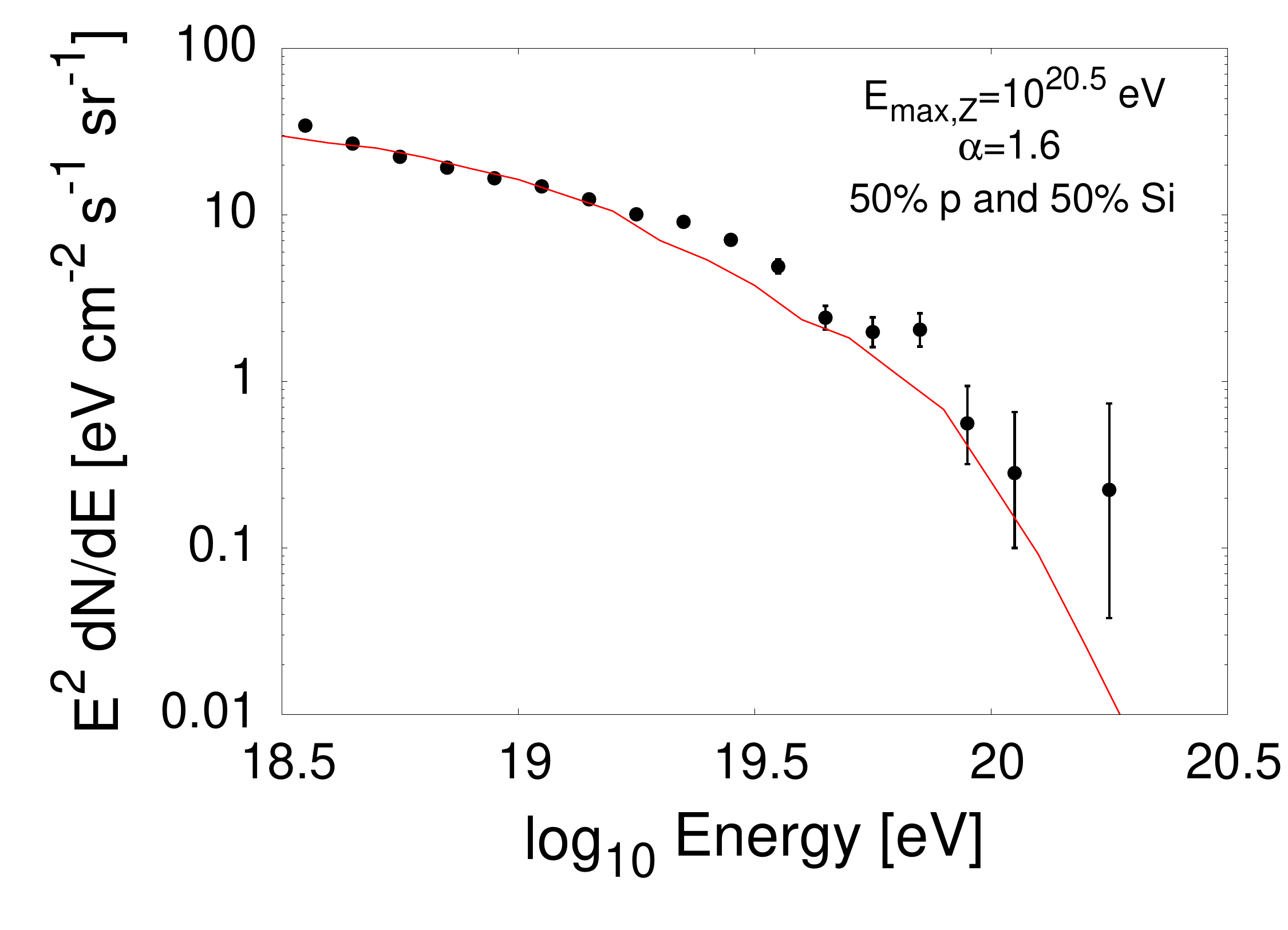}}
{\includegraphics[angle=0,width=0.32\linewidth,type=pdf,ext=.pdf,read=.pdf]{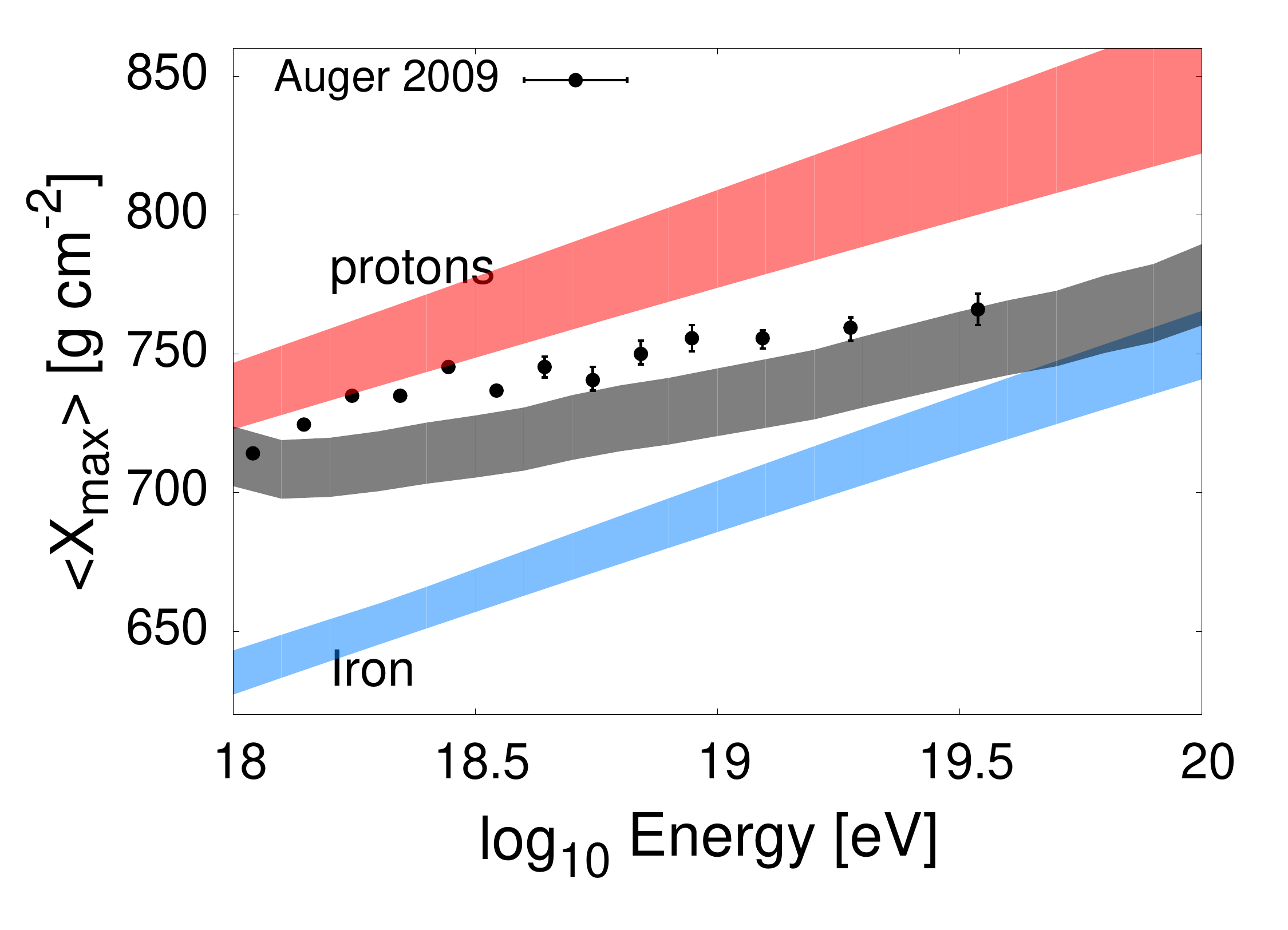}}
{\includegraphics[angle=0,width=0.32\linewidth,type=pdf,ext=.pdf,read=.pdf]{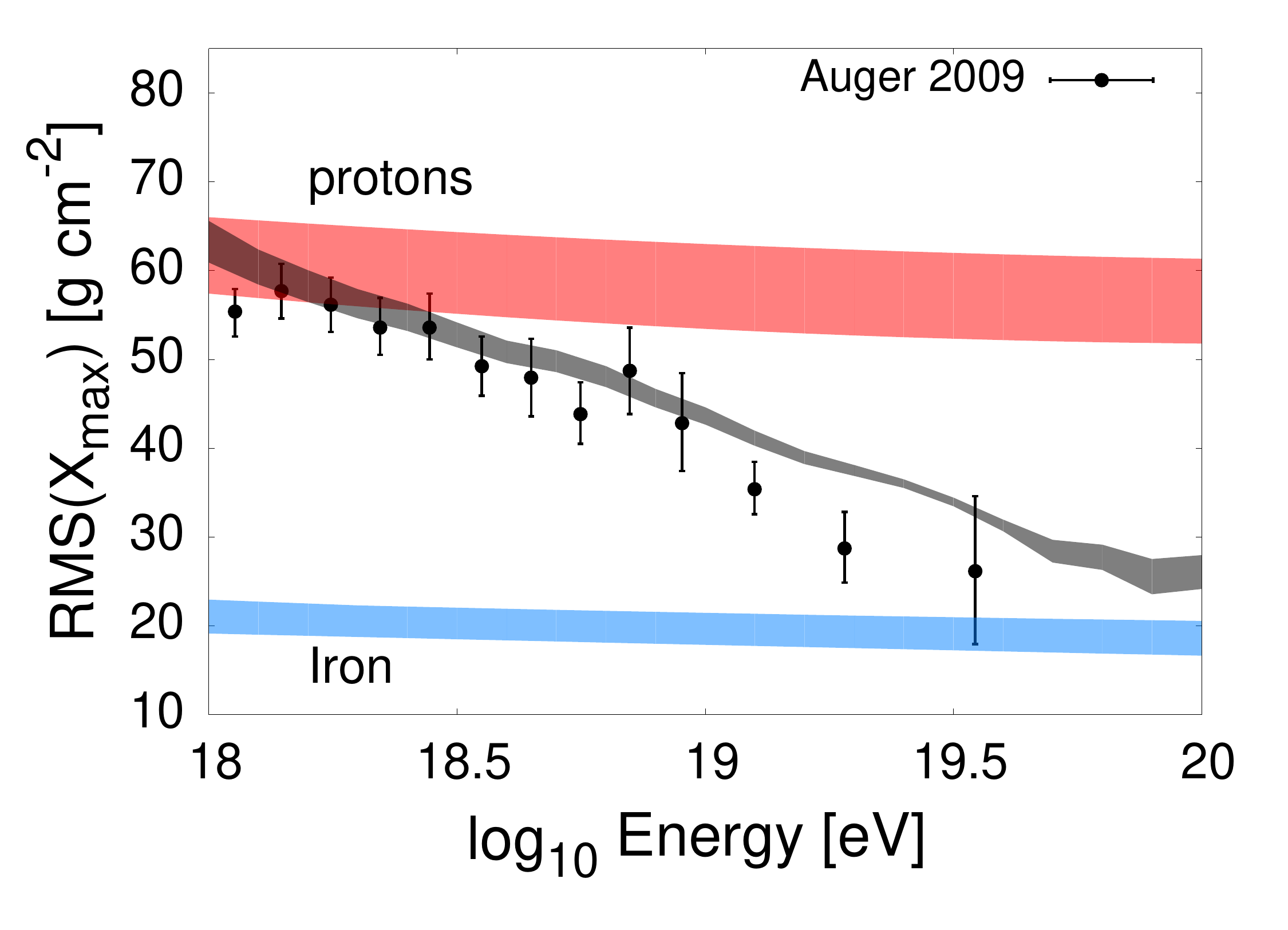}}\\
\caption{The same as in previous figures, but for selected cases in which a mixed chemical composition is injected from cosmic ray accelerators. From top-to-bottom, the chemical mixture at injection is given by: 80\% protons and 20\% iron, 80\% nitrogen and 20\% iron, 30\% nitrogen and 70\% silicon, 20\% protons and 80\% nitrogen, and 50\% protons and 50\% silicon, respectively. The injected spectral index, $\alpha$, was selected in each case to best match the spectral shape reported by the PAO. The ratios of different elements were chosen to provide the best fit to the combination of $\langle X_{\rm max} \rangle$ and RMS($X_{\rm max}$). In each case, we have used $E_{\rm max}$ of either $10^{20.5}$ eV or $10^{21}$ eV, depending on which yields the best fit, 
where $E_{\rm max}$ refers to the cutoff energy for iron nuclei, which is related to the cutoff for a species with electric charge $Z$ by $E_{{\rm max}, Z} = (Z/26)\times E_{\rm max}$. See text for more details.}
\label{mixed2}
\end{center}
\end{figure*}

In Fig.~\ref{mixed2}, we show results for some selected cases of mixed chemical composition. In particular, we consider combinations of protons, nitrogen, silicon, or iron, and, as before, have selected in each case the injected spectral index, energy cutoff, and fractional mixture that provides the best fit to the PAO spectrum, $\langle X_{\rm max} \rangle$ and ${\mathrm{RMS}}(X_{\rm max})$ measurements. We find that while either the $\langle X_{\rm max} \rangle$ or the ${\mathrm{RMS}}(X_{\rm max})$ data can be accommodated without much difficulty, it is again challenging to find an injected spectrum and composition that is in agreement with both $\langle X_{\rm max} \rangle$ and ${\mathrm{RMS}}(X_{\rm max})$ simultaneously. In particular, if $\langle X_{\rm max} \rangle$ is well fit, then the predictions for ${\mathrm{RMS}}(X_{\rm max})$ tend to exceed the measured values. Similarly, if the measurements of ${\mathrm{RMS}}(X_{\rm max})$ are well fit, the predictions for $\langle X_{\rm max} \rangle$ tend to be well below the measured values. The only exception we find to this conclusion is the case in which the injected spectrum consists of a mixture of nitrogen nuclei (80\%) and protons (20\%), which (using the EPOS hadronic simulation) can provide a reasonable fit to both $\langle X_{\rm max} \rangle$ and ${\mathrm{RMS}}(X_{\rm max})$ measurements. In this case, however, the spectral shape drops far more rapidly than measured by the PAO, leading to a poor fit. The mixtures of chemical species which we have found lead to the best overall fit to the PAO's spectrum, $\langle X_{\rm max} \rangle$, and ${\mathrm{RMS}}(X_{\rm max})$ measurements (80\% nitrogen and 20\% iron, 80\% silicon and 20\% nitrogen, and 90\% silicon and 10\% protons) accommodate the data only about as well as in the case of a pure silicon injection spectrum. Thus, the consideration of a mixed, two component, chemical composition from cosmic ray sources does not significantly improve the agreement with the PAO data.

\section{The Impact of Intergalactic Magnetic Fields}
\label{resultsb}

A charged cosmic ray moving through a uniform magnetic field will be deflected by an angle $\theta_0 = L_{\rm coh}/R_L$, where $L_{\rm coh}$ is the size of the region being traveled through (the coherence length) and $R_{L}$ is the Larmor radius of the particle. Over the course of its propagation, a typical UHECR will travel through a large number of regions with differing magnetic orientations, which over a distance $D$ leads to an overall deflection of $\theta \approx \theta_0 \sqrt{D/L_{\rm coh}}$.  For nano-Gauss scale fields with coherence lengths of 1~Mpc, this yields
\begin{eqnarray}
\theta \approx 17^{\circ} \bigg(\frac{10^{20}\,{\rm eV}}{E}\bigg) \bigg(\frac{D}{10\,{\rm Mpc}}\bigg)^{0.5} \bigg(\frac{L_{\rm coh}}{{\rm Mpc}}\bigg)^{0.5}\bigg(\frac{B}{{\rm nG}}\bigg) \bigg(\frac{Z}{10}\bigg),
\end{eqnarray}
where $E$ is the energy of the cosmic ray, $Z$ is its charge, and $B$ is the strength of the intergalactic magnetic field. Such deflections cause cosmic rays to propagate over longer effective distances than the straight line distance, with the path extension, in the small angle limit, being approximately given by:
\begin{widetext}
\begin{eqnarray}
\frac{D_{\rm eff}}{D}\approx 1+\frac{\theta^2}{2}\approx 1 + \bigg[0.04 \, \bigg(\frac{10^{20}\,{\rm eV}}{E}\bigg)^2  \bigg(\frac{D}{10\,{\rm Mpc}}\bigg)  \bigg(\frac{L_{\rm coh}}{{\rm Mpc}}\bigg) \bigg(\frac{B}{{\rm nG}}\bigg)^2 \bigg(\frac{Z}{10}\bigg)^2\bigg].
\end{eqnarray}
\end{widetext}

\begin{figure*}[!]
\begin{center}
{\includegraphics[angle=0,width=0.32\linewidth,type=pdf,ext=.pdf,read=.pdf]{Fe_Spectrum_Emax_21_alpha_1.6}}
{\includegraphics[angle=0,width=0.32\linewidth,type=pdf,ext=.pdf,read=.pdf]{Fe_Xmax_Emax_21_alpha_1.6}}
{\includegraphics[angle=0,width=0.32\linewidth,type=pdf,ext=.pdf,read=.pdf]{Fe_RMS_Emax_21_alpha_1.6}}\\

{\includegraphics[angle=0,width=0.32\linewidth,type=pdf,ext=.pdf,read=.pdf]{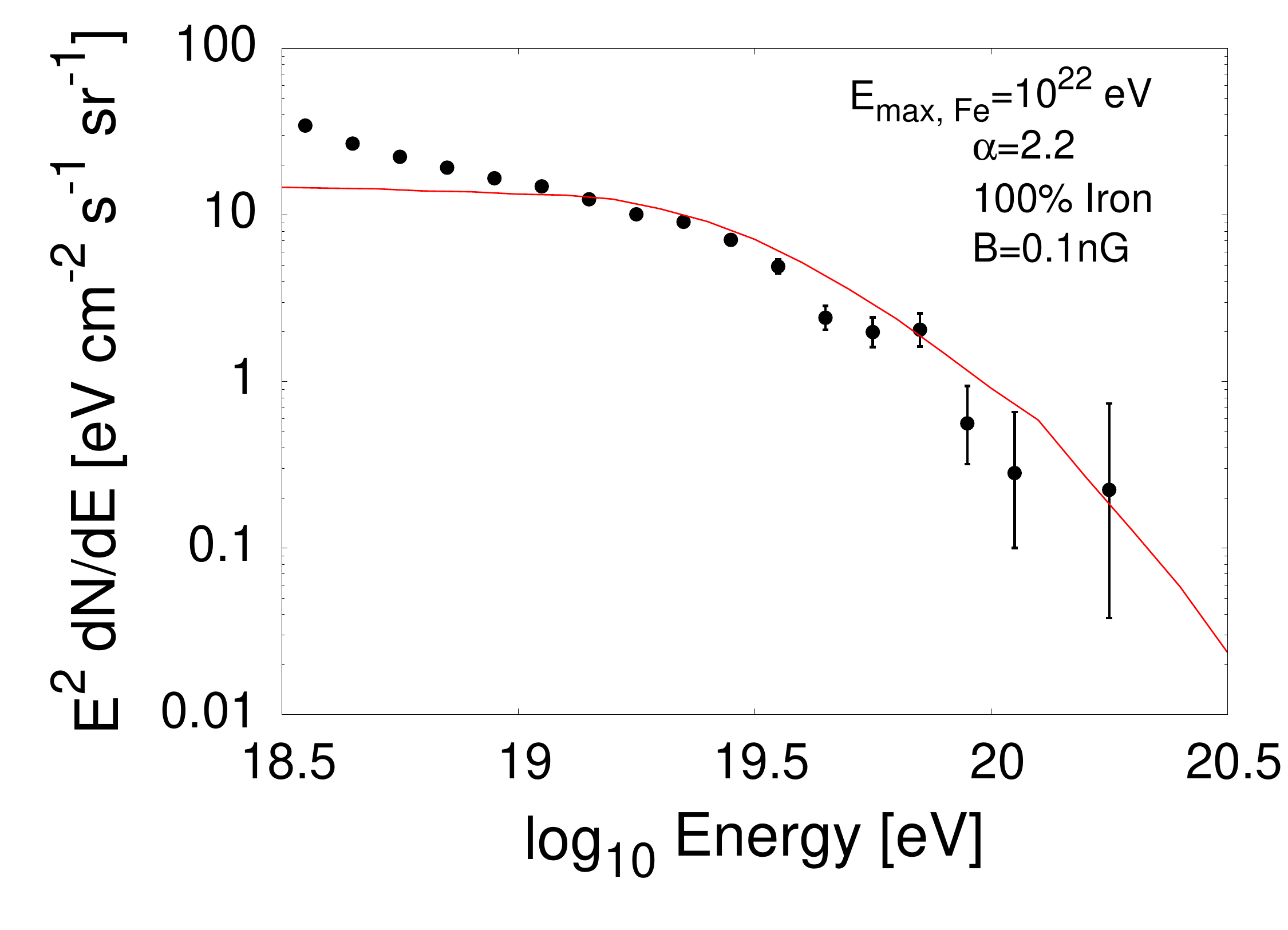}}
{\includegraphics[angle=0,width=0.32\linewidth,type=pdf,ext=.pdf,read=.pdf]{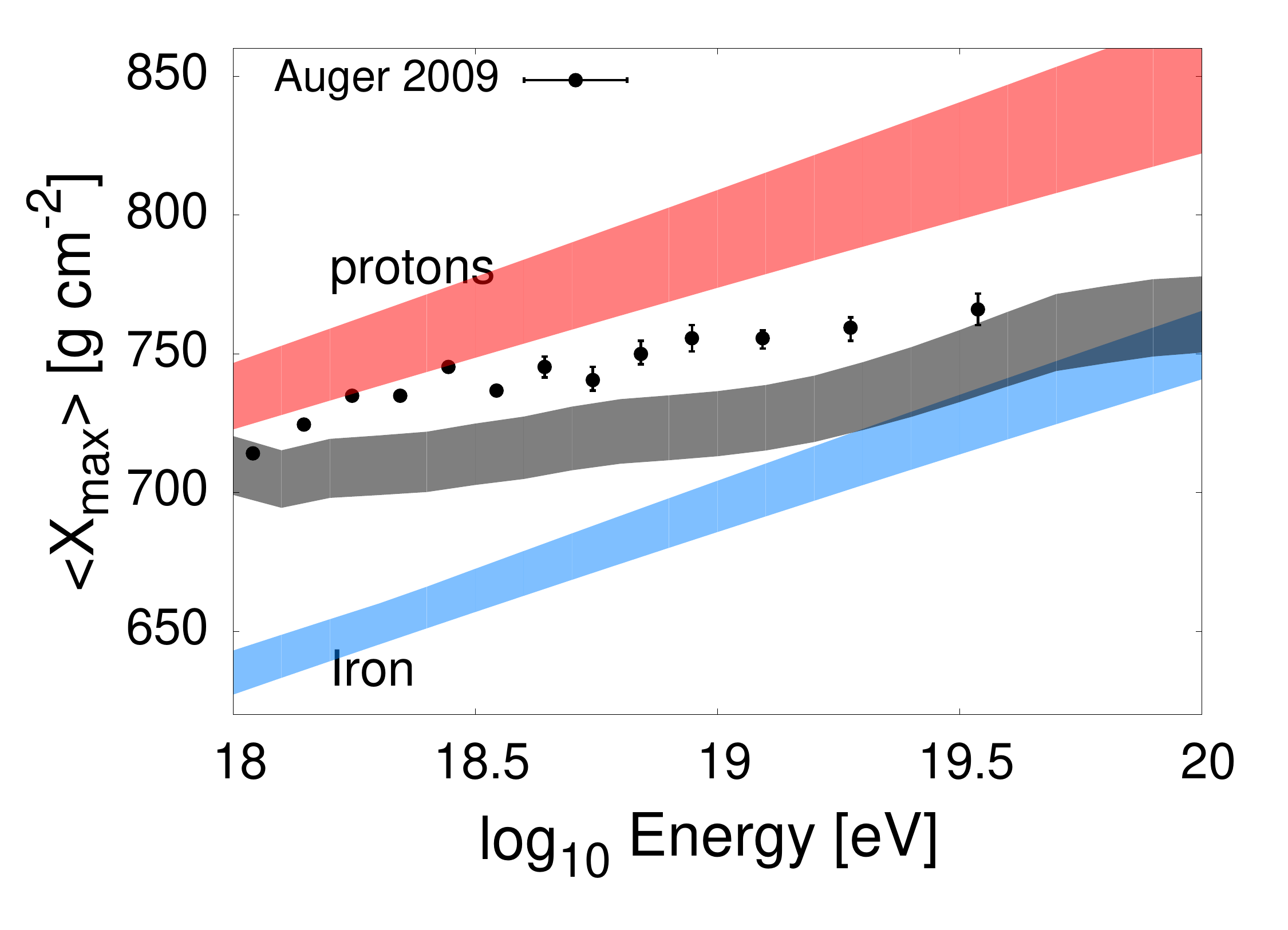}}
{\includegraphics[angle=0,width=0.32\linewidth,type=pdf,ext=.pdf,read=.pdf]{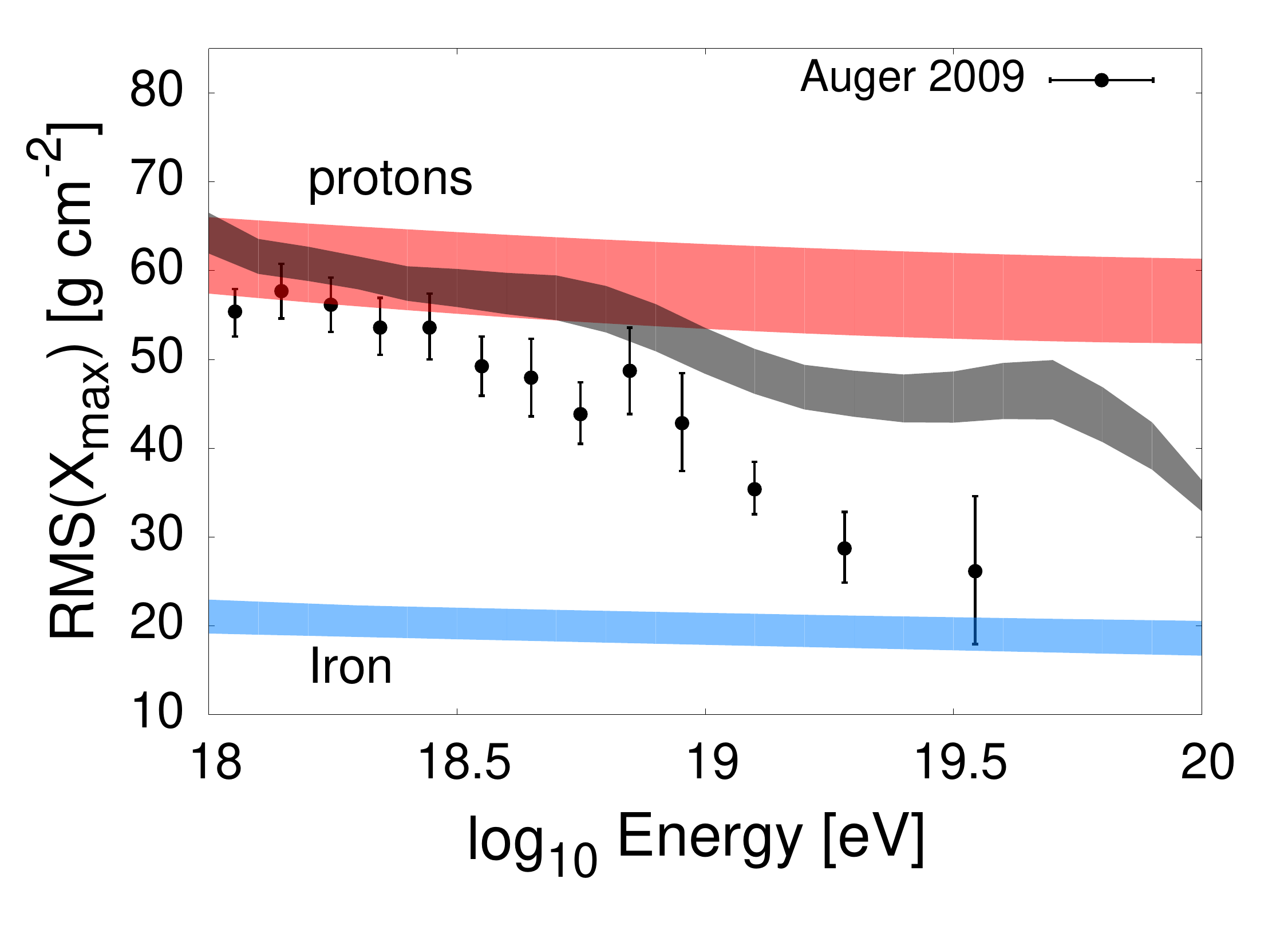}}\\

{\includegraphics[angle=0,width=0.32\linewidth,type=pdf,ext=.pdf,read=.pdf]{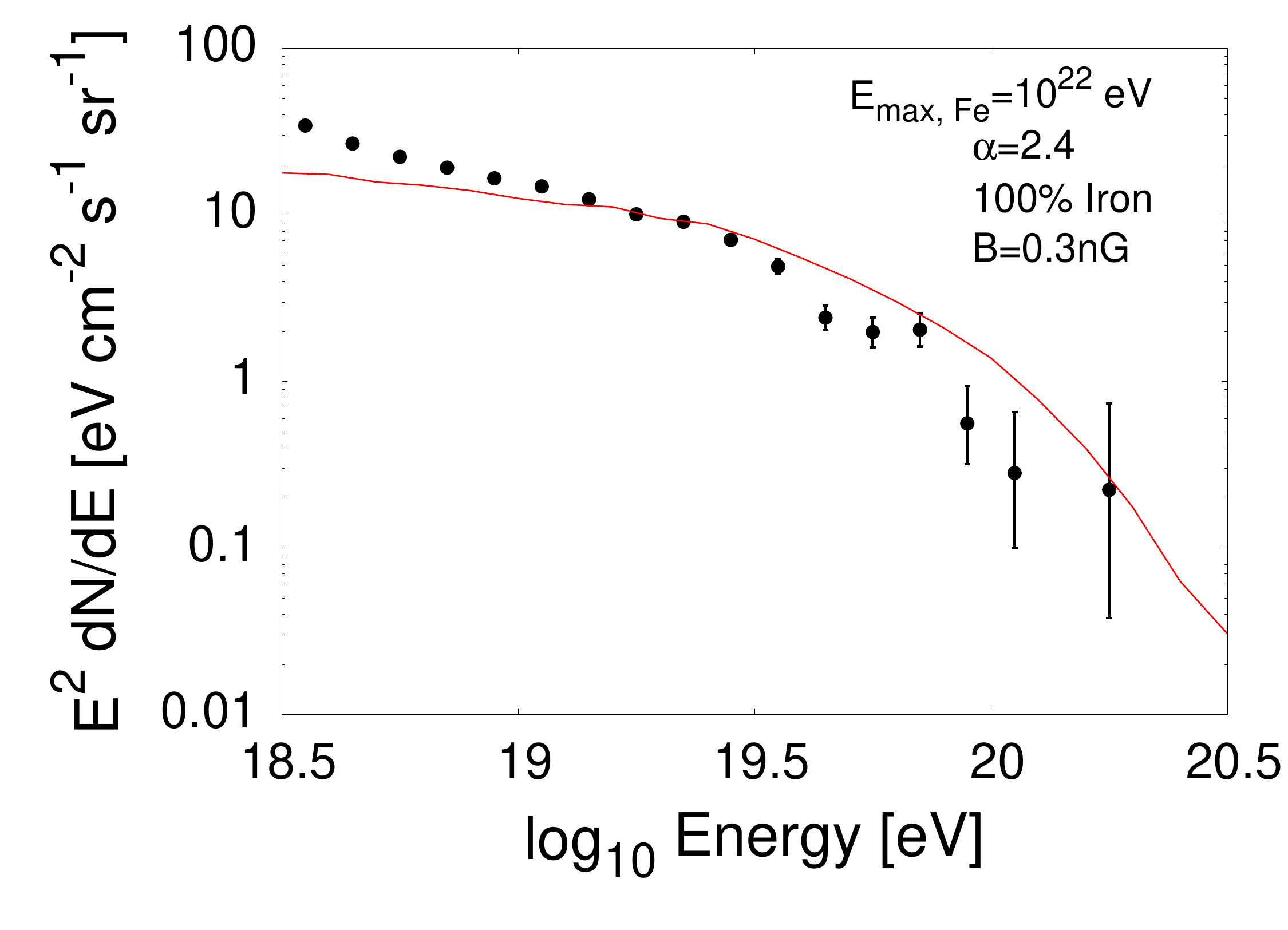}}
{\includegraphics[angle=0,width=0.32\linewidth,type=pdf,ext=.pdf,read=.pdf]{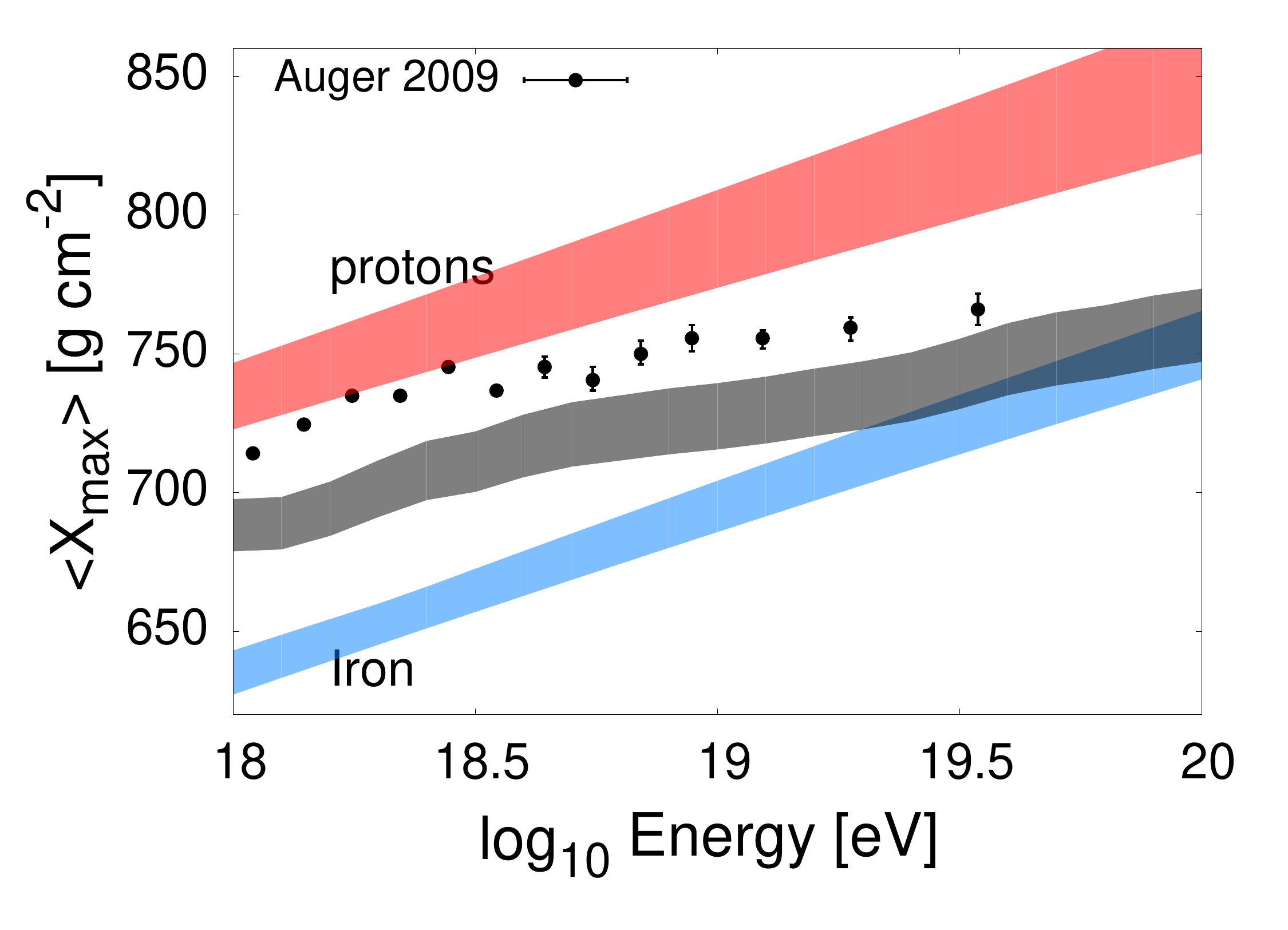}}
{\includegraphics[angle=0,width=0.32\linewidth,type=pdf,ext=.pdf,read=.pdf]{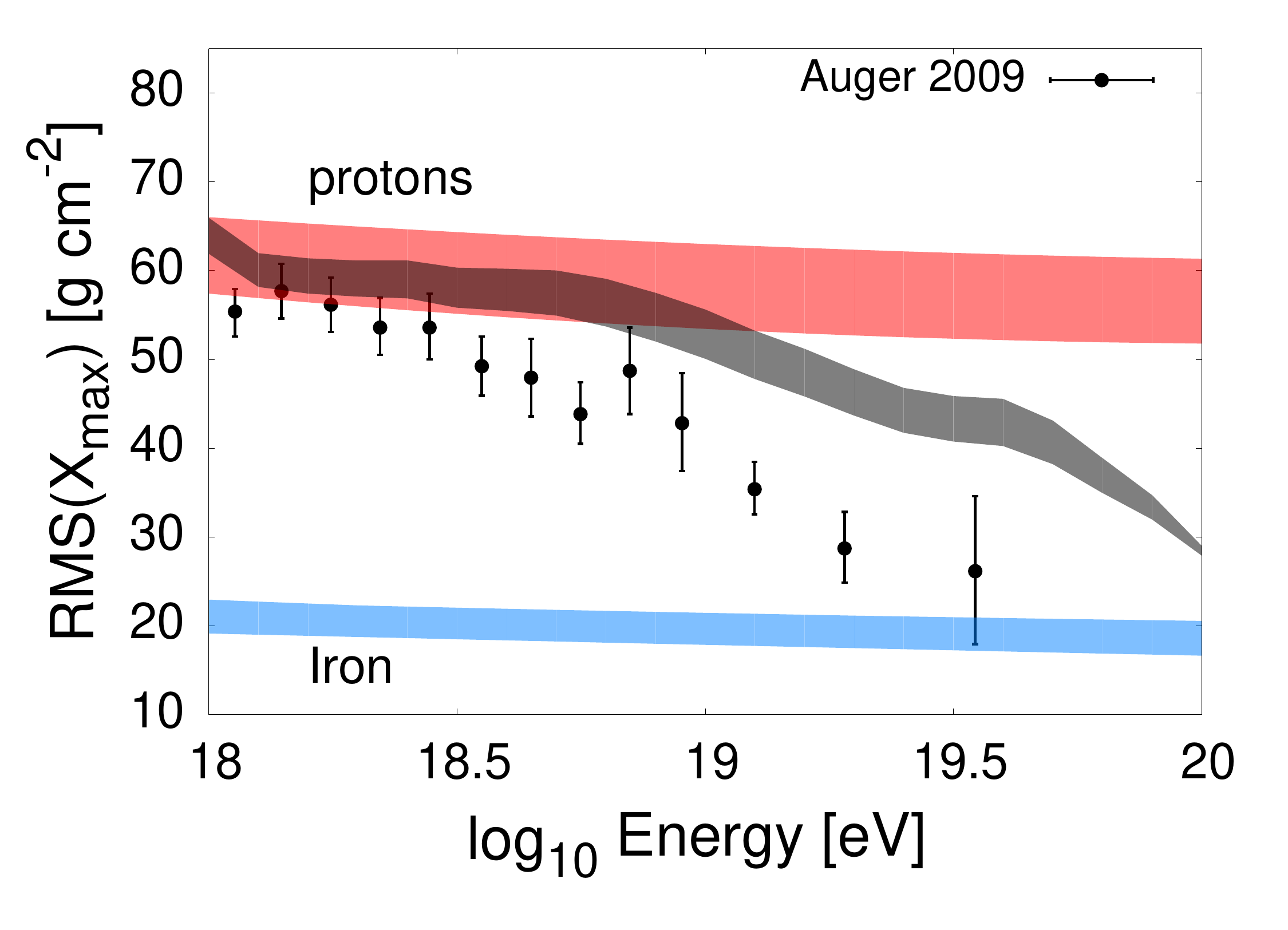}}\\
\caption{Results for an all-iron injection spectrum with negligible magnetic fields (top), 0.1 nano-Gauss  (middle), and 0.3 nano-Gauss (bottom) intergalactic magnetic fields. For the non-zero B-field cases, a coherence length of the magnetic field of 1~Mpc has been used.}
\label{bfield1}
\end{center}
\end{figure*}

\begin{figure*}[!]
\begin{center}
{\includegraphics[angle=0,width=0.32\linewidth,type=pdf,ext=.pdf,read=.pdf]{Si_Spectrum_Emax_21_alpha_1.8}}
{\includegraphics[angle=0,width=0.32\linewidth,type=pdf,ext=.pdf,read=.pdf]{Si_Xmax_Emax_21_alpha_1.8}}
{\includegraphics[angle=0,width=0.32\linewidth,type=pdf,ext=.pdf,read=.pdf]{Si_RMS_Emax_21_alpha_1.8}}\\
\vspace{-0.2cm}
{\includegraphics[angle=0,width=0.32\linewidth,type=pdf,ext=.pdf,read=.pdf]{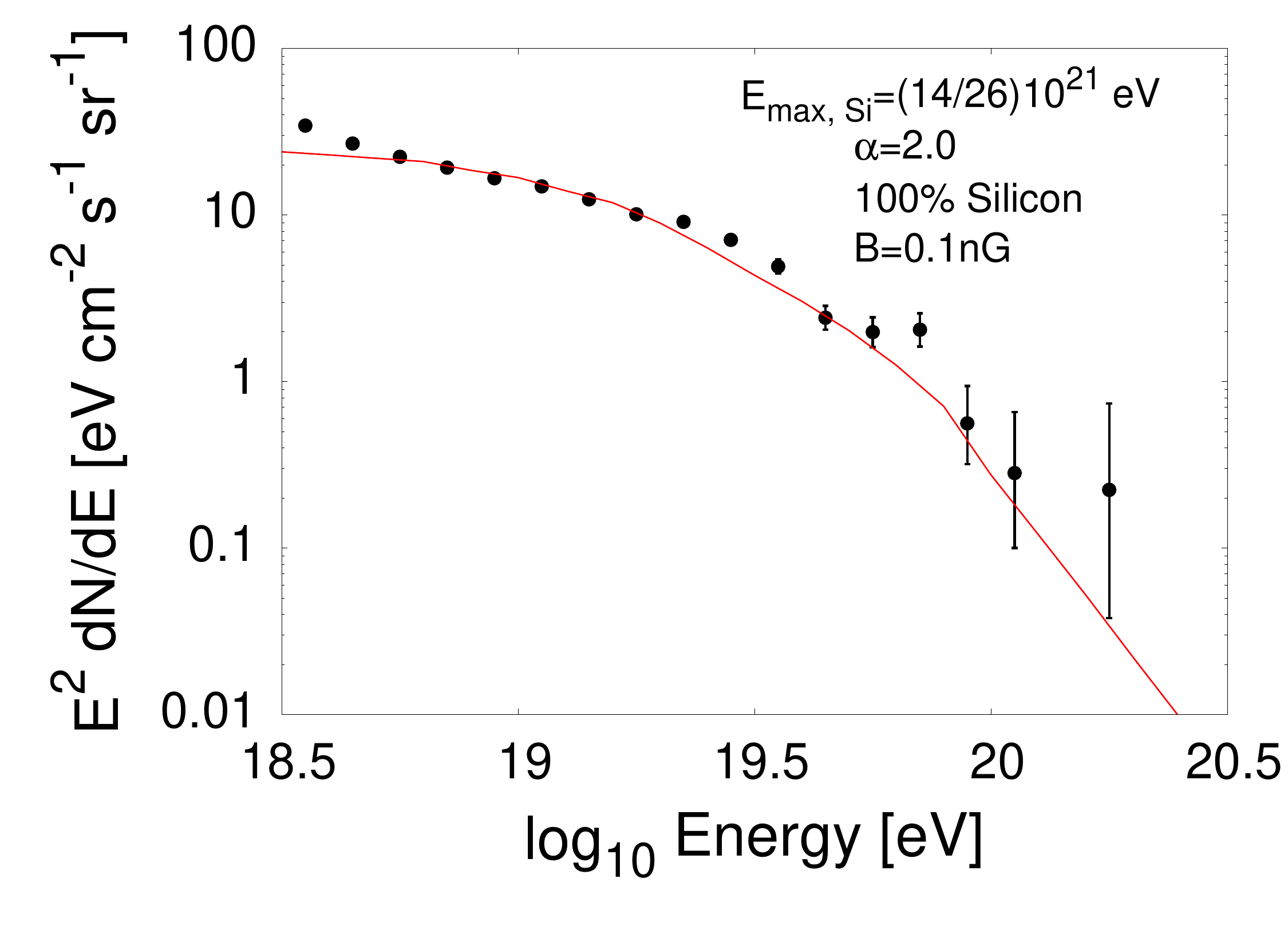}}
{\includegraphics[angle=0,width=0.32\linewidth,type=pdf,ext=.pdf,read=.pdf]{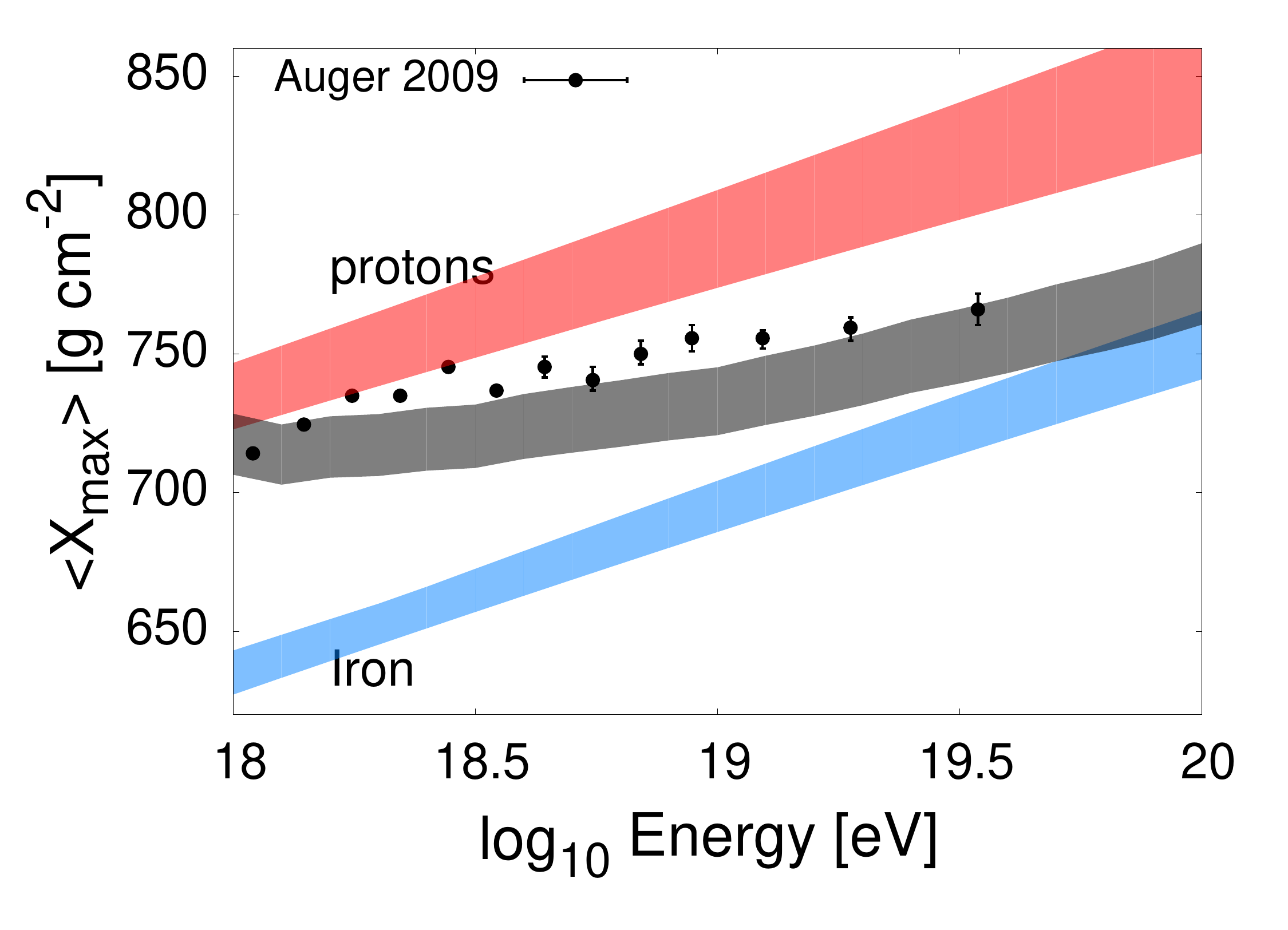}}
{\includegraphics[angle=0,width=0.32\linewidth,type=pdf,ext=.pdf,read=.pdf]{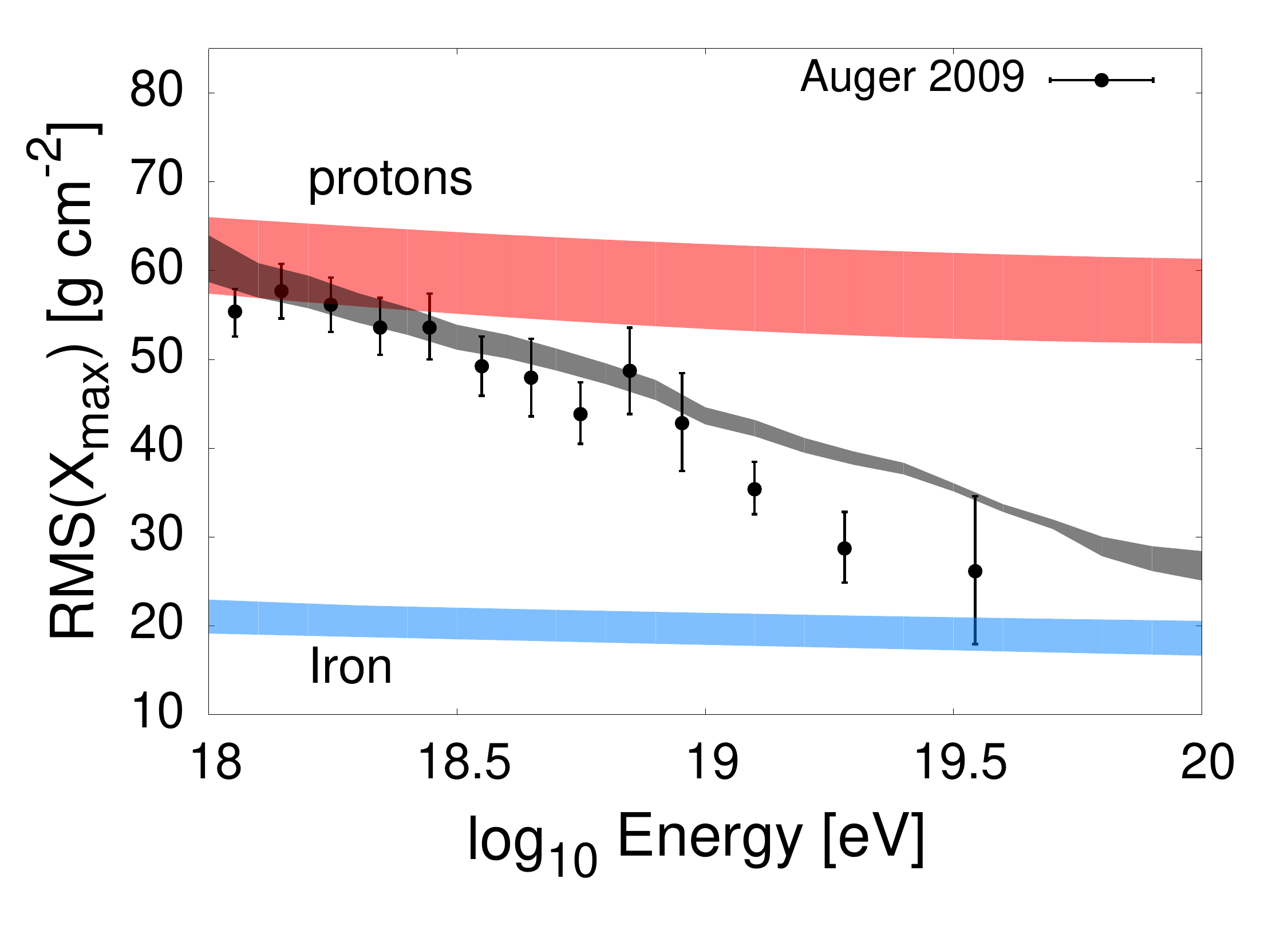}}\\
\vspace{-0.2cm}
{\includegraphics[angle=0,width=0.32\linewidth,type=pdf,ext=.pdf,read=.pdf]{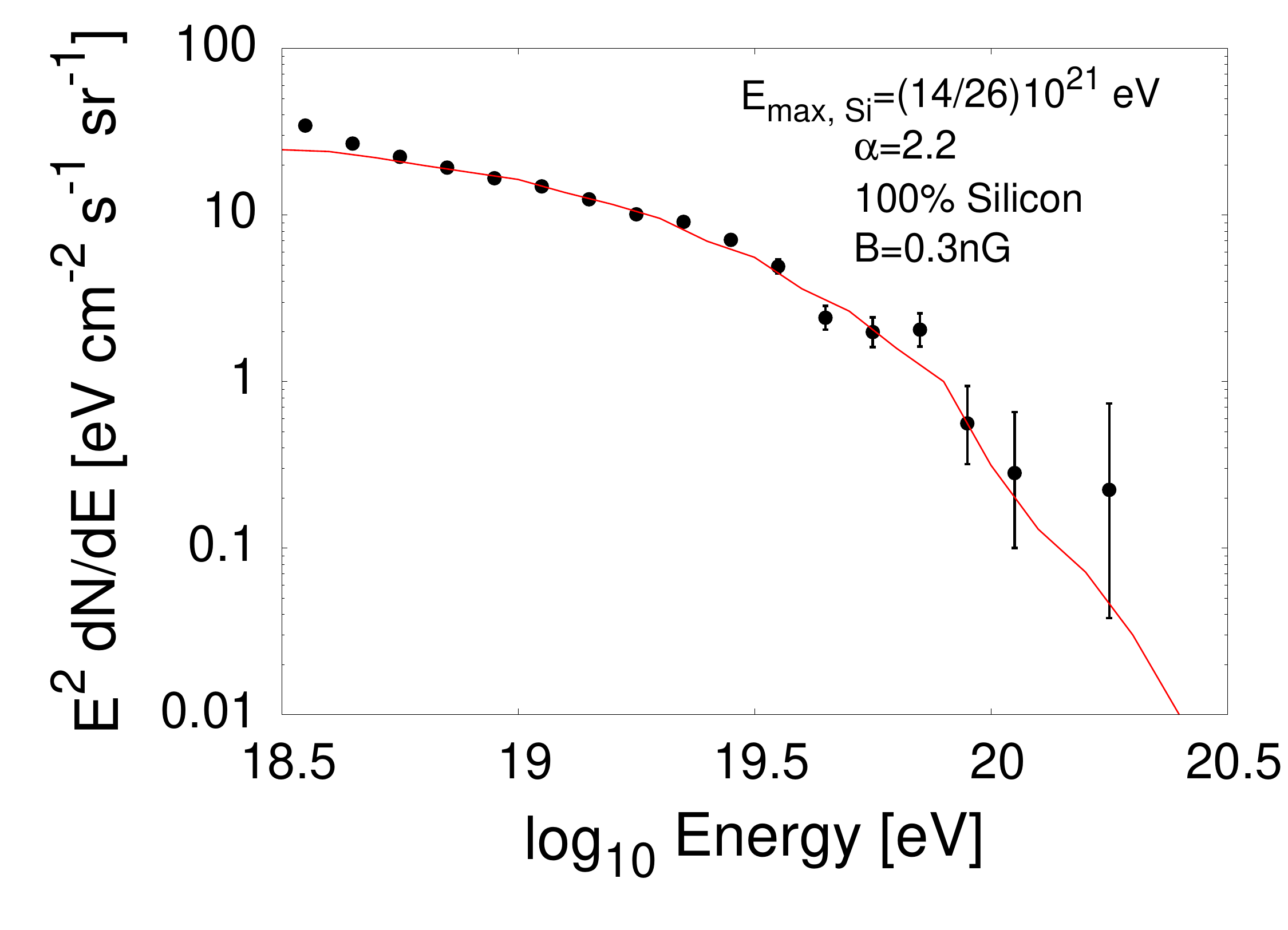}}
{\includegraphics[angle=0,width=0.32\linewidth,type=pdf,ext=.pdf,read=.pdf]{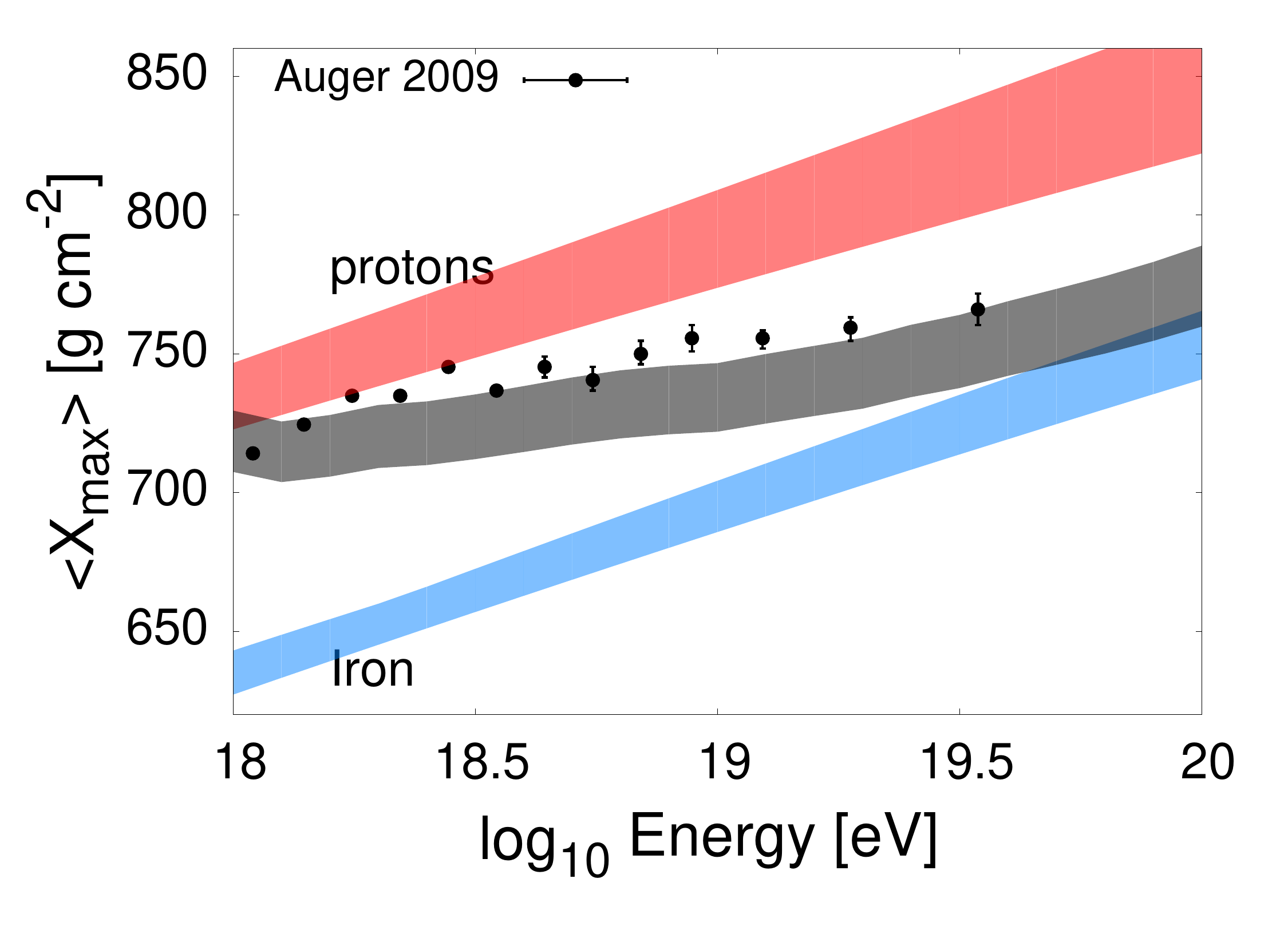}}
{\includegraphics[angle=0,width=0.32\linewidth,type=pdf,ext=.pdf,read=.pdf]{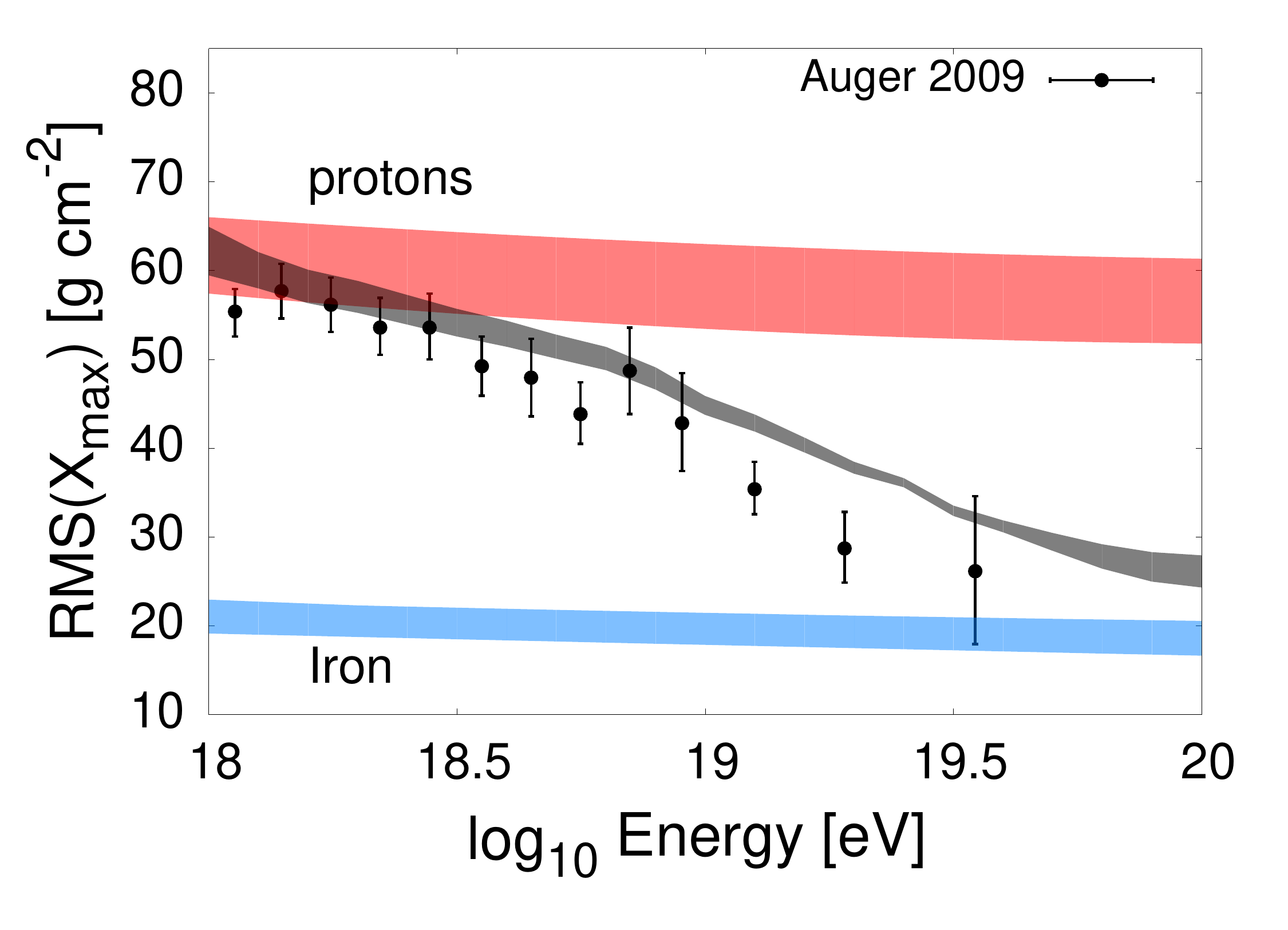}}\\
\vspace{-0.3cm}
\caption{Results for an all-silicon injection spectrum with 0.0 nG (top), 0.1 nG  (middle), and 0.3 nG (bottom) intergalactic magnetic fields. For the non-zero B-field cases, a coherence length of the magnetic field of 1~Mpc has been used.}
\label{bfield2}

{\includegraphics[angle=0,width=0.32\linewidth,type=pdf,ext=.pdf,read=.pdf]{N_Spectrum_Emax_20.5_alpha_1.6}}
{\includegraphics[angle=0,width=0.32\linewidth,type=pdf,ext=.pdf,read=.pdf]{N_Xmax_Emax_20.5_alpha_1.6}}
{\includegraphics[angle=0,width=0.32\linewidth,type=pdf,ext=.pdf,read=.pdf]{N_RMS_Emax_20.5_alpha_1.6}}\\
\vspace{-0.2cm}
{\includegraphics[angle=0,width=0.32\linewidth,type=pdf,ext=.pdf,read=.pdf]{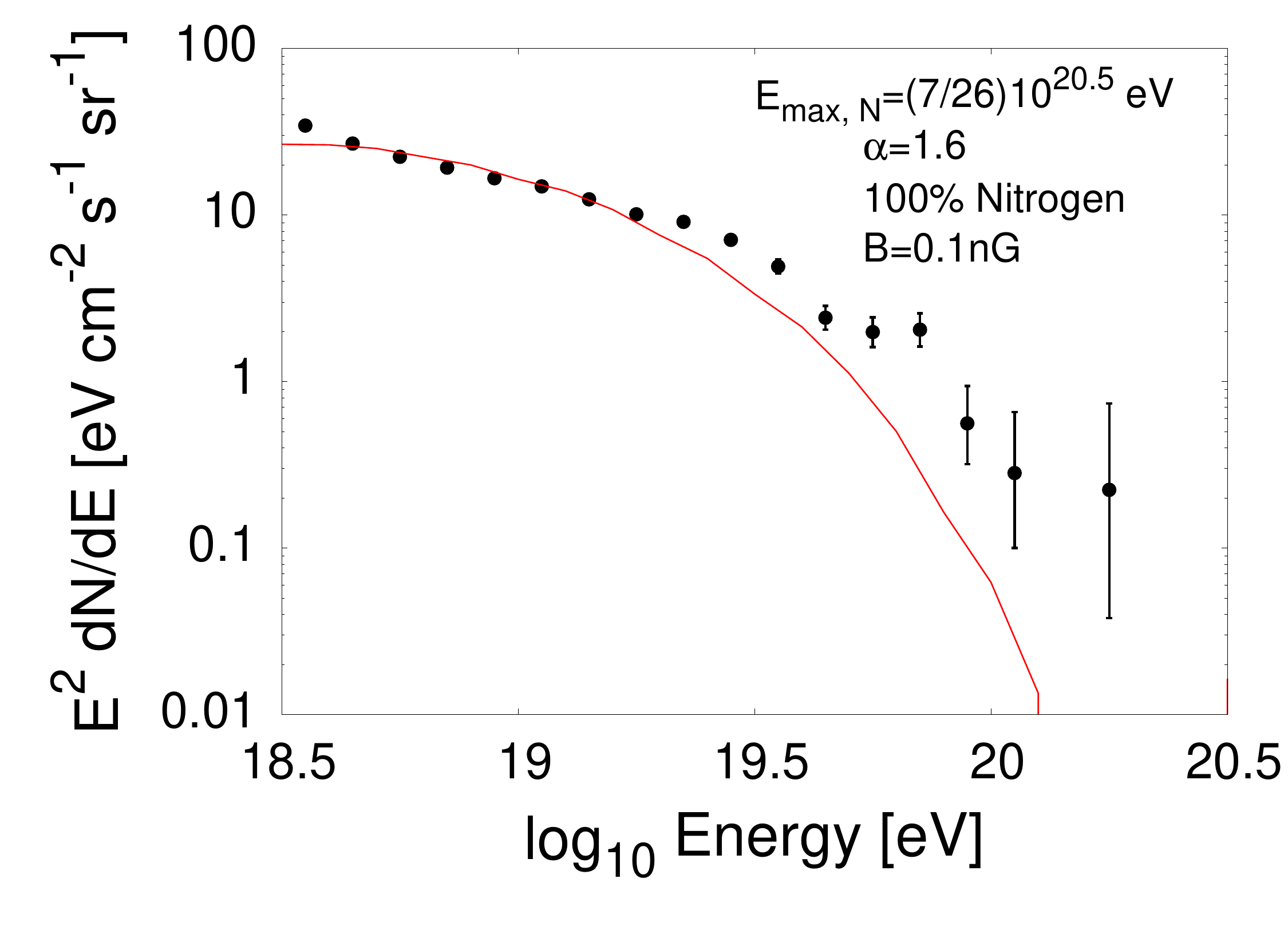}}
{\includegraphics[angle=0,width=0.32\linewidth,type=pdf,ext=.pdf,read=.pdf]{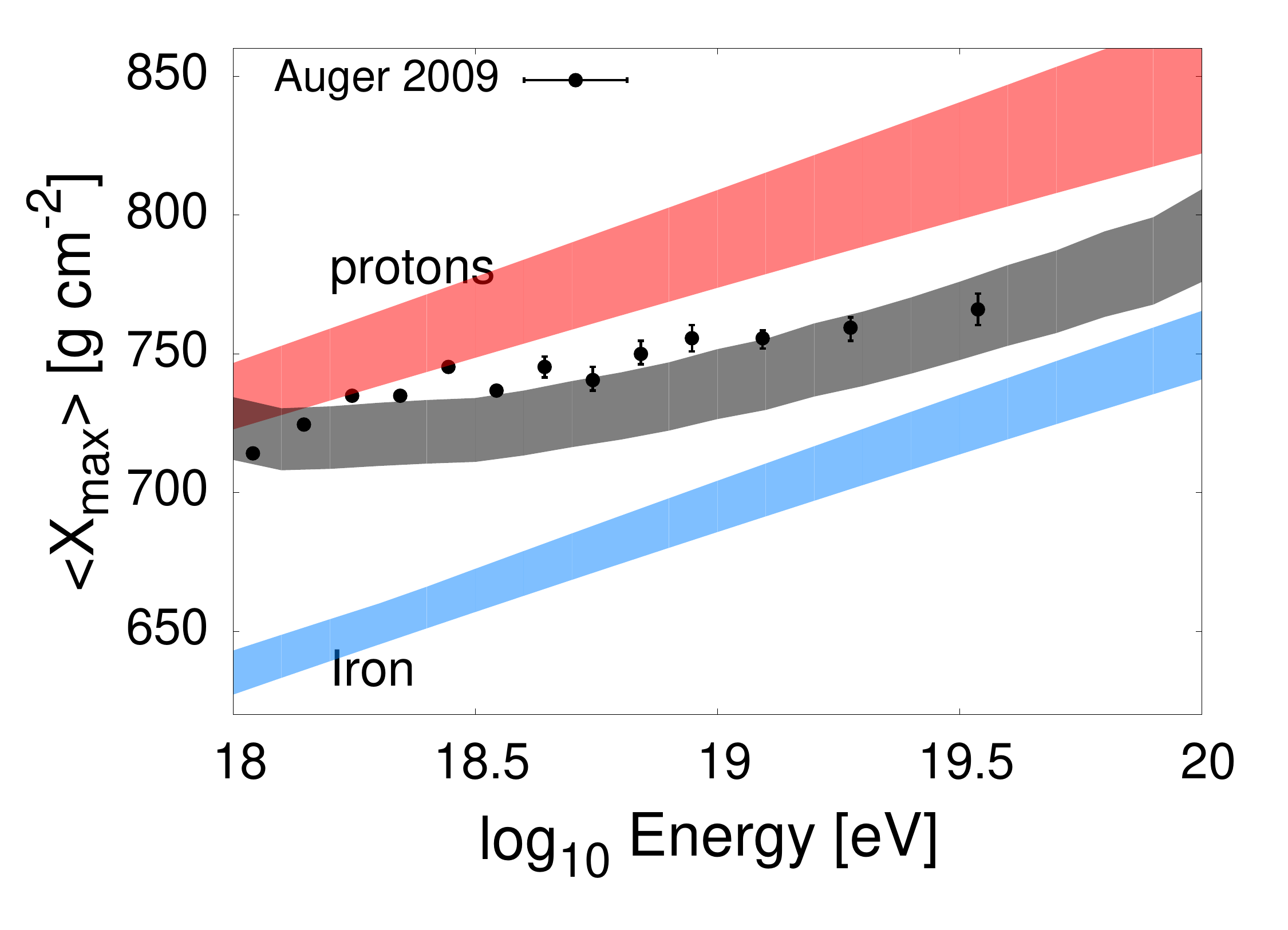}}
{\includegraphics[angle=0,width=0.32\linewidth,type=pdf,ext=.pdf,read=.pdf]{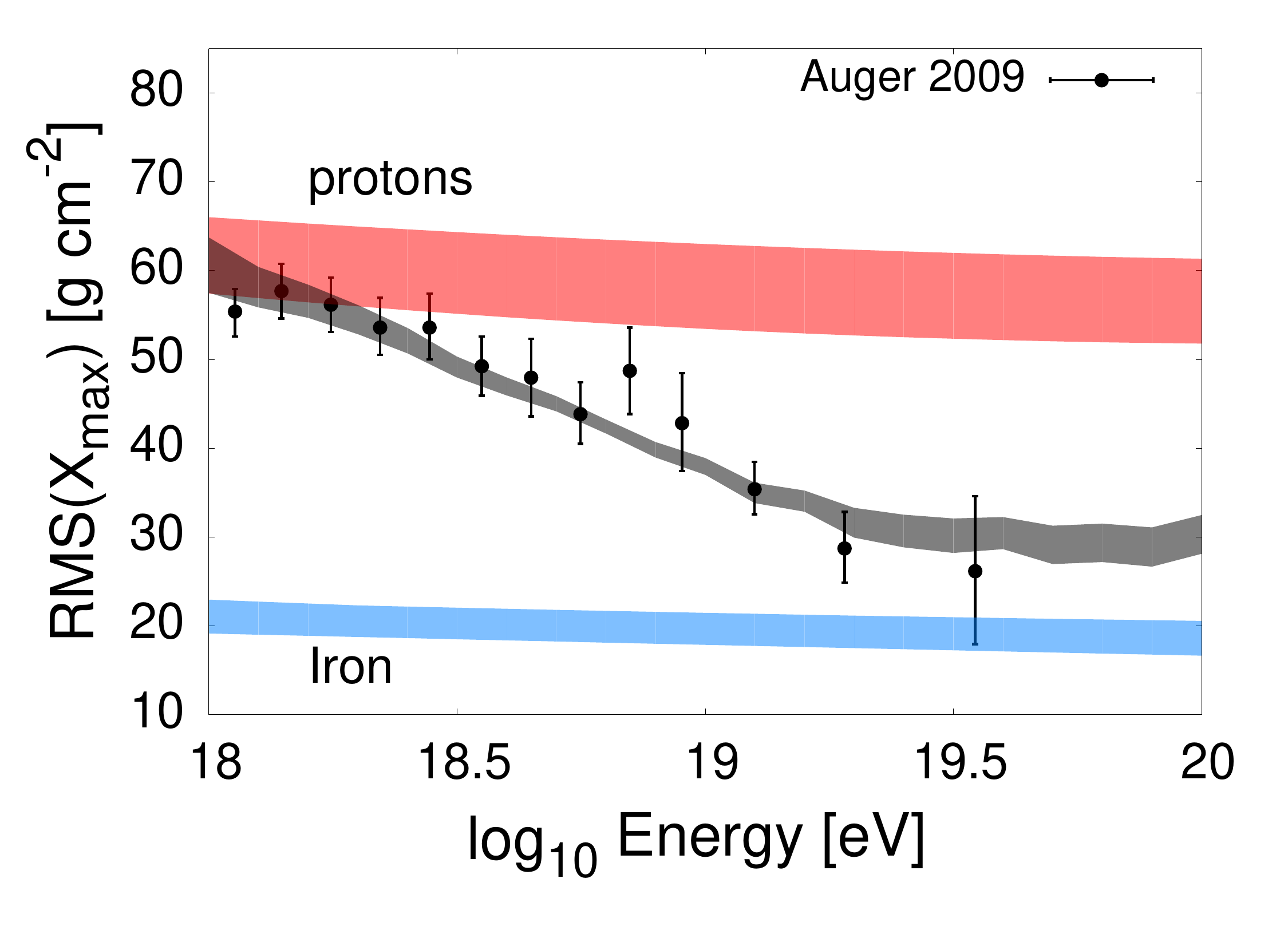}}\\
\vspace{-0.2cm}
{\includegraphics[angle=0,width=0.32\linewidth,type=pdf,ext=.pdf,read=.pdf]{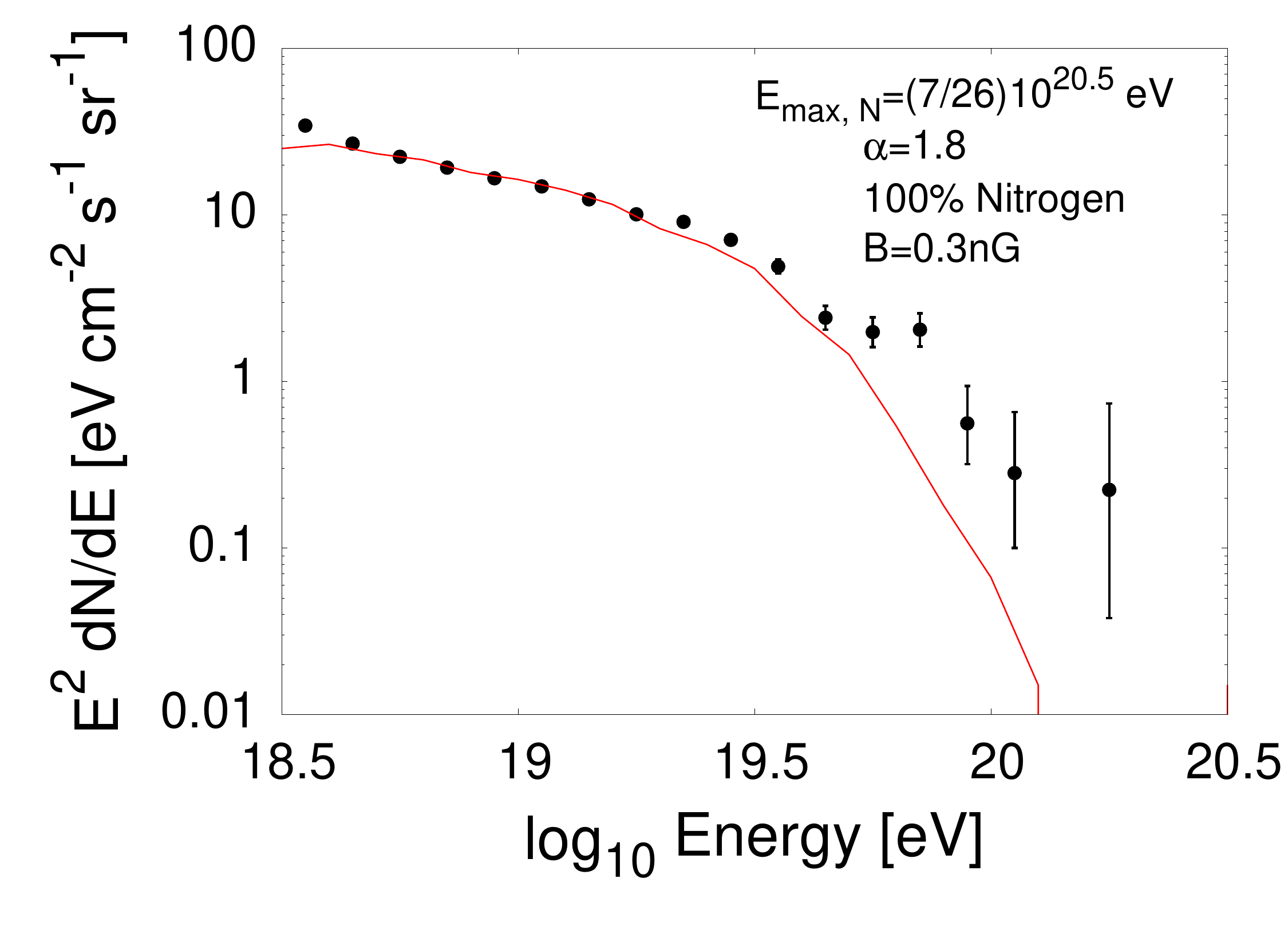}}
{\includegraphics[angle=0,width=0.32\linewidth,type=pdf,ext=.pdf,read=.pdf]{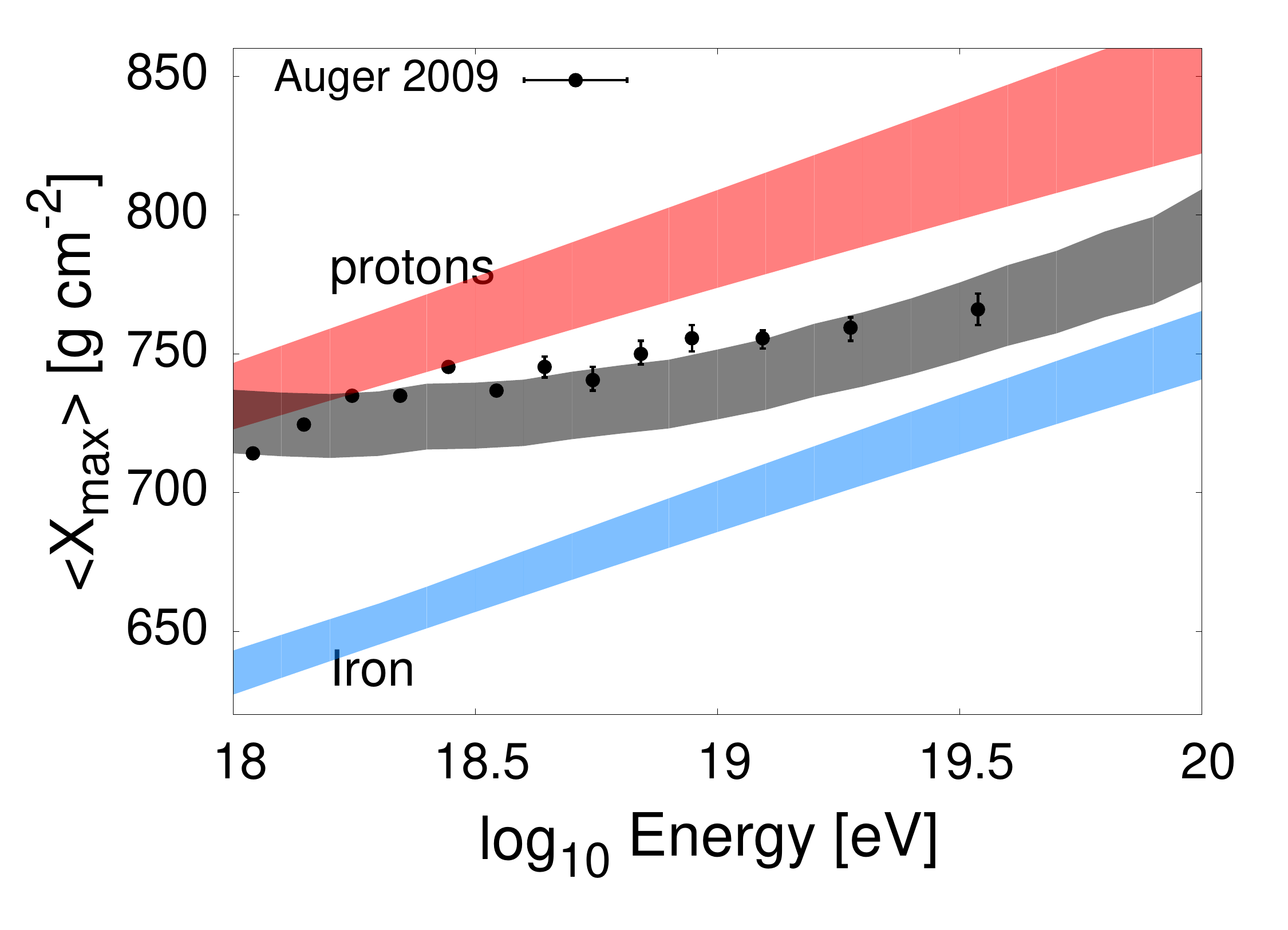}}
{\includegraphics[angle=0,width=0.32\linewidth,type=pdf,ext=.pdf,read=.pdf]{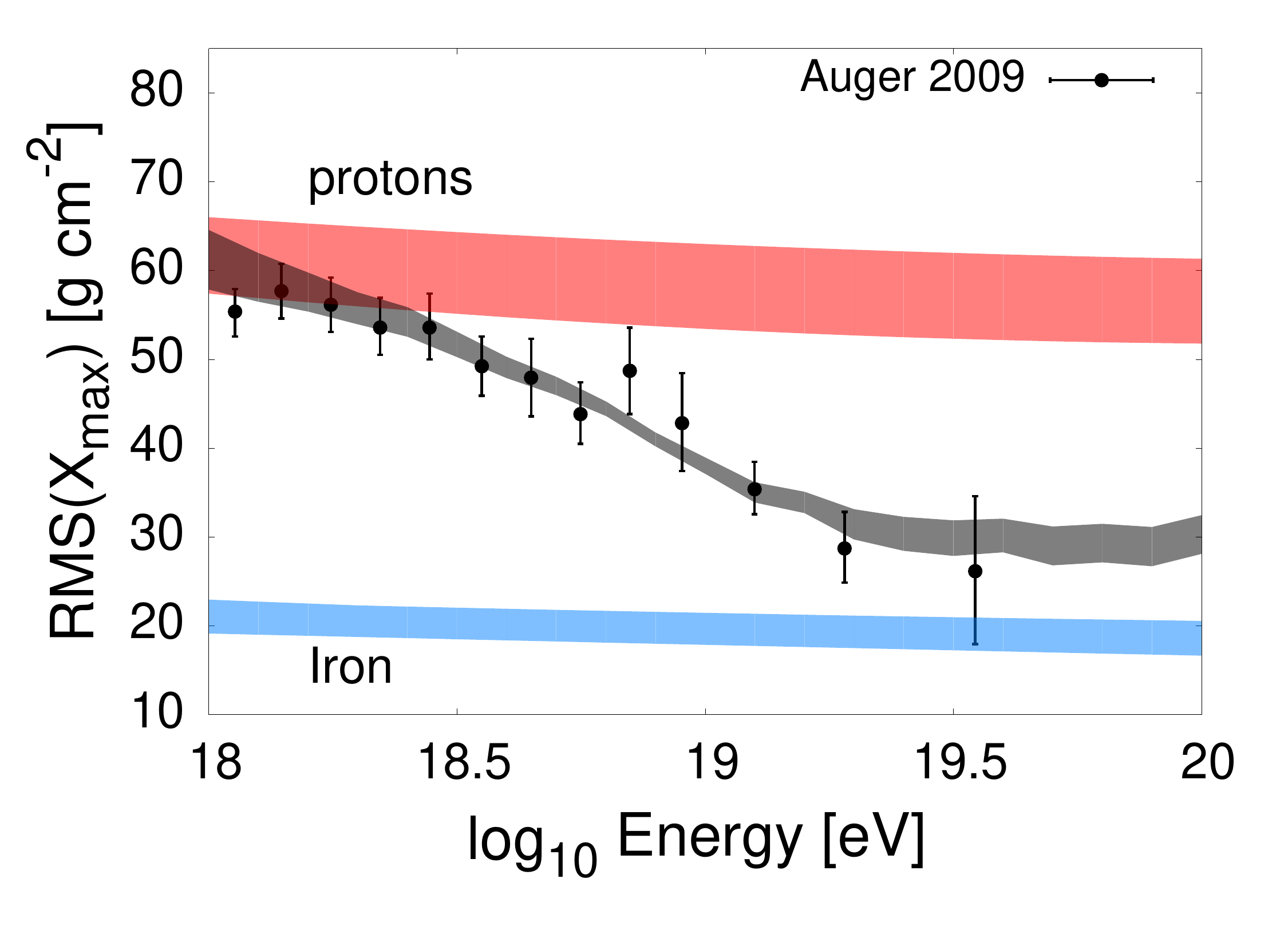}}\\
\vspace{-0.3cm}
\caption{Results for an all-nitrogen injection spectrum with 0.0 nG (top), 0.1 nG  (middle), and 0.3 nG (bottom) intergalactic magnetic fields. For the non-zero B-field cases, a coherence length of the magnetic field of 1~Mpc has been used.}
\label{bfield3}
\end{center}
\end{figure*}

Magnetic deflections impact particles with the greater electric charge and/or lower energy more than other cosmic rays, and thus effect the resulting spectral shape and composition that are predicted to reach Earth~\cite{magnetic}. In Figs.~\ref{bfield1},~\ref{bfield2}, and~\ref{bfield3}, we show our results for all-iron, all-silicon, and all-nitrogen accelerating sources, respectively, and for intergalactic magnetic fields of 0.0, 0.1, and 0.3~nG, with coherent scales of 1 Mpc for the non-zero B-field cases.  In the case of all-iron injection, the tension between the PAO's $\langle X_{\rm max} \rangle$ and $\mathrm{RMS}(X_{\rm max})$ measurements is reduced to some degree by the effects of the magnetic fields, but the spectral fit becomes worse as the field strength is increased. In the all-silicon and all-nitrogen cases, quite good agreement can be found with the PAO data. In particular, for $\sim$($B$/0.3~nG)$\times$($L_{\rm coh}$/1~Mpc)$^{1/2}$, we find an overall $\chi^2$ per degree-of-freedom of 1.45 and 1.14 for the all-silicon and all-nitrogen cases, respectively (using EPOS, which yields a better fit than the other hadronic simulations). These represent far better fits than we have found without magnetic fields.

\begin{figure*}[!]
\begin{center}
{\includegraphics[angle=0,width=0.32\linewidth,type=pdf,ext=.pdf,read=.pdf]{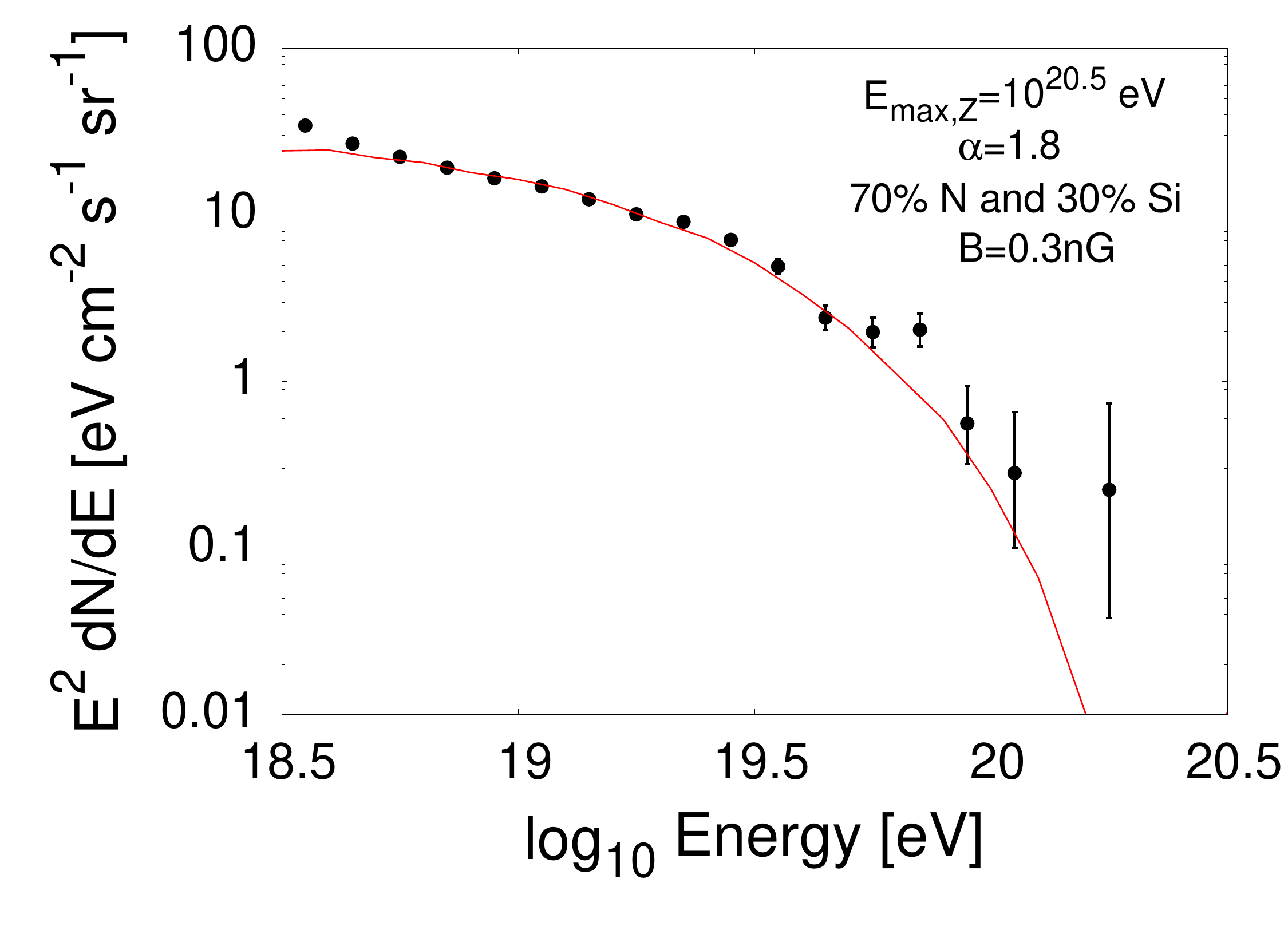}}
{\includegraphics[angle=0,width=0.32\linewidth,type=pdf,ext=.pdf,read=.pdf]{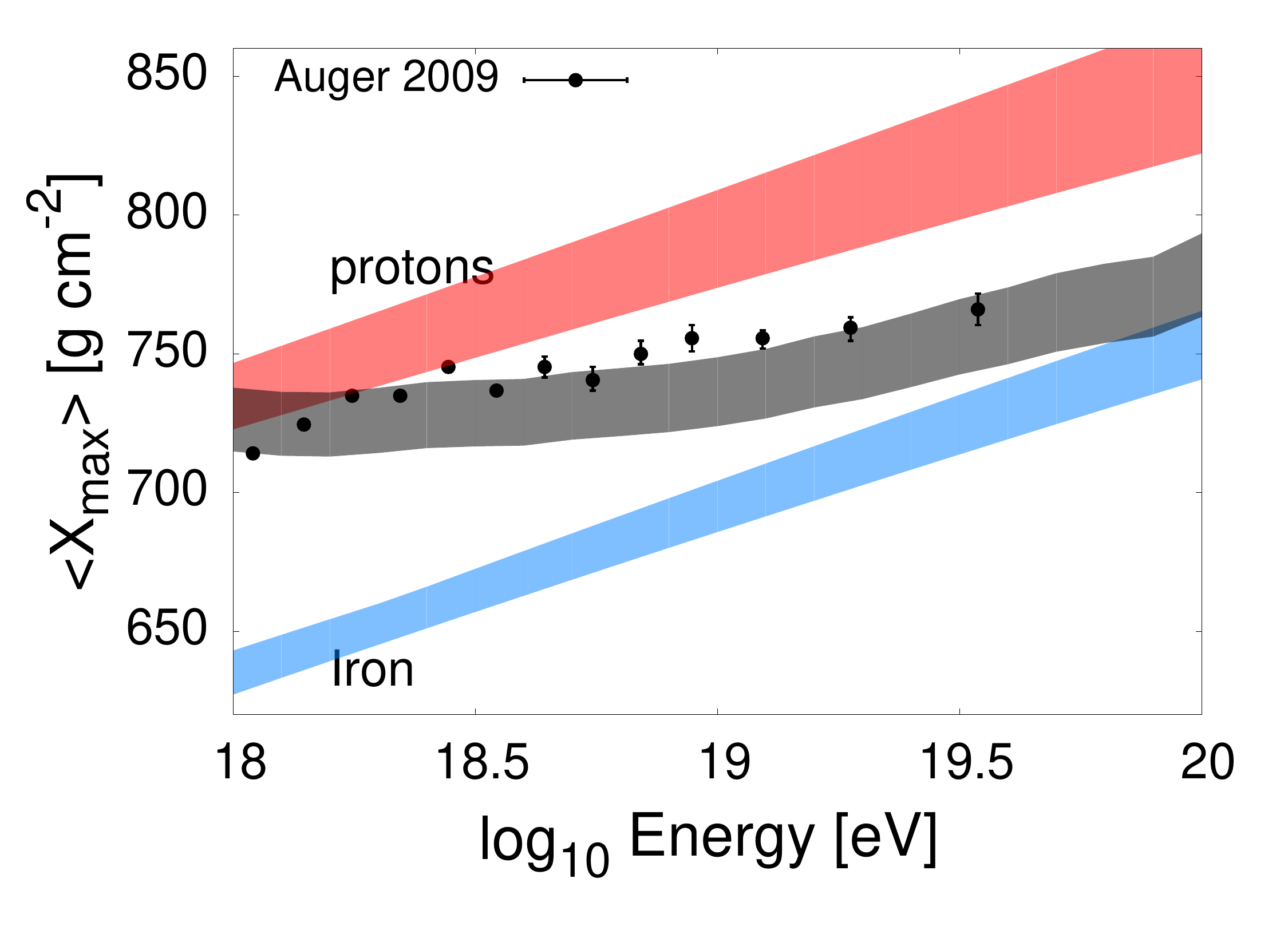}}
{\includegraphics[angle=0,width=0.32\linewidth,type=pdf,ext=.pdf,read=.pdf]{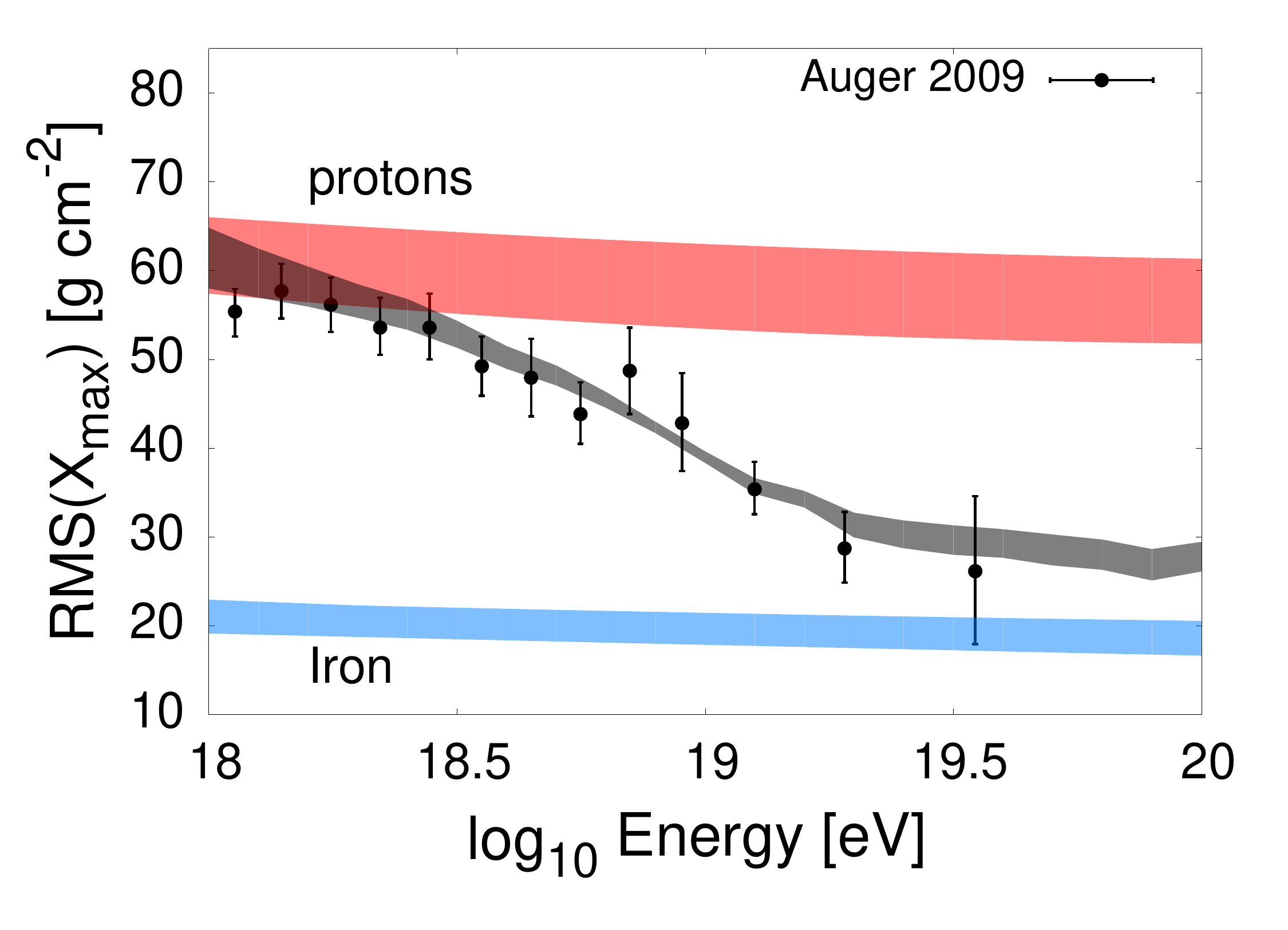}}\\
{\includegraphics[angle=0,width=0.32\linewidth,type=pdf,ext=.pdf,read=.pdf]{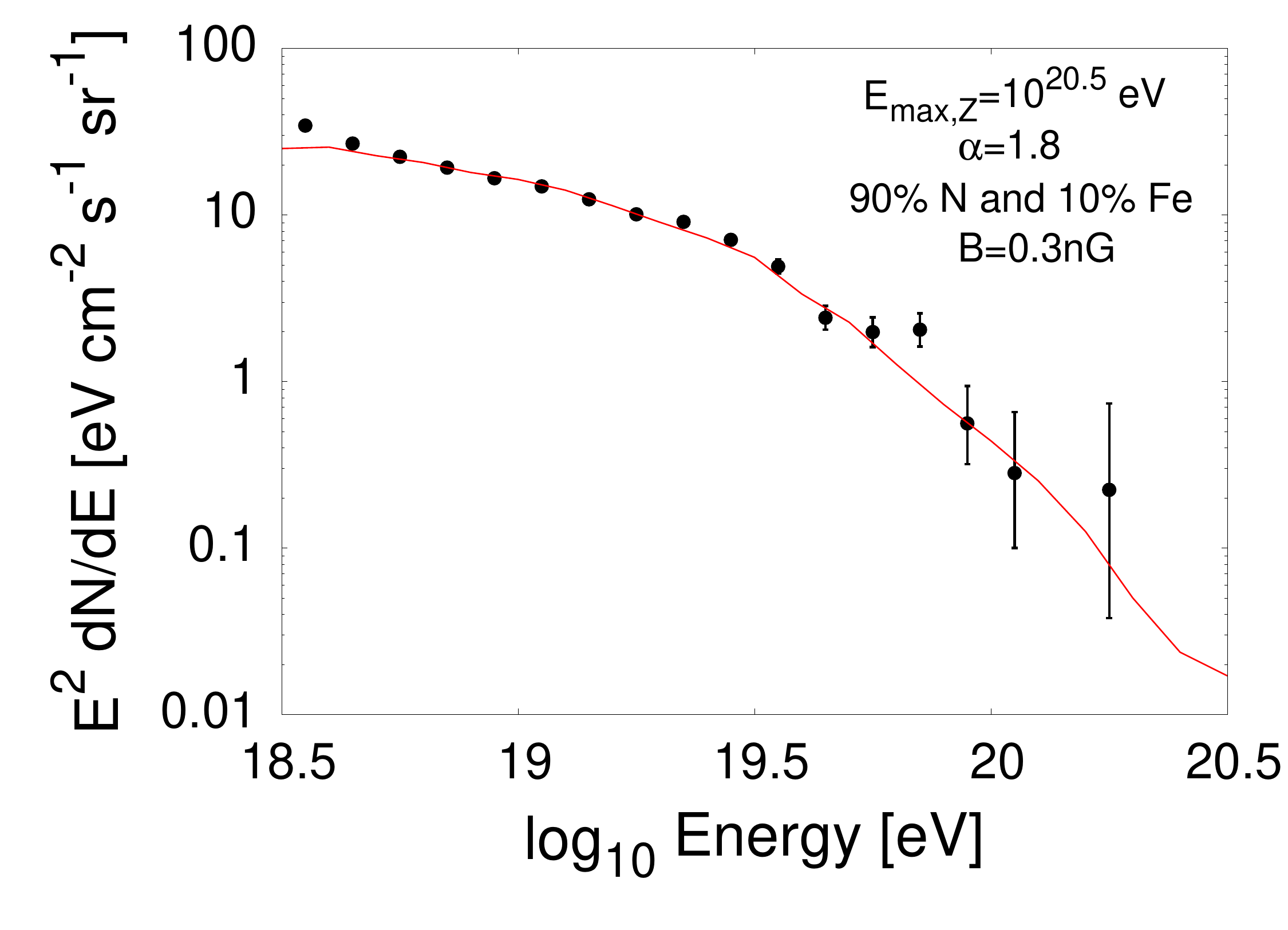}}
{\includegraphics[angle=0,width=0.32\linewidth,type=pdf,ext=.pdf,read=.pdf]{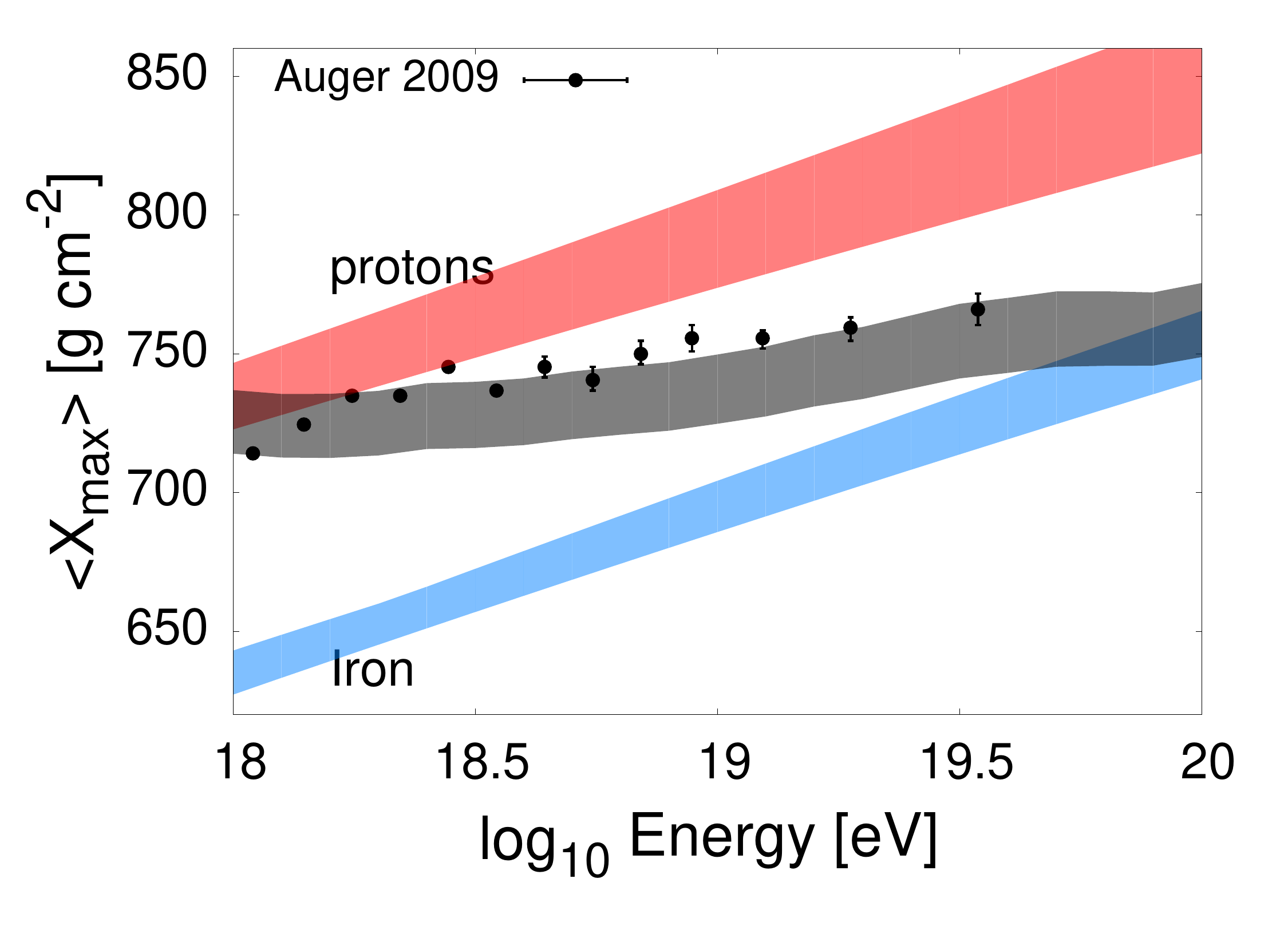}}
{\includegraphics[angle=0,width=0.32\linewidth,type=pdf,ext=.pdf,read=.pdf]{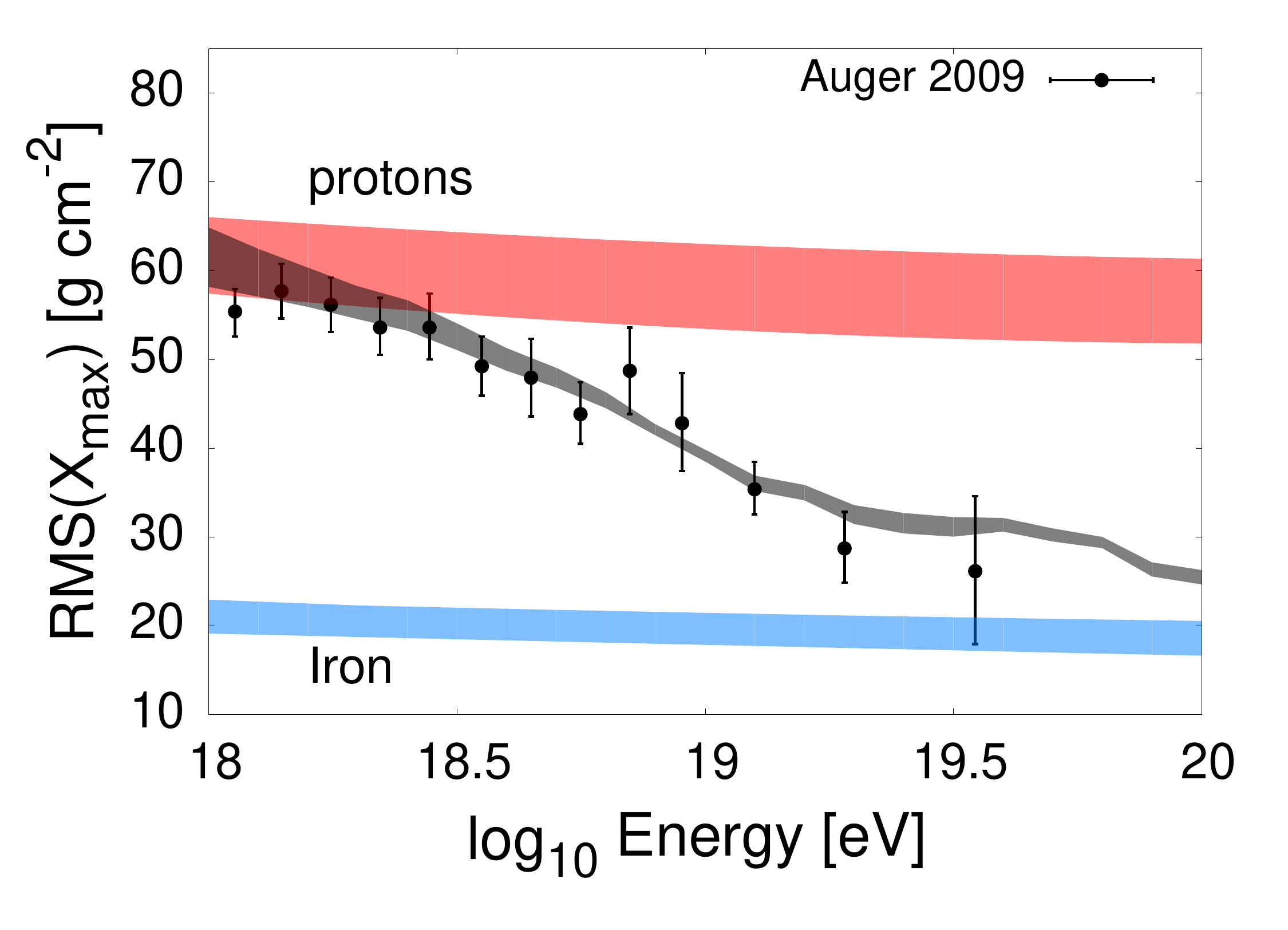}}\\
\caption{The same as in previous figures, but for injection mixtures of 70\% nitrogen and 30\% silicon (top) or 90\% nitrogen and 10\% iron (bottom), for 0.3~nG intergalactic magnetic fields with coherence lengths of 1~Mpc. These cases each provide excellent fits to the spectrum, $\langle X_{\rm max} \rangle$, and $\mathrm{RMS}(X_{\rm max})$ measurements from the PAO. See text for more details.}
\label{bmix}
\end{center}
\end{figure*}

There are also mixtures of injected nuclei that can provide excellent fits to the PAO's data if the effects of intergalatic magnetic fields are included. In Fig.~\ref{bmix}, we show results for the two mixtures that we found to provide the best overall fits to the PAO's data. As expected, a mixture of injected nitrogen and silicon provides a very good fit ($\chi^2$ per degree-of-freedom less than 1.0 for mixtures with between 10\% and 50\% silicon). We also find that injected nitrogen with a small ($\sim 10\%$) admixture of iron yields a very good fit with 0.3 nano-Gauss magnetic fields with coherence lengths of 1~Mpc. In each case considered here, including an admixture of protons in the injected flux from UHECR sources only worsens the overall quality of the fit.

To some extent, the key aspects of these results follow from the consideration of the unfolded arriving UHECR composition at Earth from the PAO data. As an example case to demonstrate this, we consider the 
$\sim$10$^{19.3}$~eV $\langle X_{\rm max} \rangle$ and $\mathrm{RMS}(X_{\rm max})$ PAO data point. For the EPOS model, this point is found to sit on the line expected for nitrogen nuclei for both $\langle X_{\rm max} \rangle$ and $\mathrm{RMS}(X_{\rm max})$. The possibility that a mixed composition may be also consistent with the data is then tested through the consideration of fractional admixtures, in multiples of 1/5, of protons, helium, nitrogen, and iron. Only the case of a small admixture of 20\% helium is found to still be able to provide agreement for both the $\langle X_{\rm max} \rangle$ and $\mathrm{RMS}(X_{\rm max})$ data (staying within the error bars of their respective data point), no other possible admixtures are found to provide such agreement with this data point in both data sets. Thus, in order to ensure such a composition arrives, the injected composition must be both relatively narrow (in charge distribution) and heavier than nitrogen.

\section{Discussion and Conclusions}
\label{conclusions}

In this article, we have studied the propagation of ultra-high energy cosmic ray nuclei in an attempt to find scenarios that provide a good fit to the observations of the Pierre Auger Observatory (PAO), including the measured spectrum, average depth of shower maximum ($\langle X_{\rm max} \rangle$), and the RMS fluctuations of the depth of shower maximum ($\mathrm{RMS}(X_{\rm max})$). 

We have found that for all values of injected maximum energy and spectral index considered for the UHECR sources, the PAO's data, for energies above 10$^{18.5}$~eV, is best accommodated in cases in which a homogeneous distribution of sources inject largely intermediate mass nuclei, ranging from roughly nitrogen to silicon. This result is found for all hadronic models considered, with the best fit results being found using the EPOS model. Furthermore, within this same range of source parameters, the presence of $\sim$($B$/0.3~nG)$\times$($L_{\rm coh}$/1~Mpc)$^{1/2}$ intergalactic magnetic fields lead to far better fits than are found in the absence of cosmic ray deflection. No good fits to the PAO's measurements were found without significant intergalactic magnetic fields.\footnote{Although magnetic fields in the vicinity of the sources could also plausibly account for this conclusion.} Along with these results we note that for the EPOS model, whose results are responsible for our best fits, no good fits were found with the in any scenario considered in which the sources of the UHECRs inject a significant ($\gsim 10\%$) fraction of protons. Should the PAO data points all be shifted up in energy by approximately 20\%, as has been suggested \cite{Engel:2007cm}, the best fit results, for all cases quoted here, are found to shift towards a heavier nuclei composition.

These results have a number of interesting and important consequences. Firstly, if the cosmic ray spectrum consists largely of intermediate mass nuclei at the highest energies, as our results suggest is the case, one would not expect any significant correlation between the arrival directions and sources of such particles. For a $\mu$G-scale Galactic Magnetic Field with a correlation lenth of $\sim$1 kpc, a $6\times 10^{19}$ eV nucleus with $Z \approx 10$ would be deflected by on the order of 10$^{\circ}$, in addition to any deflection resulting from intergalactic fields. This conclusion, perhaps, suggesting difficulties for the prospects of UHECR astronomy, for all but the most bright local sources.

Under the assumption that the UHECRs consist uniquely of protons, a ``guaranteed'' flux of UHE neutrinos is predicted to be produced as a result of their propagation. This signal, known as the cosmogenic neutrino flux~\cite{cosmogenic}, is potentially within the reach of experiments such as IceCube~\cite{icecube}, ANITA~\cite{anita}, and the PAO itself~\cite{paoneutrino}. If the the UHECR spectrum consists of heavy nuclei, however, the cosmogenic flux can be considerably suppressed~\cite{cosmogenicnuclei}. In the scenarios that we have found to fit the PAO's data, the cosmogenic neutrino spectrum is predicted to be suppressed by an order of magnitude or more relative to the all-proton case~\cite{suppress}. The neutrino flux from the sources of the UHECRs themselves is also expected to be suppressed in these scenarios~\cite{Anchordoqui:2007tn}.

The authors note that since the submission of this work, new results on the UHECR composition from $\langle X_{\rm max} \rangle$ measurements have been published from the HiRes experiment \cite{Abbasi:2009nf}. These results appear at first sight to sit in contradiction to those from the PAO. Along with the apparent contradiction between these data sets, other recent results from the Yakutsk experiment fail to resolve this confusion, with different analysis of the data arriving at different results for the UHECR composition \cite{Glushkov:2007gd,Knurenko:2007yv}, an issue whose origin may lie in the different methods used to asign energy to the events \cite{private}. Due to the strength of the conclusions that have been possible from PAO results, the conflict between these sets of results are considered a pressing issue which we hope the experimental community will thoroughly investigate.

\vspace{1.0cm}

{\it Acknowledgements:} We would like to thank both Subir Sarkar and Felix Aharonian for their invaluable contributions to this work. We would also like to thank Michael Unger for useful discussions. DH is supported by the US Department of Energy, including grant DE-FG02-95ER40896, and by NASA grant NAG5-10842.


\begin{thebibliography}{99}

\bibitem{Stecker:1969fw}
F.~W.~Stecker,
Phys.\ Rev.\ {\bf 180} (1969) 1264.

\bibitem{Puget:1976nz}
J.~L.~Puget, F.~W.~Stecker and J.~H.~Bredekamp,
Astrophys.\ J.\ {\bf 205} (1976) 638.

\bibitem{Stecker:1998ib}
F.~W.~Stecker and M.~H.~Salamon,
Astrophys.\ J.\ {\bf 512} (1999) 521.

\bibitem{us}
  D.~Hooper, S.~Sarkar and A.~M.~Taylor,
  Astropart.\ Phys.\  {\bf 27} (2007) 199
  [arXiv:astro-ph/0608085]; 
  D.~Hooper, S.~Sarkar and A.~M.~Taylor,
  Phys.\ Rev.\  D {\bf 77} (2008) 103007
  [arXiv:0802.1538 [astro-ph]].

\bibitem{debate}
  D.~Allard, A.~V.~Olinto and E.~Parizot,
  arXiv:astro-ph/0703633;
  D.~Allard, E.~Parizot and A.~V.~Olinto,
  Astropart.\ Phys.\  {\bf 27}, 61 (2007)
  [arXiv:astro-ph/0512345];
  L.~O.~Drury, J.~P.~Meyer and D.~C.~Ellison,
  arXiv:astro-ph/9905008;
  L.~N.~Epele and E.~Roulet,
  JHEP {\bf 9810}, 009 (1998)
  [arXiv:astro-ph/9808104];
  T.~Yamamoto, K.~Mase, M.~Takeda, N.~Sakaki and M.~Teshima,
  Astropart.\ Phys.\  {\bf 20}, 405 (2004)
  [arXiv:astro-ph/0312275];
  D.~Allard, E.~Parizot, E.~Khan, S.~Goriely and A.~V.~Olinto,
  Astron.\ Astrophys.\  {\bf 443}, L29 (2005)
  [arXiv:astro-ph/0505566].

\bibitem{magnetic}
  E.~Armengaud, G.~Sigl and F.~Miniati,
  Phys.\ Rev.\  D {\bf 72}, 043009 (2005)
  [arXiv:astro-ph/0412525];
  G.~Sigl and E.~Armengaud,
  JCAP {\bf 0510}, 016 (2005)
  [arXiv:astro-ph/0507656];
  G.~Bertone, C.~Isola, M.~Lemoine and G.~Sigl,
  Phys.\ Rev.\  D {\bf 66}, 103003 (2002)
  [arXiv:astro-ph/0209192];



\bibitem{hillas}
A.~M.~Hillas,
Ann.\ Rev.\ Astron.\ Astrophys.\ {\bf 22} (1984) 425.



\bibitem{anisotropy}
  J.~Abraham {\it et al.}  [Pierre Auger Collaboration],
  Science {\bf 318}, 938 (2007)
  [arXiv:0711.2256 [astro-ph]];
  J.~Abraham {\it et al.}  [Pierre Auger Collaboration],
  Astropart.\ Phys.\  {\bf 29}, 188 (2008)
  [Erratum-ibid.\  {\bf 30}, 45 (2008)]
  [arXiv:0712.2843 [astro-ph]].

\bibitem{berezinsky}
  V.~Berezinsky, A.~Z.~Gazizov and S.~I.~Grigorieva,
  Phys.\ Lett.\  B {\bf 612}, 147 (2005)
  [arXiv:astro-ph/0502550];
  R.~Aloisio, V.~Berezinsky, P.~Blasi, A.~Gazizov, S.~Grigorieva and B.~Hnatyk,
  Astropart.\ Phys.\  {\bf 27}, 76 (2007)
  [arXiv:astro-ph/0608219];
  R.~Aloisio, V.~Berezinsky, P.~Blasi and S.~Ostapchenko,
  arXiv:0706.2834 [astro-ph].


\bibitem{spectrum}
  J.~Abraham {\it et al.}  [The Pierre Auger Collaboration],
  arXiv:0906.2189;
  J.~Abraham {\it et al.}  [Pierre Auger Collaboration],
  Phys.\ Rev.\ Lett.\  {\bf 101}, 061101 (2008)
  [arXiv:0806.4302 [astro-ph]].


\bibitem{spectrumcomposition}
  D.~Allard, N.~G.~Busca, G.~Decerprit, A.~V.~Olinto and E.~Parizot,
  JCAP {\bf 0810}, 033 (2008)
  [arXiv:0805.4779 [astro-ph]].



\bibitem{xmax}
J.~Abraham {\it et al.}  [The Pierre Auger Collaboration],
Presentations for the 31st International Cosmic Ray Conference, Lodz, Poland (2009),
arXiv:0906.2319.

\bibitem{ungerSocor} 
M.~Unger, talk at SOCoR workshop, Trondheim, (2009).

\bibitem{Blumenthal}
G.~Blumenthal,
Phys.\ Rev.\ D {\bf 1} (1970) 6.

\bibitem{gzk}
 K.~Greisen,
 Phys.\ Rev.\ Lett.\  {\bf 16}, 748 (1966);
 G.~T.~Zatsepin and V.~A.~Kuzmin,
 JETP Lett.\  {\bf 4}, 78 (1966)
 [Pisma Zh.\ Eksp.\ Teor.\ Fiz.\  {\bf 4}, 114 (1966)].




\bibitem{Khan:2004nd}
E.~Khan {\it et al.},
Astropart.\ Phys.\ {\bf 23} (2005) 191.





\bibitem{ms}
M.~A.~Malkan and F.~W.~Stecker,
Astrophys.\ J.\  {\bf 555}, 641 (2001)
arXiv:astro-ph/0009500;

\bibitem{other}
  F.~Aharonian {\it et al.}  [HEGRA Collaboration],
  Astron.\ Astrophys.\  {\bf 403}, 523 (2003)
  [arXiv:astro-ph/0301437];
  A.~Franceschini, H.~Aussel, C.~J.~Cesarsky, D.~Elbaz and D.~Fadda,
  Astron.\ Astrophys.\  {\bf 378}, 1 (2001)
  [arXiv:astro-ph/0108292].

\bibitem{Stecker:2007zj}
  F.~W.~Stecker, M.~G.~Baring and E.~J.~Summerlin,
  Astrophys.\ J.\  {\bf 667} (2007) L29
  [arXiv:0707.4676 [astro-ph]].

\bibitem{PC}
Michael Unger, private communications.



\bibitem{conex} 
T.~Bergmann {\it et al.},
Astropart.\ Phys.\ {\bf 26} (2007) 420.

\bibitem{qgsjet}
N.N. Kalmykov and S.S.\ Ostapchenko, Phys.\ Atom.\ Nucl.\ {\bf 56} (1993), 346;
S.S. Ostapchenko, Nucl.\ Phys.\ Proc.\ Suppl.\ {\bf 151} (2006), 143;
N.~N.~Kalmykov, S.~S.~Ostapchenko and A.~I.~Pavlov,
Nucl.\ Phys.\ Proc.\ Suppl.\  {\bf 52B}, 17 (1997).



\bibitem{sibyll}
  E.~J.~Ahn, R.~Engel, T.~K.~Gaisser, P.~Lipari and T.~Stanev,
  arXiv:0906.4113 [hep-ph];
  R.~S.~Fletcher, T.~K.~Gaisser, P.~Lipari and T.~Stanev,
  Phys.\ Rev.\  D {\bf 50}, 5710 (1994).


\bibitem{epos}
  T.\ Pierog and K.\ Werner, Phys.\ Rev.\ Lett.\ {\bf 101}
  (2008), 171101;
  K.~Werner and T.~Pierog,
  AIP Conf.\ Proc.\  {\bf 928}, 111 (2007)
  [arXiv:0707.3330 [astro-ph]].

\bibitem{Ulrich:2009yq}
  R.~Ulrich, R.~Engel, S.~Muller, F.~Schussler and M.~Unger,
  arXiv:0906.3075 [astro-ph.HE].

\bibitem{Ulrich:2009hm}
  R.~Ulrich, R.~Engel, S.~Muller, T.~Pierog, F.~Schussler and M.~Unger,
  arXiv:0906.0418 [astro-ph.HE].

\bibitem{Wibig:2009zza}
  T.~Wibig,
  Phys.\ Lett.\  B {\bf 678} (2009) 60.

\bibitem{Wibig:2009zz}
  T.~Wibig,
  Phys.\ Rev.\  D {\bf 79} (2009) 094008.


\bibitem{Aloisio:2009sj}
  R.~Aloisio, V.~Berezinsky and A.~Gazizov,
  arXiv:0907.5194 [astro-ph.HE].

\bibitem{Engel:2007cm}
  R.~Engel  [Pierre Auger Collaboration],
  arXiv:0706.1921 [astro-ph].

\bibitem{cosmogenic}
  V.~S.~Berezinsky and G.~T.~Zatsepin,
  Phys.\ Lett.\ B {\bf 28} 423 (1969);
  Yad.\ Fiz.\  {\bf 11}, 200 (1970);
  F.~W.~Stecker,
  Astrophys.\ J.\  {\bf 228}, 919 (1979);
  C.~T.~Hill and D.~N.~Schramm,
  Phys.\ Lett.\  B {\bf 131}, 247 (1983);
  R.~Engel, D.~Seckel and T.~Stanev,
  Phys.\ Rev.\  D {\bf 64}, 093010 (2001)
  [arXiv:astro-ph/0101216];
  Z.~Fodor, S.~D.~Katz, A.~Ringwald and H.~Tu,
  JCAP {\bf 0311}, 015 (2003)
  [arXiv:hep-ph/0309171].



\bibitem{icecube}
 H.~Landsman, L.~Ruckman and G.~S.~Varner  [IceCube Collaboration],
{\it Prepared for 30th International Cosmic Ray Conference (ICRC 2007), Merida, Yucatan, Mexico, 3-11 Jul 2007};
M.~Ackermann {\it et al.}  [IceCube Collaboration],
  Astrophys.\ J.\  {\bf 675}, 1014 (2008)
  [arXiv:0711.3022 [astro-ph]].

\bibitem{anita}
  P.~W.~Gorham {\it et al.}  [ANITA collaboration],
  Phys.\ Rev.\ Lett.\  {\bf 103}, 051103 (2009)
  [arXiv:0812.2715 [astro-ph]].

\bibitem{paoneutrino}
  J.~Abraham {\it et al.}  [The Pierre Auger Collaboration],
  Phys.\ Rev.\ Lett.\  {\bf 100}, 211101 (2008)
  [arXiv:0712.1909 [astro-ph]].


\bibitem{cosmogenicnuclei}
  D.~Hooper, A.~Taylor and S.~Sarkar,
  Astropart.\ Phys.\  {\bf 23}, 11 (2005)
  [arXiv:astro-ph/0407618];
  M.~Ave, N.~Busca, A.~V.~Olinto, A.~A.~Watson and T.~Yamamoto,
  Astropart.\ Phys.\  {\bf 23}, 19 (2005)
  [arXiv:astro-ph/0409316];
  D.~Allard {\it et al.},
  JCAP {\bf 0609}, 005 (2006)
  [arXiv:astro-ph/0605327].

\bibitem{suppress}
  L.~A.~Anchordoqui, H.~Goldberg, D.~Hooper, S.~Sarkar and A.~M.~Taylor,
  Phys.\ Rev.\  D {\bf 76}, 123008 (2007)
  [arXiv:0709.0734 [astro-ph]].

\bibitem{Anchordoqui:2007tn}
  L.~A.~Anchordoqui, D.~Hooper, S.~Sarkar and A.~M.~Taylor,
  Astropart.\ Phys.\  {\bf 29} (2008) 1
  [arXiv:astro-ph/0703001].

\bibitem{Abbasi:2009nf}
  R.~U.~Abbasi {\it et al.},
  arXiv:0910.4184 [Unknown].

\bibitem{Glushkov:2007gd}
  A.~V.~Glushkov, I.~T.~Makarov, M.~I.~Pravdin, I.~E.~Sleptsov, D.~S.~Gorbunov, G.~I.~Rubtsov and S.~V.~Troitsky,
  JETP Lett.\  {\bf 87} (2008) 190
  [arXiv:0710.5508 [astro-ph]].

\bibitem{Knurenko:2007yv}
  S.~P.~Knurenko, A.~V.~Sabourov and I.~Y.~Sleptsov,
  arXiv:0711.2130 [astro-ph].

\bibitem{private}
  A.~V.~Glushkov, private communication






























\end{thebibliography}
\end{document}